\documentclass{JFM-FLM_Au}

\usepackage{bm}
\usepackage{url}

\usepackage{amsmath}
\allowdisplaybreaks

\usepackage{upgreek}

\renewcommand{\Delta}{\Updelta}

\usepackage{xcolor}
\usepackage{soul}
\usepackage{hyperref}

\soulregister{\ref}{7}
\soulregister{\pageref}{7}
\soulregister{\eqref}{7}
\soulregister{\citet}{7}
\soulregister{\citep}{7}

\newtheorem{definition}{Definition}

\newtheorem{problem}{Problem}

\lefttitle{Z. Hao and J. Jim\'enez}
\righttitle{Journal of Fluid Mechanics}

\title{Restarts of bursts in turbulence in a log-minimal channel}

\author{Zengrong Hao\aff{1} \and Javier Jim\'enez\aff{1}}

\affiliation{\aff{1}School of Aeronautics and Space Engineering, Universidad Polit\'ecnica de Madrid, 28040 Madrid, Spain}

\corresau{Zengrong Hao, z.hao@upm.es}

\begin{document}
\maketitle

\begin{abstract}
Recent evidence on the sustainment of wall-normal-velocity bursts in wall-bounded turbulence \citep[e.g.,][] {Jimenez2022} challenges the classical streak-dependent picture, suggesting that the problem should be approached relying on no \textit{a priori} knowledge regarding other flow structures. This paper discusses the restarts of bursts in a log-minimal channel within the framework of a linearised Navier-Stokes system with forcing terms encapsulating the nonlinear effects of all other structures. Two generic issues are addressed. The first concerns the conditions for burst-restart-like solutions for the forced linearised system itself. We formulate optimisation problems to understand the `minimal requirements' for burst restarting. The solutions illustrate three conceptual periods in a typical restarting process, distinguished by the behaviour of spanwise vorticity structures: breakup, counter-rotating catch-up, and co-rotating catch-up. External forces promote this process by breaking up forward-inclined vortices and merging co-rotating, catching-up vortices. A quantity termed linearly available energy (LAE) is accordingly proposed to parameterise the restarting process. The second issue concerns the contributory features to the observed burst restarts in real turbulence. We show that an essential role of nonlinearity in restarting a burst is to increase a decaying state's LAE to a level sufficient for the onset of the subsequent burst. Flow patterns extracted during the restarting stage exhibit breakup and merging effects, both facilitated by nonlinearity. This suggests that the two effects observed in both the linearised models and real turbulence are manifestations of real flow structures that cause burst restarts.

\end{abstract}



\section{Introduction}
\label{section:Introduction}

Fluctuations of the wall-normal velocity are one of the hallmarks of wall-bounded turbulence. They are directly responsible for the transfer of mass, momentum, and other flow-carried information across different wall distances. These fluctuations are characterised by intense events that occur intermittently in time and sporadically in space. Such events are typically termed `bursts', considering that each individual event is relatively short-lived and spatially local compared with the structures of streamwise velocity, which are organised as long-lived, streamwise elongated streaks \citep{Jimenez2005}.

Our understanding of the mechanisms of bursts has improved significantly in the past decade. In minimal simulation channels whose sizes are comparable to the typical length scales of bursts in the logarithmic layer, \citet{Jimenez2013} found that the bursts have many statistical properties consistent with the classical Orr process \citep{Orr1907}. In general, the latter states that in a parallel shear flow, an initially backward-inclined disturbance is gradually tilted forward by the shear, during which the cross-shear velocity component (wall-normal component for wall-bounded flows) of the disturbance is sequentially amplified and attenuated over time, peaking when the disturbance is normal to the shear. The change in the velocity amplitudes during this forward-tilting process is a natural result of the continuity constraint, and the process is essentially linear in the sense that the disturbance under consideration interacts only with the mean shear flow but not with other disturbances. The connection between this linearised Orr process and the bursts in wall-bounded turbulence was further confirmed by \citet{Jimenez2015} considering individual burst events in log-minimal channels. He showed that the development of a strong burst event largely follows the amplitude-inclination relationship indicated by the linearised Orr process. The above analyses were later extended by \citet{Encinar2020} to large-scale channels where bursts arise sporadically in space, and the relevance of the linearised Orr process to the burst events was also proven. 

With this encouraging understanding of the burst mechanism, which interprets the development of a burst event as a shear-driven forward-tilting process, the question naturally arises as to how a burst is restarted -- an apparent backward-tilting process -- to close a burst cycle and thus sustain wall-bounded turbulence. Historically, this question has usually been regarded as part of the more general problem termed the self-sustaining process, around which many theories have been proposed from various perspectives. 

Many analyses have shown that flows with long streaks of streamwise velocity can develop certain types of instabilities \citep{Jimenez1994,Hamilton1995,Andersson2001,Schoppa2002}, thereby generating bursting quasi-streamwise vortices or rollers. This vortex/roll-streak structure is qualitatively consistent with some sets of invariant solutions of the Navier-Stokes equations \citep{Nagata1990,Waleffe1997,Waleffe2001,Kawahara2001,Jimenez2001}. Linearised methods such as resolvent analysis \citep{Mckeon2010,Moarref2013,Mckeon2017,Bae2021} approach the self-sustaining mechanism by identifying external forcing modes that can effectively excite energetic motions via the linearised Navier-Stokes operator. They showed that the most amplified motions have roll-streak structures. Many reduced-order models for the self-sustaining process also assume the central role of the streaks. Examples include the quasi-linear models \citep{Farrell2012,Thomas2014,Farrell2016} and their generalised form \citep{Marston2016,Hernandez2022a,Hernandez2022b} and the constrained energy transfer models in \citet{Lozano2021}, all indicating that some type of interaction between streaky mean flow and smaller-scale fluctuations is needed to sustain bursts in wall-bounded turbulence.

Despite the above progress, the prospect of a general self-sustaining theory centring on streaks can be questioned. In fact, numerical studies in recent years have demonstrated that even if all long streaks are artificially damped, the bursts of wall-normal velocity are essentially sustained as in natural turbulence \citep{Jimenez2022}. This finding, with increasing evidence \citep[e.g.,][]{Martinez2023,Martinez2025,Maeyama2023}, challenges the traditional picture that regards both streaks and bursts as mutually induced ingredients in an integrated self-sustaining process. Instead, it suggests a division of the process into two separate sub-processes: the bursting cycle itself, whose individual developments and restarts do not directly depend on the presence of streaks, and the emergence of streaks, which are by-products of collections of bursts. Targeted at the second sub-process, \citet{Jimenez2025} recently investigated the regeneration of long streaks from initial streak-free fields and speculated on the regeneration mechanism. The present paper is targeted at the first sub-process, in particular, the restarts of bursts, a problem that should be approached without relying on specific \textit{a priori} knowledge -- or bias -- regarding other structures.

Methodologically, such a structure-neutral approach to the problem of burst restarting can be developed from two complementary ends of a methodological spectrum, with the ultimate goal of bridging the gap between them. One end of the spectrum is the full Navier-Stokes equations. Based on them, one can implement Monte Carlo experiments across different prescribed perturbations to search for those that affect burst restarts. In recent years, some Monte Carlo methodologies have been proposed to identify the perturbations that have `significant consequences' in general for two-dimensional turbulence \citep{Jimenez2018b,Jimenez2020}, three-dimensional isotropic turbulence \citep{Encinar2023}, and wall-bounded turbulence \citep{Osawa2024}. An adaptation of these general methodologies can be expected in the future to search for the perturbations  specifically relevant to burst restarts in wall-bounded turbulence. 

The other end of the spectrum consists of simplified models. In particular, the well-established linearised Orr process, which essentially interprets the development of a burst event, can be used as a starting point to approach the restart of a burst through extensions. Just as burst development can be regarded as a solution to a linearised Navier-Stokes system, a burst cycle can likewise be regarded as a solution to the same linearised system, but with external forcing that acts primarily on the restart phase of the cycle. In real turbulence governed by the full Navier-Stokes equations, such forcing is a term representing the net effect of nonlinearity. This term does not by itself identify the nonlinear mechanism for the burst cycle; however, within the forced linearised system, it is a manifestation of the nonlinearity and an immediate cause of burst restarts, thereby providing a basis for inferring the underlying nonlinear mechanism. Under this concept, the mechanism of burst restarting can be approached by addressing the following three generic questions at different levels.

The first question is: for the forced linearised system itself, what conditions are required to yield solutions that exhibit burst restarts? This question concerns the `minimal requirements' for burst restarting and can be addressed through properly posed optimisation problems in mathematics. Such optimal solutions, although apparently far removed from real flows, filter out `non-essential' features and, through cautious interpretation, provide insight into the most relevant elements underlying a phenomenon of interest. Previous examples associating optimal linearised solutions with flow physics include \citet{Butler1993,delAlamo2006,Jimenez2013}, which modelled burst development by seeking the maximum unforced transient growth; \citet{Hwang2010,Mckeon2010}, which identified long-lived structures by seeking the most amplified mode under time-harmonic forces; and \citet{Ballouz2024,Nishimoto2025}, which represented features in non-stationary flows using the most amplified mode under time-local forces. 
In the present paper, focusing on a log-minimal channel, we pose several optimisation problems in section \ref{section:Requirements} to obtain insights into the `minimal requirements' for burst restarts.

The second question is: for real turbulence, what specific features of flows and nonlinear terms contribute to the observed burst restarts? The observed flows and nonlinear terms in real turbulence can be regarded as `sufficient' to induce the observed burst restarts, but they are probably not `necessary' in their full complexity. The features that genuinely contribute to burst restarts are instead obscured by the chaotic behaviour of turbulence. In the present paper, 
building on the insights in \S\ref{section:Requirements}, 
we discuss in \S\ref{section:RealTurb} such `contributory features' to burst restarts in real turbulence by extracting relevant information from turbulence data.

The third question is: what structures of real turbulence manifest the above `contributory features' and are therefore causal structures for burst restarts in real turbulence? This question, whose answer will essentially bridge the gap between the forced linearised framework and the Navier-Stokes equations, is left for future work.

The remainder of the present paper begins with an introduction to the database of minimal-channel turbulence under consideration and the corresponding forced linearised system in \S\ref{section:Methodology}. This is followed by discussions of burst restarts in the forced linearised system in \S\ref{section:Requirements} and in real turbulence in \S\ref{section:RealTurb}, which address the first and second questions introduced above, respectively. Our conclusions and prospective efforts to address the third question are presented in \S\ref{section:Conclusions}.

\section{Turbulence database and the forced linearised system}
\label{section:Methodology}

Subsection \ref{subsection:Database} describes the minimal-channel turbulence database, from which the time-averaged mean velocity profile will be used in \S\ref{section:Requirements} and the full turbulence data will be used in \S\ref{section:RealTurb}. Subsection \ref{subsection:Linearisation} then introduces the forced linearised system corresponding to the minimal-channel turbulence, which forms the basis for \S\ref{section:Requirements}.

\subsection{Turbulence database}
\label{subsection:Database}

Our analysis focuses on the turbulent flow in a log-minimal channel, for which the database was provided in \citet{Jimenez2023}. The flow is between two parallel walls placed at $y=0$ and $y=2h$, respectively, and is driven by a pressure gradient in the streamwise ($x$) direction. The friction Reynolds number is $\Rey_\tau=hu_\tau/\nu=950$, where $u_\tau$ is the friction velocity and $\nu$ is the molecular kinematic viscosity of the fluid. The flow is spatially periodic in the streamwise and spanwise ($z$) directions, with periods $L_x=\pi h/2$ and $L_z=\pi h/4$, respectively, which are approximately the minimal sizes to sustain turbulence in the range $y/h\approx0.2$ - $0.6$. The unit vectors in $x$, $y$, and $z$ are denoted by $\boldsymbol{e}_x$, $\boldsymbol{e}_y$, and $\boldsymbol{e}_z$, respectively. The average of the streamwise velocity over $x$, $z$, and time $t$ is denoted by $U=U(y)$, and the bulk velocity is $U_\textit{bulk}=\left(\int_{0}^{h}U\mathrm{d}y\right)/h$. The fluctuation velocity vector is denoted by $\boldsymbol{u}=u\boldsymbol{e}_x+v\boldsymbol{e}_y+w\boldsymbol{e}_z$ and the corresponding vorticity is $\boldsymbol{\omega}=\bnabla\times\boldsymbol{u}=\omega_x\boldsymbol{e}_x+\omega_y\boldsymbol{e}_y+\omega_z\boldsymbol{e}_z$. The direct numerical simulation code uses dealiased Fourier discretisation in $x$ and $z$ and Chebyshev discretisation in $y$, as in \citet{Kim1987}. The number of dealiased Fourier modes in $x$ and $z$ is $N_x=N_z=192$, and the number of Chebyshev collocation points in $y$ is $N_y=385$. The database contains flow fields from $t=0$ to $t\approx650\,h/u_\tau$, with a sampling time interval of approximately $0.025\,h/u_\tau$. Due to the lack of large scale structures in the central part of this channel, the flows below and above the centre plane are regarded as independent realisations, resulting in around $5\times10^4$ snapshots. More details can be found in \citet{Jimenez2023}.

A variable $\xi=\xi(x,y,z,t)$ in this spatially periodic channel can be expanded as a Fourier series,
\begin{equation}
\label{eq:defFourier}
    \xi(x,y,z,t) = \sum_{n_x,n_z} \hat{\xi}_{n_x,n_z}(y,t) \exp\left[\,{\mathrm{i}\,(k_xx+k_zz)}\,\right],
\end{equation}
where the wavenumbers are $k_x=2\pi n_x/L_x$ and $k_z=2\pi n_z/L_z$, with integer $n_x$ and $n_z$. The magnitude of a wavevector $\boldsymbol{k}=k_x\boldsymbol{e}_x+k_z\boldsymbol{e}_z$ is denoted by $k=|\boldsymbol{k}|=\sqrt{k_x^2+k_z^2}$. For a Fourier mode $(n_x,n_z)$, we denote the coordinate parallel to the wavevector by $x_{/\!/}$ and the coordinate perpendicular to the $x_{/\!/}$-$y$ plane by $x_\bot$. The corresponding velocity components satisfy $\hat{u}_{/\!/}=-\partial_y\hat{v}/(\mathrm{i}k)$ and $\hat{u}_\bot=-\hat{\omega}_y/(\mathrm{i}k)$. 
The energy norm $|\!| \cdot |\!|_E$ for a Fourier mode with the velocity vector $\hat{\boldsymbol{u}}$ is defined by averaging the kinetic energy over the entire channel $y\in[0,2h]$, i.e.,
\begin{eqnarray}
\label{eq:defEnergyNorm}
    |\!|\,\hat{\boldsymbol{u}}\,|\!|_E & = & \sqrt{\frac{1}{2h}\int_{0}^{2h}\left(\,|\hat{u}|^2+|\hat{v}|^2+|\hat{w}|^2\,\right)\mathrm{d}y} \nonumber\\
    & = & \sqrt{\frac{1}{2h}\int_{0}^{2h}\left(\,|\hat{v}|^2 + \left\vert\frac{\partial_y\hat{v}}{k}\right\vert^2 + \left\vert\frac{\hat{\omega}_y}{k}\right\vert^2\,\right)\mathrm{d}y}\,.
\end{eqnarray}
 
The wall-parallel dimensions of this log-minimal channel are comparable to the characteristic length scales of a single log-layer burst event. Such an event can therefore be represented by the lowest-order Fourier modes, including the straight mode $(n_x,n_z)=(1,0)$ and the oblique mode $(n_x,n_z)=(1,1)$ (or $(1,-1)$). These two modes are accordingly selected as simplified representations of bursts in the present study. 
This simplification should not be taken to imply that the underlying mechanism of these channel-scale bursts is intrinsically different from that of smaller-scale wall-normal fluctuations (whether or not they are also termed `bursts'). \citet{Encinar2020} showed that, over a broad range of scales, the amplification-decaying stage of wall-normal fluctuations (represented by local wave packets rather than by higher-order Fourier modes) can be reasonably interpreted by the linearised Orr mechanism at the statistical level, although smaller-scale fluctuations may be less coherent at the level of individual events \citep{Jimenez2015,Jimenez2023}.

Following \citet{Jimenez2023}, we are particularly interested in the summary properties in the band $y\in[y_\mathrm{low},y_\mathrm{upp}]$ with $y_\mathrm{low}\approx40\nu/u_\tau$ and $y_\mathrm{upp}\approx0.6h$, which excludes the viscous sublayer and the central region where turbulence is not well represented in this small channel. Within this band, a Fourier variable $\hat{\xi}(y)$ has the average intensity defined as 
\begin{equation}
\label{eq:defIntensity}
    I_{\xi} = \sqrt{ \frac{1}{y_\mathrm{upp}-y_\mathrm{low}}\int_{y_\mathrm{low}}^{y_\mathrm{upp}} |\,\hat{\xi}\,|^2 \,\mathrm{d}y }
\end{equation}
and the average inclination angle defined as
\begin{equation}
\label{eq:defIncline}
    \mathit{\Psi}_{\xi} = \tan^{-1} \left[ \frac{-\mathrm{Im}\left(\int_{y_\mathrm{low}}^{y_\mathrm{upp}}\hat{\xi}^*\partial_y\hat{\xi} \,\mathrm{d}y\right)}{k_x\int_{y_\mathrm{low}}^{y_\mathrm{upp}}|\,\hat{\xi}\,|^2\,\mathrm{d}y} \right],
\end{equation}
where `$*$' denotes the complex conjugation for a scalar or the Hermitian conjugation for a tensor, and `$\mathrm{Im}$' stands for the imaginary part. This angle is positive for forward-inclined structures and negative for backward-inclined ones. The time scale corresponding to the mean shear rate in this band is $\mathcal{S}^{-1}=(y_\mathrm{upp}-y_\mathrm{low})/(U(y_\mathrm{upp})-U(y_\mathrm{low}))\approx0.074\,h/u_\tau$.

Following \citet{delAlamo2009}, we write the harmonic wave in physical space corresponding to a Fourier variable $\hat{\xi}(y,t)$ as $\xi(x,y,z,t)$ $=$ $\hat{\xi}(y,t)\exp[\mathrm{i}(k_x x\!+\!k_z z)]$ $=$ $A(y,t)\exp[\mathrm{i}(k_x x \!+\! k_z z \!+\! \beta(y,t))]$, with $A=|\,\hat{\xi}\,|$ and $\beta=\arg\hat{\xi}$. If $k_x>0$, for fixed $y$ and $z$, a point with constant phase $k_x x \!+\! k_z z \!+\! \beta(y,t)$ moves with speed $\mathrm{d}_tx = - \partial_t\beta/k_x$, in which $\partial_t\beta$ satisfies $\hat{\xi}^*\partial_t\hat{\xi}=A\partial_tA+\mathrm{i}A^2\partial_t\beta$. We accordingly define the instantaneous advection speed for $\hat{\xi}$ as
\begin{equation}
\label{eq:defAdvSpd}
    a_\xi = \frac{-\mathrm{Im}\left(\,\hat{\xi}^*\partial_t\hat{\xi}\,\right)}{k_x\,|\,\hat{\xi}\,|^2},\quad\textup{if }\ k_x>0 \:\textup{ and }\: |\,\hat{\xi}\,|>0.
\end{equation}
The $y$-profile of the advection speed $a_\xi(y)$ can help illustrate some characteristic shear deformation of the flow structures associated with $\xi$, especially when $a_\xi(y)$ considerably deviates from the mean velocity profile $U(y)$.

\begin{figure}
  \centerline{\includegraphics[width=0.70\textwidth]{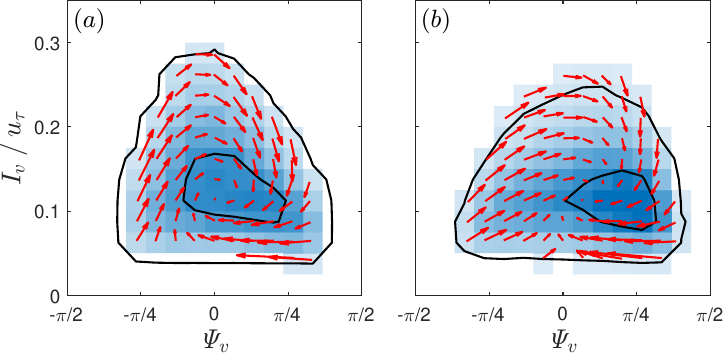}}
  \caption{Mean displacement of turbulence states over a time interval of approximately $0.05\,h/u_\tau$ in the $I_v$-$\mathit{\Psi}_v$ subspace. Conditioned on each cell, the arrow connects the average current state with the average future state after the time interval. The blue shade represents the probability density, with the black contours containing 30\% and 95\% of the probability mass. Each cell within the  95\% contour contains at least 100 samples. Panel $(a)$ is for the spatial Fourier mode $\left(n_x,n_z\right)=(1,0)$ and $(b)$ is for $\left(n_x,n_z\right)=(1,1).$}
\label{fig:PFO}
\end{figure}

The burst cycle in this log-minimal channel is reflected in the mean displacement of states over a short time interval in the coarse-grained, two-dimensional $I_v$-$\mathit{\Psi}_v$ subspace, as shown in figure \ref{fig:PFO}. A trend of clockwise displacement can be observed for both the straight and oblique modes. The displacement also tends to spiral towards the centre of the probability distribution, especially for the oblique mode, which reflects the non-deterministic property of coarse-grained dynamics, as discussed in \citet{Jimenez2023}. In general, the upper-left edge of the distribution, which is characterised by increases in $I_v$ with increasing $\mathit{\Psi}_v$ for $\mathit{\Psi}_v\lesssim0$, represents the amplification stage of bursts, while the upper-right edge, where $I_v$ decreases with increasing $\mathit{\Psi}_v$ for $\mathit{\Psi}_v\gtrsim0$, represents its decaying stage. Both stages are forward-tilting processes and are essentially interpreted by the linearised Orr mechanism \citep{Jimenez2013,Jimenez2015}. The lower edge of the probability distribution features decreases in $\mathit{\Psi}_v$ from $\mathit{\Psi}_v>0$ to $\mathit{\Psi}_v<0$ at a low level of $I_v$, leading to states that are ready to burst again. This apparent backward-tilting process is termed the restart stage and is the main concern of this paper.

\subsection{The forced linearised system}
\label{subsection:Linearisation}

The present study decomposes the turbulent flow field into the $x$-$z$-$t$-averaged mean profile and the remaining fluctuations and treats the interaction between the two constituents as part of the linearised dynamics. The interactions among the fluctuations, which include instantaneous fluctuations of $x$-$z$-averaged mean profiles, are all treated as nonlinear effects. For the fluctuation velocity $\boldsymbol{u}$ in the channel, the Navier-Stokes equations read
\begin{subeqnarray}
\label{eq:NSorg}
    && \left(\partial_t+U\partial_x\right)\boldsymbol{u} = -v\partial_yU\boldsymbol{e}_x -\bnabla p + \nu\nabla^2\boldsymbol{u} + \boldsymbol{f}, \quad \boldsymbol{f}=-\bnabla\cdot\left(\boldsymbol{u}\boldsymbol{u}-\overline{\boldsymbol{u}\boldsymbol{u}}\right), \\
    && \bnabla\cdot\boldsymbol{u} = 0,
\end{subeqnarray}
with the boundary conditions
\begin{equation}
\label{eq:NSorgBC}
    \boldsymbol{u}=\boldsymbol{0}\quad\text{at } y=0 \text{ and } y=2h,
\end{equation}
where $p$ is the fluctuation pressure, the overline `$\overline{\ \cdot\ }$' denotes the average over $x$, $z$, and $t$, and $\boldsymbol{f}$ is the nonlinear term. In the linearised system, $\boldsymbol{f}$ is treated as an external force. Equation (\ref{eq:NSorg}$a$) can be reworked \citep{Drazin1981} into the forced Orr-Sommerfeld equation for $v$,
\begin{equation}
\label{eq:OSorg}
    \left(\partial_t+U\partial_x\right)\nabla^2v = \partial_{y\!y}U\partial_xv + \nu\nabla^4v + \nabla^2f_y^s\,,
\end{equation}
and the forced Squire equation for $\omega_y$,
\begin{equation}
\label{eq:SQorg}
    \left(\partial_t+U\partial_x\right)\omega_y = -\partial_yU\partial_zv + \nu\nabla^2\omega_y + \left(\bnabla\times\boldsymbol{f}^s\right)_y,
\end{equation}
with the boundary conditions
\begin{equation}
\label{eq:OSSQBC}
    v=\partial_yv=\omega_y=0\quad\text{at } y=0 \text{ and } y=2h,
\end{equation}
where $\boldsymbol{f}^s=f^s_x\boldsymbol{e}_x+f^s_y\boldsymbol{e}_y+f^s_z\boldsymbol{e}_z$ is the solenoidal part of $\boldsymbol{f}$ with $\bnabla\cdot\boldsymbol{f}^s=0$.

For a spatial Fourier mode with $\boldsymbol{k}=k_x\boldsymbol{e}_x+k_z\boldsymbol{e}_z$, (\ref{eq:OSorg}-\ref{eq:SQorg}) can be written as
\begin{subeqnarray}
\label{eq:OSSQFourier}
    \left(\partial_t+\mathrm{i}k_x U\right)\hat{\nabla}^2\hat{v} & = & \ \mathrm{i}k_x \partial_{y\!y}U\,\hat{v} + \nu\hat{\nabla}^4\hat{v} \ \:\: + \quad \hat{\nabla}^2\hat{f}_y^s\quad\: , \\
    \left(\partial_t+\mathrm{i}k_x U\right)\,\hat{\omega}_y \, & = & -\mathrm{i}k_z \partial_yU\, \hat{v} + \nu\hat{\nabla}^2\hat{\omega}_y + (\hat{\bnabla}\times\hat{\boldsymbol{f}}^s)_y\,,
\end{subeqnarray}
where $\hat{\bnabla}=\mathrm{i}\boldsymbol{k}+\boldsymbol{e}_y\partial_y$, $\hat{\nabla}^2 = \hat{\bnabla}\cdot\hat{\bnabla} =\partial_{y\!y}-k^2$, and $\hat{\nabla}^4=\hat{\nabla}^2\hat{\nabla}^2$. It is noteworthy that the left-hand side of (\ref{eq:OSSQFourier}$a$) can be interpreted as mean-flow advection of the vorticity component $\hat{\omega}_\bot$, as follows from the identity $\hat{\omega}_\bot=\mathrm{i}k\hat{v}-\partial_y\hat{u}_{/\!/}=\hat{\nabla}^2\hat{v}/(\mathrm{i}k)$. Thus, in the absence of external forcing, the magnitude of $\hat{\omega}_\bot$ at a given $y$ can change only through the first two terms on the right-hand side. These two terms represent the deformation of the mean shear by wall-normal velocity fluctuations and the decay of $\hat{\omega}_\bot$ by viscosity, respectively, and generally have no significant effect in the logarithmic layer of wall-bounded turbulence \citep[see][\S6.1]{Jimenez2018a}. This suggests that $|\hat{\omega}_\bot(y)|$ in the logarithmic layer largely remains constant over time in an unforced linearised flow, which is key to understanding bursts.

A compact form of (\ref{eq:OSSQFourier}) can be written as
\begin{equation}
\label{eq:OSSQconciseFourier}
    \partial_t\hat{q} = \mathcal{L}\,\hat{q} + \hat{\varphi}\,,
\end{equation}
in which 
\begin{eqnarray}
\label{eq:defOperators}
    && \hat{q} = \left[
    \begin{array}{l}
        \,\hat{v} \\
        \hat{\omega}_y 
    \end{array}
    \right], \quad 
    \hat{\varphi} = \left[
    \begin{array}{c}
        \hat{f}_y^s  \\
        (\hat{\bnabla}\times\hat{\boldsymbol{f}}^s)_y 
    \end{array}
    \right], \quad
    \mathcal{L} = \left[
    \begin{array}{cc}
        \mathcal{L}_\mathrm{OS} & 0 \\
        \mathcal{L}_\mathrm{IH} & \mathcal{L}_\mathrm{SQ}
    \end{array}
    \right], \nonumber\\
    && \mathcal{L}_\mathrm{OS} = \hat{\nabla}^{-2}\!\left[\mathrm{i}k_x\!\left(\partial_{y\!y}U\!-\!U\hat{\nabla}^2\right)\right] + \nu\hat{\nabla}^2, \nonumber\\
    && \mathcal{L}_\mathrm{SQ} = -\mathrm{i}k_xU + \nu\hat{\nabla}^2, \quad
    \mathcal{L}_\mathrm{IH} = -\mathrm{i}k_z\partial_yU,
\end{eqnarray}
where $\hat{\nabla}^{-2}$ denotes the corresponding solution operator subject to homogeneous boundary conditions at $y=0$ and $y=2h$. Treating the external force $\hat{\varphi}$ as a prescribed term, we can formally write the solution to the initial value problem governed by (\ref{eq:OSSQconciseFourier}) as
\begin{equation}
\label{eq:GenSolForm}
    \hat{q}(y,t) = e^{\mathcal{L}t}\hat{q}(y,0) + \int_0^t e^{\mathcal{L}(t-\tau)}\hat{\varphi}(y,\tau)\,\mathrm{d}\tau\,.
\end{equation}
In the special case $\hat{\varphi}=0$, (\ref{eq:GenSolForm}) reduces to 
\begin{equation}
\label{eq:GenSolFormFree}
    \hat{q}(y,t) = e^{\mathcal{L}t}\hat{q}(y,0)\,.
\end{equation}
In the special case with a time-harmonic force $\hat{\varphi}(y,t) = \hat{\varphi}_\mathrm{sp}(y)\,e^{-\mathrm{i}k_x a_f t}$ where $a_f\in\mathbb{R}$ is the force's advection speed and the subscript `sp' stands for the spatial shape, if the system is exponentially stable, the solution (\ref{eq:GenSolForm}) as $t\to\infty$ is
\begin{equation}
\label{eq:GenSolFormResv}
    \hat{q}(y,t) = \mathcal{R}(a_f)\,\hat{\varphi}_\mathrm{sp}(y) \,e^{-\mathrm{i}k_x a_f t}\,,
\end{equation}
where $\mathcal{R}$ is the resolvent defined as
\begin{equation}
\label{eq:defResolvent}
    \mathcal{R}(a_f) = -\left(\mathcal{L}+\mathrm{i}k_x a_f \mathcal{I}\right)^{-1},
\end{equation}
with $\mathcal{I}$ the identity operator.

As discussed in \cite{Schmid2001}, the linear operator $\mathcal{L}$ satisfies an eigenequation 
\begin{eqnarray}
\label{eq:eigenEquation}
    \mathcal{L}\,\mathsfbi{\hat{g}} = -\mathrm{i}k_x\mathsfbi{\hat{g}}\,\mathsfbi{c}\,,
\end{eqnarray}
where 
\begin{equation}
\label{eq:eigenModes}
    \quad\mathsfbi{\hat{g}} = \left[
    \begin{array}{c}
        \mathsfbi{\hat{g}}^v \\ \mathsfbi{\hat{g}}^\omega
    \end{array}
    \right] = \left[
    \begin{array}{ccc}
        \hat{v}_1(y) & \hat{v}_2(y) & \cdots \\
        \hat{\omega}_{y1}(y) & \hat{\omega}_{y2}(y) & \cdots
    \end{array}
    \right]
\end{equation}
is the set of eigenmodes and $\mathsfbi{c} = \mathrm{diag}\left[\:c_1 \quad c_2 \quad \cdots\:\right]$ is the corresponding set of complex phase speeds. The eigenmodes can be divided into two classes $\mathsfbi{\hat{g}}=[\,\mathsfbi{\hat{g}}_\mathrm{OS}^{\vphantom{*}}\quad\mathsfbi{\hat{g}}_\mathrm{SQ}^{\vphantom{*}}]$, where
\begin{equation}
\label{eq:OSeigvs}
    \mathsfbi{\hat{g}}_\mathrm{OS}^{\vphantom{*}} = \left[
    \begin{array}{c}
        \mathsfbi{\hat{g}}^v_\mathrm{OS} \\
        \mathsfbi{\hat{g}}^{\omega}_\mathrm{OS}
    \end{array}
    \right] = \left[
    \begin{array}{ccc}
        \hat{v}_\mathrm{OS1}(y) & \hat{v}_\mathrm{OS2}(y) & \cdots \\
        \hat{\omega}_{y\mathrm{OS1}}(y) & \hat{\omega}_{y\mathrm{OS2}}(y) & \cdots
    \end{array}
    \right]
\end{equation}
assembles the Orr-Sommerfeld modes and
\begin{equation}
\label{eq:SQeigvs}
    \mathsfbi{\hat{g}}_\mathrm{SQ}^{\vphantom{*}} = \left[
    \begin{array}{c}
        0 \\
        \mathsfbi{\hat{g}}^{\omega}_\mathrm{SQ}
    \end{array}
    \right] = \left[
    \begin{array}{ccc}
        0 & 0 & \cdots \\
        \hat{\omega}_{y\mathrm{SQ1}}(y) & \hat{\omega}_{y\mathrm{SQ2}}(y) & \cdots
    \end{array}
    \right]
\end{equation}
assembles the Squire modes. In this paper, we use the mean velocity profile $U(y)$ provided by the turbulence database and the molecular viscosity $\nu$ to obtain $\mathcal{L}$, and numerically solve (\ref{eq:eigenEquation}) using the method in \citet{Schmid2001}. All the complex phase speeds $c_i,\,i=1,2,\cdots$ have negative imaginary parts, i.e., the base flow is exponentially stable \citep{Reynolds1967}. Based on our preliminary tests, the highly dissipative eigenmodes with $\mathrm{Im}(c_i)\lesssim-1.4\,U_\textit{bulk}$ have essentially no effect on the results and are thus discarded, and the number of retained eigenmodes is $N_\mathrm{OS}\approx N_\mathrm{SQ}\approx 210$. Using these eigenmodes, a state $\hat{q}(y)$ and an external force $\hat{\varphi}(y)$ can be expanded as
\begin{subeqnarray}
\label{eq:EigenExpansion}
    \hat{q} = \mathsfbi{\hat{g}}\,\boldsymbol{\kappa} &=&  \mathsfbi{\hat{g}}_\mathrm{OS}^{\vphantom{*}}\,\boldsymbol{\kappa}_\mathrm{OS}^{\vphantom{*}} + \mathsfbi{\hat{g}}_\mathrm{SQ}^{\vphantom{*}}\,\boldsymbol{\kappa}_\mathrm{SQ}^{\vphantom{*}}\,, \\
    \hat{\varphi} = \mathsfbi{\hat{g}}\boldsymbol{\chi} &=& \mathsfbi{\hat{g}}_\mathrm{OS}^{\vphantom{*}}\boldsymbol{\chi}_\mathrm{OS}^{\vphantom{*}} + \mathsfbi{\hat{g}}_\mathrm{SQ}^{\vphantom{*}}\boldsymbol{\chi}_\mathrm{SQ}^{\vphantom{*}}\,,
\end{subeqnarray}
respectively, where $\boldsymbol{\kappa}_\mathrm{OS}^{\vphantom{*}},\boldsymbol{\chi}_\mathrm{OS}^{\vphantom{*}}\in\mathbb{C}^{N_\mathrm{OS}}$ and $\boldsymbol{\kappa}_\mathrm{SQ}^{\vphantom{*}},\boldsymbol{\chi}_\mathrm{SQ}^{\vphantom{*}}\in\mathbb{C}^{N_\mathrm{SQ}}$ are column vectors of coefficients. The governing equation (\ref{eq:OSSQconciseFourier}) can thus be written as
\begin{equation}
\label{eq:governEqnCoef}
    \partial_t\boldsymbol{\kappa} = -\mathrm{i}k_x\mathsfbi{c}\,\boldsymbol{\kappa} + \boldsymbol{\chi}\,.
\end{equation}

For a state $\hat{q}=[\hat{v}\quad\hat{\omega}_y]^\mathrm{T}$, the corresponding band-wise intensity and inclination angle, defined as (\ref{eq:defIntensity}) and (\ref{eq:defIncline}), are denoted by $I_\xi\!\left[\hat{q}\right]$ and $\mathit{\Psi}_\xi\!\left[\hat{q}\right]$, respectively. Specifically, the intensity of the wall-normal velocity, which is the objective function for most optimisation problems in this paper, can be written as
\begin{subeqnarray}
\label{eq:defIntensityMat}
    && I_v\left[\hat{q}\right] = \sqrt{ \boldsymbol{\kappa}_\mathrm{OS}^*\mathsfbi{M}\,\boldsymbol{\kappa}_\mathrm{OS}^{\vphantom{*}} } = |\!|\mathsfbi{M}^{1/2}\boldsymbol{\kappa}_\mathrm{OS}^{\vphantom{*}}|\!|_2\,, \\
    && \textup{with }\ \mathsfbi{M} = \frac{1}{y_\mathrm{upp}-y_\mathrm{low}}\int_{y_\mathrm{low}}^{y_\mathrm{upp}} \mathsfbi{\hat{g}}^{v*}_\mathrm{OS}\,\mathsfbi{\hat{g}}^v_\mathrm{OS} \,\mathrm{d}y\,,
\end{subeqnarray}
where $\mathsfbi{M}$ is a Hermitian positive definite matrix in $\mathbb{C}^{N_\mathrm{OS}\times N_\mathrm{OS}}$, and $|\!|\cdot|\!|_2$ is the $L_2$ vector norm. 
The energy norm (\ref{eq:defEnergyNorm}) can be written as
\begin{subeqnarray}
\label{eq:defEnergyNormMat}
    && |\!|\,\hat{q}\,|\!|_E = \sqrt{ \boldsymbol{\kappa}_\mathrm{OS}^*\mathsfbi{E}\,\boldsymbol{\kappa}_\mathrm{OS}^{\vphantom{*}} + \boldsymbol{\kappa}^*\mathsfbi{F}\,\boldsymbol{\kappa} } \,, \\
    && \textup{with }\  \mathsfbi{E} = \frac{1}{2h}\int_{0}^{2h} \left(\mathsfbi{\hat{g}}^{v*}_\mathrm{OS}\,\mathsfbi{\hat{g}}^v_\mathrm{OS} + \frac{1}{k^2}\partial_y\mathsfbi{\hat{g}}^{v*}_\mathrm{OS}\,\partial_y\mathsfbi{\hat{g}}^v_\mathrm{OS}\right) \mathrm{d}y \\
    && \textup{ and }\ \mathsfbi{F} = \frac{1}{2h}\int_{0}^{2h}  \frac{1}{k^2}\,\mathsfbi{\hat{g}}^{\omega*}\mathsfbi{\hat{g}}^\omega \,\mathrm{d}y\,,
\end{subeqnarray}
where $\mathsfbi{E}\in\mathbb{C}^{N_\mathrm{OS}\times N_\mathrm{OS}}$ and $\mathsfbi{F}\in\mathbb{C}^{(N_\mathrm{OS}+N_\mathrm{SQ}) \times (N_\mathrm{OS}+N_\mathrm{SQ})}$ are Hermitian positive definite matrices. Specifically, if we consider only the velocity components in the $x_{/\!/}$-$y$ plane, i.e., disregarding $\hat{u}_\bot$ (or equivalently, $\hat{\omega}_y$), then (\ref{eq:defEnergyNormMat}$a$) reduces to 
\begin{equation}
\label{eq:defEnergyNormMat2C}
    |\!|\,\hat{q}\,|\!|_{E_\mathrm{2C}} = \sqrt{ \boldsymbol{\kappa}_\mathrm{OS}^*\mathsfbi{E}\,\boldsymbol{\kappa}_\mathrm{OS}^{\vphantom{*}} } = |\!|\mathsfbi{E}^{1/2}\boldsymbol{\kappa}_\mathrm{OS}^{\vphantom{*}}|\!|_2\,,
\end{equation}
which we term the two-component energy norm.

\section{Burst restarts in the forced linearised system}
\label{section:Requirements}

This section generically concerns the `minimal requirements' for the existence of burst-restart-like solutions of the forced linearised system itself. Except for the mean velocity profile $U(y)$ and molecular viscosity $\nu$, no \textit{a priori} knowledge of flows and external forces is assumed.


\subsection{Baseline development of a burst event}
\label{subsection:Baseline}

Before discussing burst restarts, let us represent a typical burst development in the linearised system. We generally follow the concept in \citet{Jimenez2013,Encinar2020}, which represents a burst development by an unforced transient growth of a disturbance that realises the maximum peak intensity relative to its initial state. In the present paper, for a given spatial Fourier mode with $(n_x,n_z)$, we maximise the band-wise intensity of $v$ (defined as (\ref{eq:defIntensity}) or (\ref{eq:defIntensityMat})) during the development, subject to a unit whole-channel energy norm (defined as (\ref{eq:defEnergyNorm}) or (\ref{eq:defEnergyNormMat})) of the initial state $\hat{q}_0(y)$. The problem is posed as follows.

\begin{figure}
  \centerline{\includegraphics[width=1.0\textwidth]{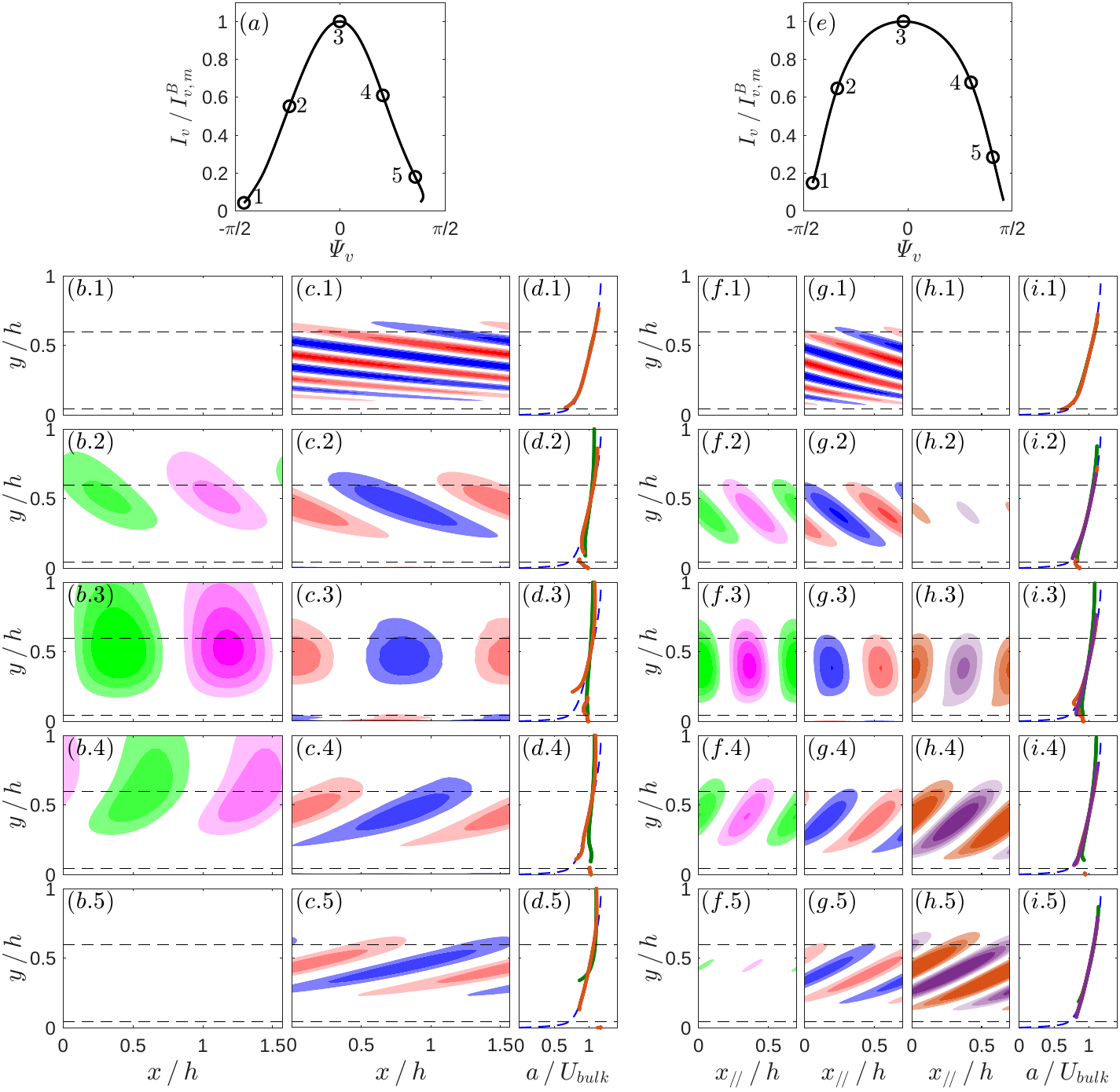}}
  \caption{Maximum unforced transient-growth solutions for the spatial Fourier modes $(n_x,n_z)=(1,0)$ in panels $(a)$-$(d)$ and $(1,1)$ in panels $(e)$-$(i)$. All flow fields are viewed in a reference frame moving at the constant streamwise speed $U_\textit{bulk}$. 
  Panels $(a,e)$: trajectories in the $I_v$-$\mathit{\Psi}_v$ subspace. Markers 1-5 correspond to the five rows of panels in $(b)$-$(d)$ and $(f)$-$(i)$. The marked times are $t=[0,\,0.64:0.24:1.36]\,h/u_\tau$ in $(a)$ and $t=[0,\,0.52:0.24:1.24]\,h/u_\tau$ in $(e)$. 
  Columns $(b,f)$: velocity $v$, with magenta and green indicating upward and downward motion, respectively. 
  Columns $(c,g)$: vorticity $\omega_\bot$, with red and blue indicating clockwise and counterclockwise rotation, respectively. 
  Column $(h)$: velocity $u_\bot$, with orange and purple indicating outward and inward motion, respectively; the $x_\bot$-component of the mean flow is prescribed outward. 
  Columns $(d,i)$: $y$-profiles of the advection speeds $a$ for $v$ (green), $\omega_\bot$ (brown), and $u_\bot$ (purple), shown as solid lines. Blue dashed lines show the mean velocity $U$. 
  The shading levels given as magnitudes are: $(b)$: $[0.68:0.28:1.52]\,I_{v,\,m}^B$; $(c)$: $[7.6:3.0:13.6]\,I_{v,\,m}^B/h$; $(f)$: $[0.66:0.26:1.44]\,I_{v,\,m}^B$; $(g)$: $[10.8:4.3:19.4]\,I_{v,\,m}^B/h$; $(h)$: $[1.98:0.80:4.38]\,I_{v,\,m}^B$.}
  \label{fig:maxGrowth}
\end{figure}

\begin{problem}[Typical burst development]
\label{problem:Baseline}
    Solve the optimisation problem
    \begin{subeqnarray}
    \label{eq:OPTmaxGrowth}
        && I_{v,\,m}^B = \max_{t\,>\,0} \left\{\, \max_{|\!|\hat{q}_0(y)|\!|_E \,=\, 1} I_v\left[\hat{q}(y,t)\right] \,\right\}, \\
        && \textup{where }\,\hat{q}(y,t) = e^{\mathcal{L}t}\hat{q}_0(y).
    \end{subeqnarray}
    The resulting optimal solution,
    \begin{equation}
    \label{eq:BSLsolution}
        \hat{q}^B(y,t)=e^{\mathcal{L}t}\hat{q}^B_0(y),
    \end{equation}
    is termed the `baseline solution'.    Note that the problem naturally results in a baseline initial state $\hat{q}_0^B$ with a zero $x_\bot$-component of the velocity, so the constraint $|\!|\hat{q}_0(y)|\!|_E =1$ in (\ref{eq:OPTmaxGrowth}$a$) is equivalent to $|\!|\hat{q}_0(y)|\!|_{E_\mathrm{2C}}=1$. The problem is numerically solved using the algorithm in \citet{Schmid2001}, the key steps of which are sumarised in appendix \ref{subsection:AlgorithmBaseline}. 
\end{problem}

The baseline solutions (\ref{eq:BSLsolution}) to problem \ref{problem:Baseline} for the Fourier modes $(n_x,n_z)=(1,0)$ and $(1,1)$ are shown in figure \ref{fig:maxGrowth}. The flow fields are displayed on the $x_{/\!/}$-$y$ plane. The evolution of each solution demonstrates a typical Orr process. Starting with a backward-inclined initial state, the disturbance is gradually tilted forward. During this process, the vorticity $\hat{\omega}_\bot(y)$ changes relatively little in magnitude, as discussed regarding equation (\ref{eq:OSSQFourier}$a$); however, the velocity $v$ experiences significant amplification and decay in sequence, peaking when the disturbance is normal to the wall. For $y\gtrsim0.2\,h$ where the disturbance magnitude is appreciable, the advection speeds $a$, defined as (\ref{eq:defAdvSpd}), for $v$, $\omega_\bot$, and $u_\bot$ essentially collapse with the mean velocity profile throughout the entire process. 

The time during which the intensity $I_v$ is greater than half of $I_{v,\,m}^B$ is $0.55\,h/u_\tau\approx7.4\,\mathcal{S}^{-1}$ for the mode $(n_x,n_z)=(1,0)$ and $0.64\,h/u_\tau\approx8.6\,\mathcal{S}^{-1}$ for $(n_x,n_z)=(1,1)$, both having orders of magnitude consistent with the findings in \citet{Jimenez2013,Encinar2020}. For the oblique mode $(n_x,n_z)=(1,1)$, the Orr process also generates a velocity $u_\bot$ as a by-product through the lift-up mechanism. This effect accumulates over time and, unlike the velocity $v$ decaying with a time scale $O(\mathcal{S}^{-1})$, $u_\bot$ decays only through viscosity, with a considerably larger time scale $O(\nu^{-1}k^{-2})$.

To realise the same peak intensity of $v$, more energy of the initial state is needed for the oblique mode than for the straight one. For $(n_x,n_z)=(1,0)$, the baseline initial state $\hat{q}_0^B$ has the energy norm $|\!|\hat{q}_0^B|\!|_E\approx0.19\,I_{v,\,m}^B$ and intensity $I_v[\hat{q}_0^B]\approx0.04\,I_{v,\,m}^B$, while for $(n_x,n_z)=(1,1)$, $\hat{q}_0^B$ has $|\!|\hat{q}_0^B|\!|_E\approx0.28\,I_{v,\,m}^B$ and $I_v[\hat{q}_0^B]\approx0.15\,I_{v,\,m}^B$. Both values of $I_v[\hat{q}_0^B]\,/\,I_{v,\,m}^B$ are of order $O\!\left((\nu k^3 k_x^{-1}\mathcal{S}^{-1})^{1/3}\right)$, a result deduced from the linearised analysis for bursts in viscous homogeneous shear flows in \citet{Jimenez2015}.

\subsection[]{Weak-sense burst restart: intensity regrowth}
\label{subsection:weakRestart}

We now turn to burst restarts. We have seen in \S\ref{subsection:Baseline} that the intensity $I_v$ of a state $\hat{q}_{\mathrm{I}}(y)$ in the decaying stage of the baseline solution (\ref{eq:BSLsolution}) will monotonically decrease with time to zero if no forces are applied afterwards. The question naturally arises regarding the possibility of injecting a relatively small perturbation $\hat{\varphi}_\mathrm{sp}(y)$ into $\hat{q}_{\mathrm{I}}(y)$ such that the subsequent unforced evolution of the perturbed state $\hat{q}_{\mathrm{I}}(y)+\hat{\varphi}_\mathrm{sp}(y)$ exhibits an increase in $I_v$ again, either immediately or at some later time. Such a perturbation is equivalent to applying an impulsive force $\hat{\varphi}(y,t) = \hat{\varphi}_\mathrm{sp}(y)\,\delta(t)$, where $\delta(t)$ is the Dirac delta function. We refer to this scenario as the weak-sense burst restart, considering that a re-burst is defined only by the band-wise intensity $I_v$ but not by other details of a state. Regarding the minimal requirement for such a weak-sense burst restart, we pose the following optimisation problem that minimises the forcing magnitude $m_f=|\!|\hat{\varphi}_\mathrm{sp}(y)|\!|_E$.

\begin{problem}[Weak-sense burst restart]
\label{problem:weakRestart}
    Let $\hat{q}_{\mathrm{I}}(y)$ be a state in the decaying stage of the baseline solution (\ref{eq:BSLsolution}). Given $m_f>0$ and $t>0$, calculate the maximum possible intensity $A=A(t,m_f;\hat{q}_{\mathrm{I}})$ defined by the optimisation problem
    \begin{subeqnarray}
    \label{eq:OPTreburst}
        && A(t,m_f;\hat{q}_{\mathrm{I}}) = \max_{|\!|\hat{\varphi}_\mathrm{sp}(y)|\!|_E\,=\,m_f} I_v\left[\hat{q}(y,t)\right], \\ 
        && \textup{where }\, \hat{q}(y,t)=e^{\mathcal{L}t}\!\left(\,\hat{q}_{\mathrm{I}}(y)+\hat{\varphi}_\mathrm{sp}(y)\,\right).
    \end{subeqnarray}
    For fixed $\hat{q}_{\mathrm{I}}$ and $m_f$, if $A(t,m_f;\hat{q}_{\mathrm{I}})$ as a function of $t$ has a local maximum at $t=t_c>0$, there exists $\hat{\varphi}_\mathrm{sp}(y)$ such that $I_v\left[\hat{q}(y,t)\right]$ increases with $t$ for some $t>0$ (i.e., $\hat{q}(y,t)$ features a `re-burst'). In such a situation, we refer to the $\hat{\varphi}_\mathrm{sp}(y)$ that yields the local maximum $A(t_c,m_f;\hat{q}_{\mathrm{I}})$ as the optimal perturbing force, and the corresponding solution $\hat{q}(y,t)$ as the optimal (weak-sense) re-burst solution. 
    The problem naturally results in an optimal perturbing force $\hat{\varphi}_\mathrm{sp}$ with a zero $x_\bot$-component. The key steps of the numerical method for this problem is summarised in appendix \ref{subsection:AlgorithmWeakRestart}.
\end{problem}

\begin{figure}
  \centerline{\includegraphics[width=0.86\textwidth]{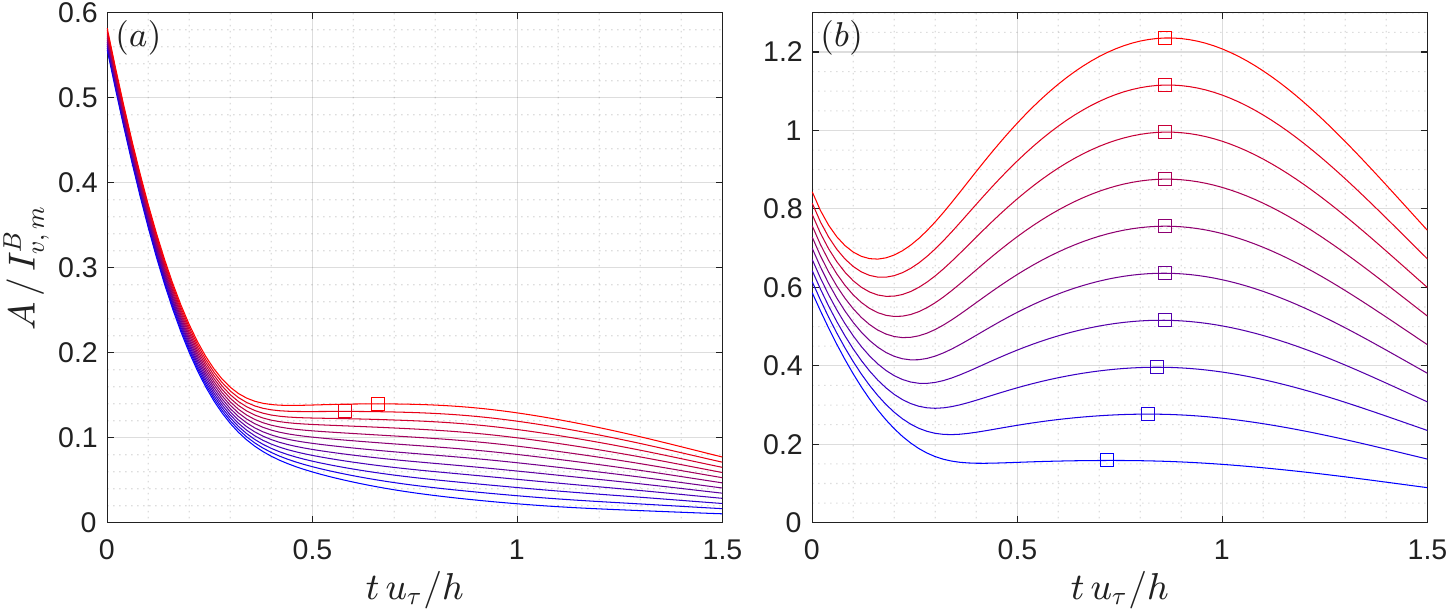}}
  \caption{Maximum possible intensity $A(t,m_f;\hat{q}_{\mathrm{I}})$ as a function of time $t$ under different force magnitudes $m_f$. The spatial Fourier mode considered is $(n_x,n_z) = (1,0)$ and the initial state $\hat{q}_{\mathrm{I}}(y)$ is selected to be the one with intensity $I_v[\hat{q}_{\mathrm{I}}] = 0.55\,I_{v,\,m}^B$ in the decaying stage of the baseline solution (\ref{eq:BSLsolution}). Colours from blue to red correspond to $m_f=[0.002:0.002:0.024]\,I_{v,\,m}^B$ for $(a)$ and $[0.026:0.023:0.233]\,I_{v,\,m}^B$ for $(b)$. For each line, the local maximum at $t=t_c$, if exists, is marked by a square.}
\label{fig:RB_AEvlp}
\end{figure}
\begin{figure}
  \centerline{\includegraphics[width=1.0\textwidth]{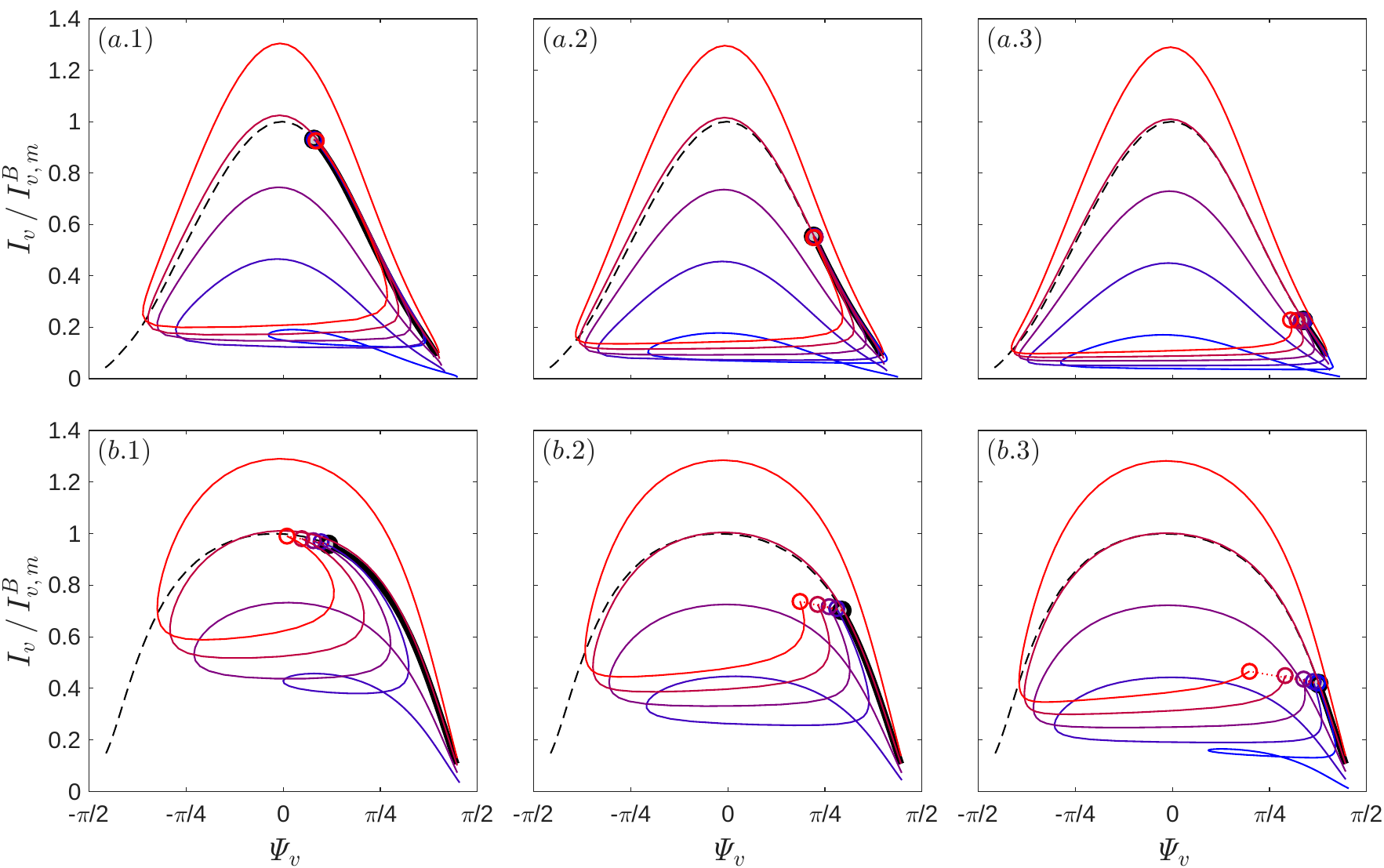}}
  \caption{Trajectories of optimal weak-sense re-burst solutions in the $I_v$-$\mathit{\Psi}_v$ subspace. $(a)$ and $(b)$ are for the spatial Fourier modes $(n_x,n_z) = (1,0)$ and $(1,1)$, respectively. Columns from left to right correspond to three selected initial states $\hat{q}_{\mathrm{I}}$ (black circles) in descending order of their intensity, $I_v[\hat{q}_{\mathrm{I}}]$. In each panel, the black line is the trajectory of the baseline solution $\hat{q}^B(t)$, including a dashed segment for the trajectory before the impulsive forces $\hat{\varphi}_\mathrm{sp}$ are exerted. Colours from blue to red represent the optimal re-burst solutions under five force magnitudes $m_f=[0.030:0.053:0.242]\,I_{v,\,m}^B$ for $(a)$ and $[0.044:0.077:0.352]\,I_{v,\,m}^B$ for $(b)$. The circles represent the states immediately after the impulsive forces, $\hat{q}_{\mathrm{I}}+\hat{\varphi}_\mathrm{sp}$, and the lines are the trajectories afterwards. Note that for $(b.1)$ and $(b.2)$, re-burst solutions for $m_f=0.044\,I_{v,\,m}^B$ do not exist and are thus not shown.}
\label{fig:RB_traj}
\end{figure}

Figure \ref{fig:RB_AEvlp} shows an example of the maximum possible intensity $A$ as a function of time under different perturbing force magnitudes $|\!|\hat{\varphi}_\mathrm{sp}|\!|_E=m_f$, given a particular initial state $\hat{q}_\mathrm{I}$ with $I_v[\hat{q}_\mathrm{I}]=0.55\,I_{v,\,m}^B$. For $m_f\lesssim0.02\,I_{v,\,m}^B$, the function $A$ monotonically decreases with $t$, suggesting that all perturbed states $\hat{q}_{\mathrm{I}}+\hat{\varphi}_\mathrm{sp}$ continue to decay afterwards with no possibility of increase. When $m_f$ reaches around $0.02\,I_{v,\,m}^B$, a local maximum of $A$ emerges at $t=t_c$, indicating that a weak-sense re-burst solution becomes possible. As $m_f$ increases further, the local maximum at $t=t_c$ becomes more apparent. For $m_f\gtrsim0.09\,I_{v,\,m}^B$, the value of $t_c$ asymptotes to $0.86\,h/u_\tau$, which is identical to the time for the baseline solution (\ref{eq:BSLsolution}) to reach its peak intensity starting from the initial state. Meanwhile, the local maximum value of $A$ increases linearly with $m_f$, reaching $1.0\,I_{v,\,m}^B$ as $m_f\approx0.19\,I_{v,\,m}^B$, which is identical to $|\!|\hat{q}_0^B|\!|_E$. This means that for large $m_f$, the optimal perturbing force resulting from problem \ref{problem:weakRestart} is $\hat{\varphi}_\mathrm{sp}(y)\approx s\,\hat{q}_0^B(y)$ with a scalar $s$, and the optimal re-burst solution in this linearised system is essentially a linear superposition of two baseline solutions triggered at different times.

In the intensity-inclination subspace, an optimal weak-sense re-burst solution exhibits several notable features in its trajectory, as shown in figure \ref{fig:RB_traj}. For most cases, the state immediately after an optimal perturbing force is close to the unperturbed state -- for the mode $(n_x,n_z)=(1,0)$, the two states are almost indistinguishable. Thereafter, the perturbed state continues to decay for some time like the unperturbed one, but then stops decaying and seemingly begins to tilt backward. Up to this point, the perturbed state has experienced a `latent period' during which no significant `aftermath effect' of the perturbing force on the state intensity is observed. After a certain time, the latent period ends, and the state begins to be amplified again, indicating a re-burst, before entering the final stage of monotonic decay. The above features in the trajectories are expected for this linearised system, in which a re-burst process can be interpreted as a linear superposition of two burst events triggered at different times. These trajectories suggest two noteworthy implications, which hold even in an unforced linearised flow: (i) an apparent backward-tilting process is possible, and (ii) states with similar intensity and inclination angles can have dramatically different future behaviour.

\begin{figure}
  \centerline{\includegraphics[width=1.0\textwidth]{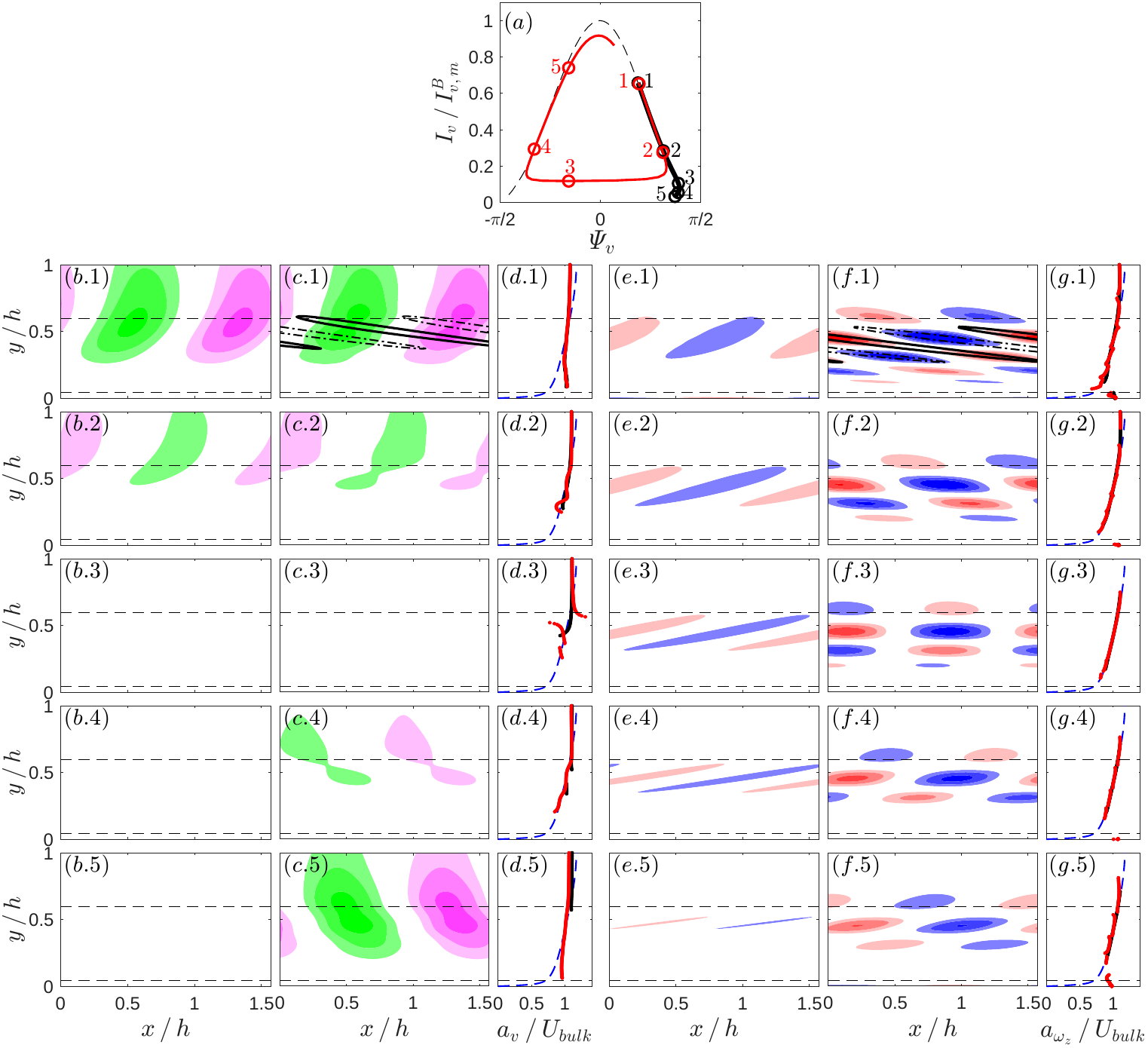}}
  \caption{Comparison of a decaying baseline solution (black lines in $(a,d,g)$ and contour shading in $(b,e)$) and an optimal weak-sense re-burst solution (red lines in $(a,d,g)$ and contour shading in $(c,f)$) for the spatial Fourier mode $(n_x,n_z) = (1,0)$. The initial state $\hat{q}_{\mathrm{I}}$ has the intensity $I_v[\hat{q}_{\mathrm{I}}]\approx0.65\,I_{v,\,m}^B$, and the re-burst solution is induced by the optimal impulsive force (with $|\!|\hat{\varphi}_\mathrm{sp}|\!|_E\approx0.17\,I_{v,\,m}^B$) that later yields a peak intensity $I_v=0.9\,I_{v,\,m}^B$. All flow fields are viewed in a reference frame moving at the constant streamwise speed $U_\textit{bulk}$. 
  Panel $(a)$: trajectories in the $I_v$-$\mathit{\Psi}_v$ subspace. Markers 1-5 denote times $t = [0:0.18:0.72]\,h/u_\tau$ following the impulse at $t=0$ and correspond to the five rows of panels in $(b)$-$(g)$. 
  Columns $(b,c)$: velocity $v$, with magenta and green indicating upward and downward motion, respectively. In $(c.1)$, solid and dashed contours indicate the upward and downward force $f^s_y$ at $t=0$, respectively. 
  Columns $(e,f)$: vorticity $\omega_z$, with red and blue indicating clockwise and counterclockwise rotation, respectively. In $(f.1)$, solid and dashed contours indicate the clockwise and counterclockwise curl of the force $(\bnabla\times\boldsymbol{f}^s)_z$ at $t=0$, respectively. 
  Columns $(d,g)$: $y$-profiles of the advection speeds $a$ for $v$ and $\omega_z$, respectively. Blue dashed lines indicate the mean velocity $U$. 
  The shading levels given as magnitudes are: $(b,c)$: $[0.66:0.26:1.18]\,I_{v,\,m}^B$; $(e,f)$: $[11.6:4.6:20.8]\,I_{v,\,m}^B/h$. The force-contour levels are: $(c.1)$: $0.05\,I_{v,\,m}^B$; $(f.1)$: $10.7\,I_{v,m}^B/h$.}
\label{fig:RB_Cont0}
\end{figure}

\begin{figure}
  \centerline{\includegraphics[width=1.0\textwidth]{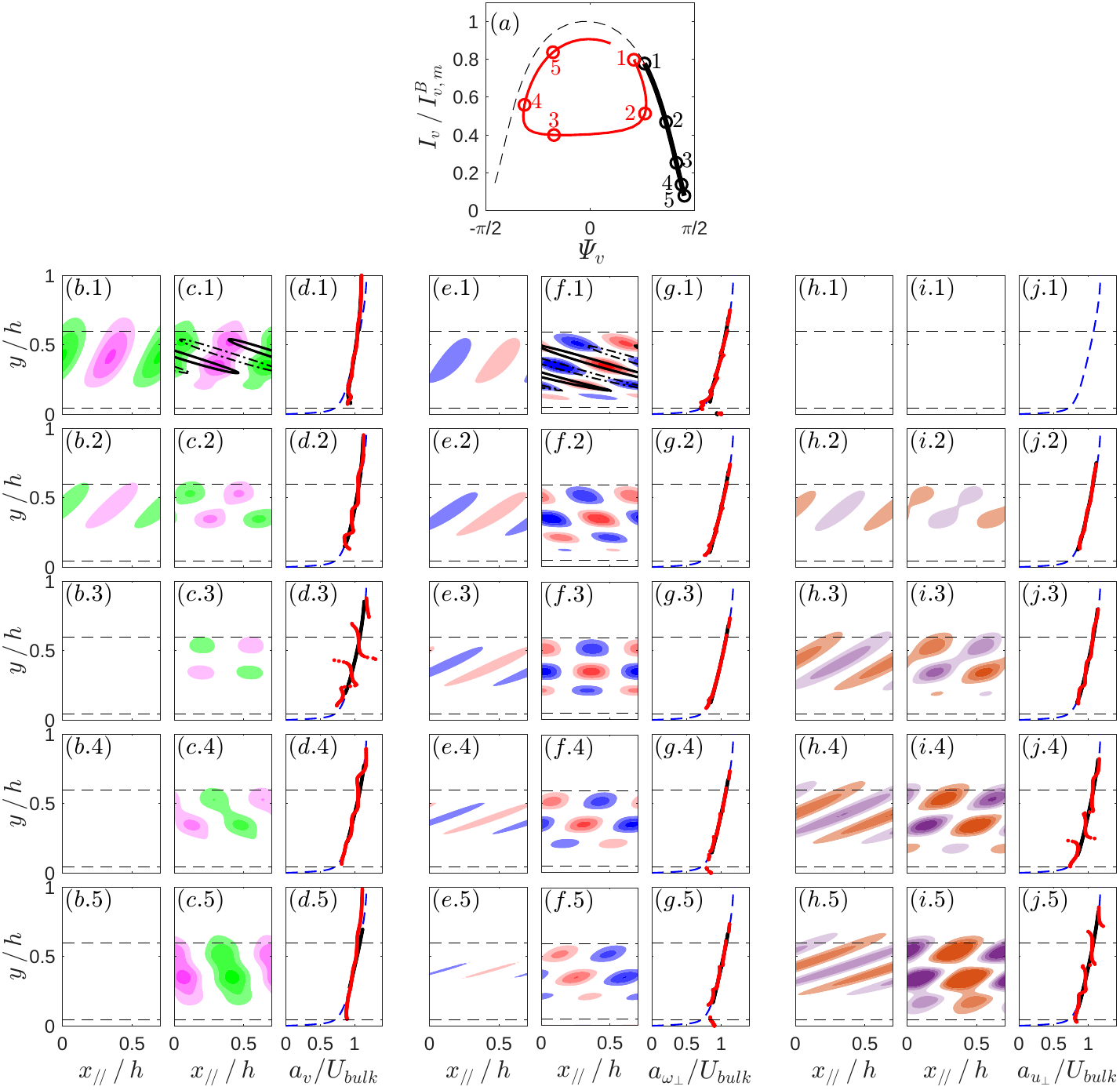}}
  \caption{Comparison of a decaying baseline solution (black lines in $(a,d,g,j)$ and contour shading in $(b,e,h)$) and an optimal weak-sense re-burst solution (red lines in $(a,d,g,j)$ and contour shading in $(c,f,i)$) for the spatial Fourier mode $(n_x,n_z) = (1,1)$. The initial state $\hat{q}_{\mathrm{I}}$ has the intensity $I_v[\hat{q}_{\mathrm{I}}]\approx 0.80\,I_{v,\,m}^B$, and the re-burst solution is induced by the optimal impulsive force (with $|\!|\hat{\varphi}_\mathrm{sp}|\!|_E\approx0.25I_{v,\,m}^B$) that later yields a peak intensity $I_v=0.9\,I_{v,\,m}^B$. All flow fields are viewed in a reference frame moving at the constant streamwise speed $U_\textit{bulk}$. 
  Panel $(a)$: trajectories in the $I_v$-$\mathit{\Psi}_v$ subspace. Markers 1-5 denote times $t = [0:0.16:0.64]\,h/u_\tau$ following the impulse at $t=0$ and correspond to the five rows of panels in $(b)$-$(j)$. 
  Columns $(b,c)$: velocity $v$, with magenta and green indicating upward and downward motion, respectively. In $(c.1)$, solid and dashed contours indicate the upward and downward force $f^s_y$ at $t=0$, respectively. 
  Columns $(e,f)$: vorticity $\omega_\bot$, with red and blue indicating clockwise and counterclockwise rotation, respectively. In $(f.1)$, solid and dashed contours indicate the clockwise and counterclockwise curl of the force $(\bnabla\times\boldsymbol{f}^s)_\bot$ at $t=0$, respectively. 
  Columns $(h,i)$: velocity $u_\bot$, with orange and purple indicating outward and inward motion, respectively. The $x_\bot$-component of the mean flow is prescribed outward, and $u_\bot=0$ is assumed at $t=0$. 
  Columns $(d,g,j)$: $y$-profiles of the advection speeds $a$ of $v$, $\omega_\bot$, and $u_\bot$, respectively. Blue dashed lines indicate the mean velocity $U$. 
  The shading levels given as magnitudes are: $(b,c)$: $[0.73:0.29:1.31]\,I_{v,\,m}^B$; $(e,f)$: $[16.7:6.7:30.1]\,I_{v,\,m}^B/h$; $(h,i)$: $[1.39:0.55:3.04]\,I_{v,\,m}^B$. The force-contour levels are: 
  $(c.1)$: $0.15\,I_{v,\,m}^B$; $(f.1)$: $15.1\,I_{v,\,m}^B/h$.} 
\label{fig:RBoblq_Cont0}
\end{figure}

The flow fields in figures \ref{fig:RB_Cont0} and \ref{fig:RBoblq_Cont0} reveal more details of the spatial structures of typical optimal weak-sense re-burst solutions. The optimal perturbing force $\hat{\varphi}_\mathrm{sp}$ in panel $(c.1)$ has a backward-inclined pattern, which is similar to that of the baseline initial state $\hat{q}_0^B$. This perturbation only slightly distorts the initial pattern of $v$ from $(b.1)$ to $(c.1)$. However, it significantly changes the initial pattern of $\omega_\bot$ from $(e.1)$ to $(f.1)$: the originally intact vortices are broken up, forming multiple layers of vortices at different wall distances. An interface occurs between adjacent layers at $y=y_\mathrm{intf}$, where $\hat{\omega}_\bot(y_\mathrm{intf})=0$. In the subsequent evolution $(f.1)$-$(f.5)$, the broken-up vortices are advected by the mean flow with little change in $|\hat{\omega}_\bot(y)|$, and the advection speed $a_{\omega_\bot}$ in $(g.1)$-$(g.5)$ is almost identical to the mean velocity $U$. The stage $(f.1)$-$(f.3)$, which corresponds to the latent period introduced above, features a vortex farther from the wall gradually catching up to a \textit{counter-rotating} vortex closer to the wall. During this stage, the advection speed $a_v$ in $(d.1)$-$(d.3)$ tends to deviate from $U$ in regions between adjacent layers of vortices, forming discontinuities. This trend towards discontinuities in $a_v$ suggests a trend towards the disconnection of $v$-structures, as illustrated in $(c.1)$-$(c.3)$. The following stage $(f.3)$-$(f.5)$, which corresponds to a re-burst, features a vortex farther from the wall gradually catching up to a \textit{co-rotating} vortex closer to the wall. During this stage, the advection speed $a_v$ in $(d.3)$-$(d.5)$ tends to be smooth again, suggesting a trend towards the reconnection of $v$-structures, as shown in $(c.3)$-$(c.5)$. 

The flow fields $(c.2)$-$(c.4)$ and $(f.2)$-$(f.4)$ illustrate an apparent backward-tilting process as $\mathit{\Psi}_v$ decreases between the decay of the preceding burst and the onset of the next. This decrease in $\mathit{\Psi}_v$ results from a breakup-and-catch-up process rather than from the uniform backward tilting of intact vortices.

For the oblique mode $(n_x,n_z)=(1,1)$, the multiple layers of vortices during the evolution $(f.1)$-$(f.5)$ in figure \ref{fig:RBoblq_Cont0} also generate multiple layers of $u_\bot$-structures in $(i.1)$-$(i.5)$ through the lift-up mechanism. Unlike the $v$-structures, the $u_\bot$-structures from different layers hardly reconnect, and the discontinuities in the advection speed $a_{u_\bot}$ can persist for a longer time, as shown in $(j.3)$-$(j.5)$. This is because $u_\bot$ has no role in the continuity constraint, and the interaction between $u_\bot$-structures from different layers is only through viscosity.

One of the most important insights provided by the optimal weak-sense re-burst solutions is that the breakup of spanwise vorticity structures during the decaying stage of a preceding burst is probably an essential feature for a burst restart. The breakup yields multiple layers of vortices, such that the subsequent advection involves the catch-up of counter-rotating and co-rotating vortices from different layers, which finally leads to a re-burst. This physical process is also consistently observed in solutions to problem \ref{problem:weakRestart} under the modified conditions tested in appendix \ref{section:Robustness}. In real turbulence, the breakup can be an essential burst-restarting mechanism provided by the nonlinear terms in the Navier-Stokes equations.

\subsection[]{Strong-sense burst restart: state recovery}
\label{subsection:strongRestart} 

The weak-sense burst-restart model in \S\ref{subsection:weakRestart} suggests the necessity of breaking up vortices for a force in the decaying stage of a preceding burst. In this model, however, the multilayer structure produced by the breakup is largely retained during the re-burst, in contrast to the intact single-layer structure during the preceding burst. Although exact structural recurrence is not expected in actual flows, this pronounced structural difference demonstrated by the weak-sense model suggests that, in real turbulence, 
additional nonlinear effects beyond the breakup may contribute to closing the burst cycle. To explore the possible role of such additional effects, this subsection further discusses burst restart in the strong sense, in which, over a time interval $T_R$, we require the force $\hat{\varphi}(y,t)$ to drive an initial decaying state $\hat{q}_{\mathrm{I}}(y)$ towards the full recovery of a target state $\hat{q}_{\mathrm{II}}(y)$ (except for a phase offset) near the beginning of the amplification stage of the baseline solution. In particular, since the primary interest of the paper is the wall-normal velocity, we consider the recovery only of the component $\hat{v}$ of a state $\hat{q}=[\hat{v}\quad\hat{\omega}_y]^\mathrm{T}$. Regarding the minimal requirement for such a strong-sense burst restart, we pose the following optimisation problem that minimises the time integral of the squared forcing magnitude $|\!|\hat{\varphi}(y,t)|\!|_E^2$.

\begin{figure}
  \centerline{\includegraphics[width=1.00\textwidth]{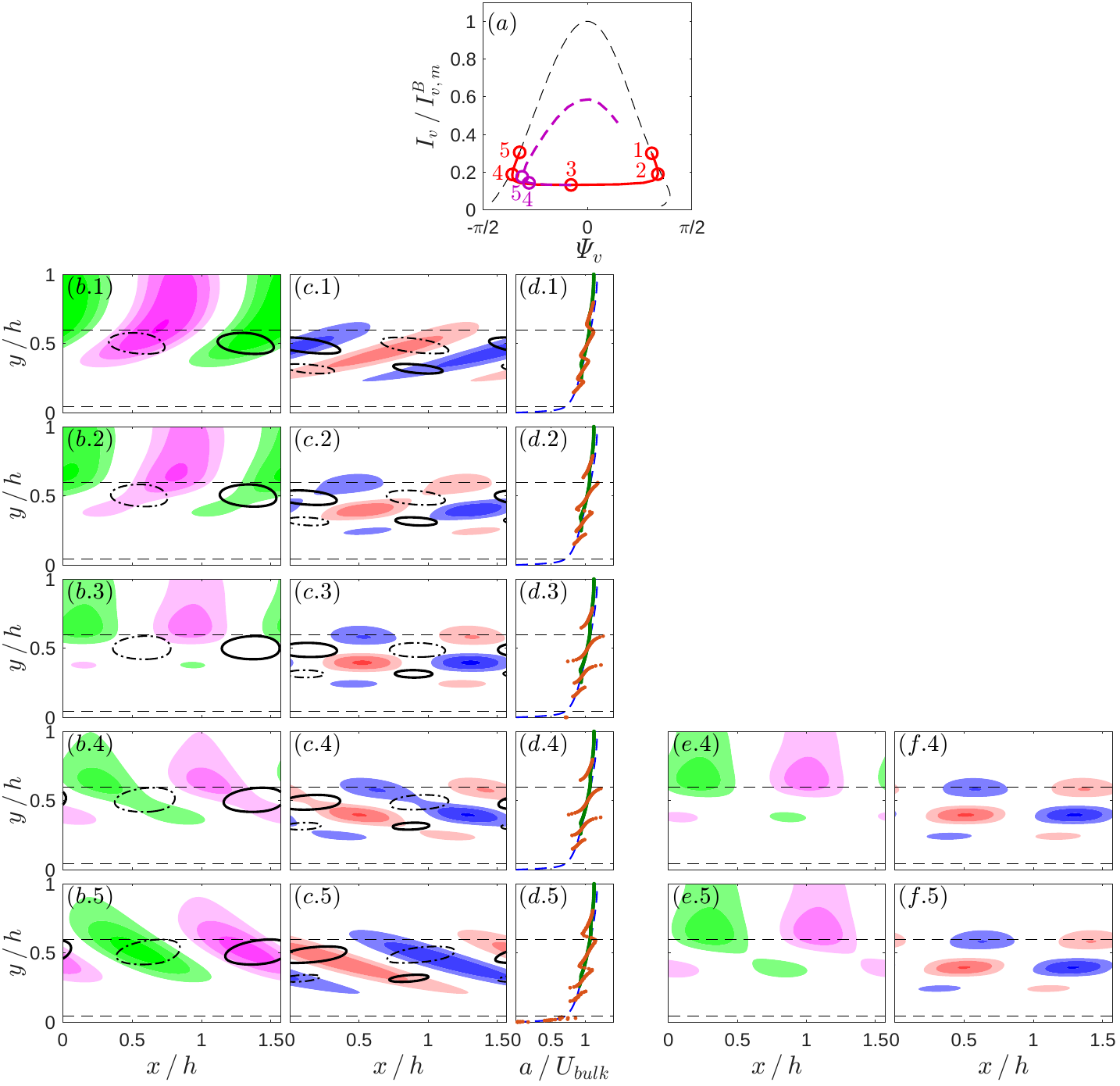}}
  \caption{An optimal recovery process ($(b,c,d)$ and the red line in $(a)$), together with a semi-recovery process ($(e,f)$ and the magenta line in $(a)$), for the spatial Fourier mode $(n_x,n_z)=(1,0)$. The initial and final states, $\hat{q}_{\mathrm{I}}$ and $\hat{q}_{\mathrm{II}}$, have the intensity $I_v[\hat{q}_{\mathrm{I}}]=I_v[\hat{q}_{\mathrm{II}}]=0.3\,I_{v,\,m}^B$, and the recovery time is $T_R=0.15\,h/u_\tau$. All flow and force fields are viewed in a reference frame moving at the constant streamwise speed $U_\textit{bulk}$. 
  Panel $(a)$: trajectories in the $I_v$-$\mathit{\Psi}_v$ subspace. Black dashed, red solid, and magenta dashed lines indicate the baseline solution, optimal recovery process, and semi-recovery process, respectively. Markers 1-5 denote times $t=[0:0.0375:0.15]\,h/u_\tau$ and correspond to the five rows of panels in $(b)$-$(f)$. 
  Columns $(b,e)$: velocity $v$, with magenta and green indicating upward and downward motion, respectively. Solid and dashed contours indicate the upward and downward force $f^s_y$, respectively. 
  Columns $(c,f)$: vorticity $\omega_z$, with red and blue indicating clockwise and counterclockwise rotation, respectively. Solid and dashed contours indicate the clockwise and counterclockwise curl of the force $(\bnabla\times\boldsymbol{f}^s)_z$, respectively. 
  Column $(d)$: $y$-profiles of the advection speeds $a$ of $f^s_y$ (green) and $(\bnabla\times\boldsymbol{f}^s)_z$ (brown). Blue dashed lines indicate the mean velocity $U$. 
  The shading levels given as magnitudes are: $(b,e)$: $[0.29:0.12:0.53]\,I_{v,\,m}^B$; $(c,f)$: $[8.1:3.2:14.5]\,I_{v,\,m}^B/h$. The force-contour levels are: $(b)$: $2.7\,I_{v,\,m}^B u_\tau/h$; $(c)$: $91\,I_{v,\,m}^B u_\tau/h^2$.}
\label{fig:ShPa_rI0030_rI1030}
\end{figure}

\begin{problem}[Strong-sense burst restart]
\label{problem:strongRestart}
    Let $\hat{q}_{\mathrm{I}}(y)=[\hat{v}_{\mathrm{I}}(y)\quad \hat{\omega}_{\mathrm{I}y}(y)]^\mathrm{T}$ and $\hat{q}_{\mathrm{II}}(y)=[\hat{v}_{\mathrm{II}}(y)\quad \hat{\omega}_{\mathrm{II}y}(y)]^\mathrm{T}$ be two states in the decaying and amplification stages, respectively, of the baseline solution (\ref{eq:BSLsolution}). Given $T_R>0$, solve the optimisation problem
    \begin{subeqnarray}
    \label{eq:OPTrecover}
        && J_v(T_R; \hat{v}_\mathrm{I},\hat{v}_\mathrm{II}) = \min_{\theta_\mathrm{off}\,\in[-\pi,\pi)}\left\{\, \min_{\hat{\varphi}(y,t)}\int_0^{T_R} |\!|\hat{\varphi}(y,t)|\!|_E^2\,\mathrm{d}t \,\right\}, \\
        && \textup{subject to } \ \ e^{\mathrm{i}\theta_\mathrm{off}}\hat{v}_{\mathrm{II}}(y) = e^{\mathcal{L}_\mathrm{OS}T_R}\hat{v}_{\mathrm{I}}(y) + \int_0^{T_R} e^{\mathcal{L}_\mathrm{OS}(T_R-t)}\hat{\varphi}(y,t)\,\mathrm{d}t, \quad
    \end{subeqnarray}
    where the minimisation over $\theta_\mathrm{off}$ is to eliminate the effect of phase offset. The result of (\ref{eq:OPTrecover}) is referred to as the `optimal recovery process'. The problem naturally results in an optimal recovering force $\hat{\varphi}$ with a zero $x_\bot$-component all the time. The numerical method for this problem is provided in appendix \ref{subsection:AlgorithmStrongRestart}.
\end{problem}

Appendix \ref{section:RecoveryTime} discusses the dependence of the obtained optimal recovery processes on the prescribed recovery times $T_R$. It indicates that the effect of the recovery force on flow structures has no qualitative differences for $T_R\lesssim 0.5\,h/u_\tau$. Therefore, the following discussion takes $T_R=0.15\,h/u_\tau$, a characteristic recovery time in real turbulence \citep{Jimenez2023}, to represent a typical optimal recovery process, which is shown in figure \ref{fig:ShPa_rI0030_rI1030}.

Unlike the force in the weak-sense re-burst in figure \ref{fig:RB_Cont0}$(f.1)$, the recovery force in figure \ref{fig:ShPa_rI0030_rI1030}$(c)$ exhibits multiple layers separated by interfaces at $y=y_\mathrm{intf}$ at which $(\bnabla\times\boldsymbol{f}^s)_z=0$. Within each layer, the advection speed $a_{(\bnabla\times\boldsymbol{f}^s)_z}$ in figure \ref{fig:ShPa_rI0030_rI1030}$(d)$ satisfies $\partial_y a_{(\bnabla\times\boldsymbol{f}^s)_z} > \partial_yU$, indicating that the recovery force tends to enhance the local shear deformation of $\omega_z$-structures. Under this multilayer recovery force, the flow evolution $(c.1)$-$(c.2)$ in figure \ref{fig:ShPa_rI0030_rI1030} features vortices in a forward-inclined state being broken up into multiple layers, which is similar to the breaking-up process $(e.1)$-$(f.1)$ in figure \ref{fig:RB_Cont0}. A main difference is that the force in figure \ref{fig:RB_Cont0}$(f.1)$ also tends to intensify the broken-up vortices, an effect that is not evident in $(c.1)$-$(c.2)$ of figure \ref{fig:ShPa_rI0030_rI1030}. The subsequent evolution $(c.2)$-$(c.3)$ in figure \ref{fig:ShPa_rI0030_rI1030} is similar to the latent period $(f.2)$-$(f.3)$ in figure \ref{fig:RB_Cont0}. However, the ensuing evolution $(c.3)$-$(c.5)$ in figure \ref{fig:ShPa_rI0030_rI1030} shows that the recovery force tends to reconnect a vortex to a co-rotating, catching-up vortex farther from the wall, forming a merged, backward-inclined vortex that is ready to burst again. This merging of vortices does not appear in the weak-sense re-burst stage $(f.3)$-$(f.5)$ in figure \ref{fig:RB_Cont0}.

To evaluate the merging effect on the restart of a burst, we consider a semi-recovery process by withdrawing the recovery force in the second half of the above recovery process. The result is presented in figure \ref{fig:ShPa_rI0030_rI1030}$(e,f)$. Without the recovery force, the broken-up vortices do not merge during the evolution $(c.3)$-$(f.4)$-$(f.5)$. There is still a subsequent re-burst, as indicated by the later trajectory (magenta) in panel $(a)$, due to the catch-up of co-rotating vortices, as discussed in \S\ref{subsection:weakRestart}, but the peak intensity of the re-burst is only around half of that in the full recovery case.

The above optimal recovery process for a strong-sense re-burst suggests that, besides the breakup effect discussed in \S\ref{subsection:weakRestart}, the merging of co-rotating, catching-up vortices during the late latent period also contributes to a burst restart. In real turbulence, such merging can be a contributory burst-restarting mechanism provided by the nonlinear terms in the Navier-Stokes equations.

\subsection{Sustaining periodic bursts}
\label{subsection:periodicBursts}

We now turn from restarting bursts over a finite time to sustaining them indefinitely. We restrict our discussion to periodic bursts (except for a phase offset), with particular emphasis on the effect of the overlap of two successive bursts. For simplicity, we assume the space-time separable form for a force $\hat{\varphi}(y,t)=\hat{\varphi}_\mathrm{sp}(y)\,\gamma(t)$.

\begin{figure}
  \centerline{\includegraphics[width=1.0\textwidth]{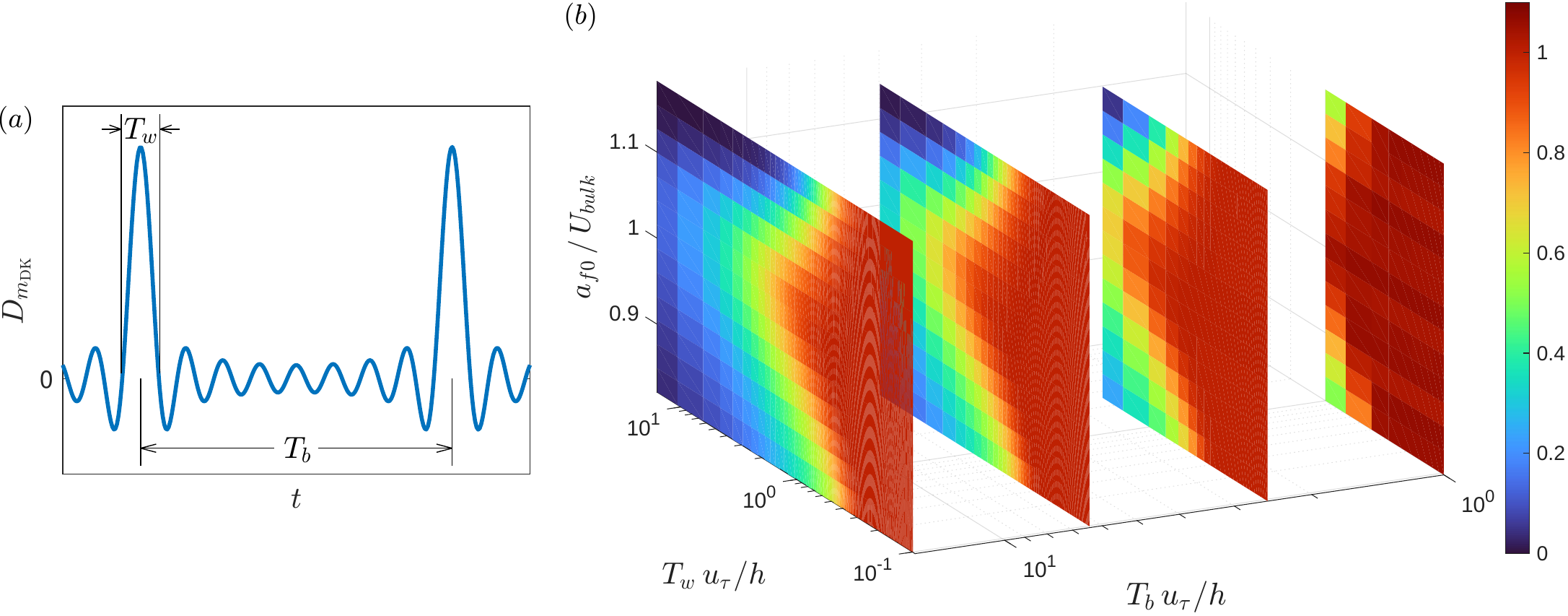}}
  \caption{$(a)$ Sketch of a Dirichlet kernel $D_{m_\mathrm{DK}}$ with $m_\mathrm{DK}=8$. $(b)$ Maximum peak intensity $I_{v,\,m}^P$ of optimal periodic burst solutions as a function of $T_b$, $T_w$, and $a_{f0}$ for the spatial Fourier mode $(n_x,n_z)=(1,0)$. The colour indicates the ratio $I_{v,\,m}^P/I_{v,\,m}^B$ at four $a_{f0}$-$T_w$ planes with $T_b=[16,\,6.3,\,2.5,\,1.0]\,h/u_\tau$. Note that the figure includes only the cases with $m_\mathrm{DK}\geq1$, i.e., $T_w\leq T_b$.}
\label{fig:Amp_Inf}
\end{figure}

The first question to be addressed is modelling the time dependence $\gamma(t)$ of a force that induces periodic bursts with a period $T_b$. It follows that $\gamma(t)$ should also have a period $T_b$. Two other characteristics are important: the time interval $T_w$ over which the force is active during each period, and the mean advection speed $a_{f0}$ of the force. One of the simplest functions having all three characteristics is a harmonic wave modulated by a Dirichlet kernel $D_{m_\mathrm{DK}}$ (as sketched in figure \ref{fig:Amp_Inf}$(a)$), i.e.,
\begin{equation}
\label{eq:forceDK}
    \gamma(t) = \frac{1}{T_b} D_{m_\mathrm{DK}}\!\left(2\pi\frac{t}{T_b}\right) e^{-\mathrm{i}k_x a_{f0} t} = \frac{1}{T_b}\sum_{\ell=-m_\mathrm{DK}}^{m_\mathrm{DK}}\exp\left[{-\mathrm{i}k_x \!\left(a_{f0} + \ell\frac{2\pi}{k_x T_b} \right)t}\right],
\end{equation}
where the integer $m_\mathrm{DK}\geq0$ specifies the time interval $T_w=T_b/m_\mathrm{DK}$, and the prefactor $1/T_b$ makes the time integral of $\gamma(t)$ over a period $T_b$ equal to unity. The form (\ref{eq:forceDK}) reduces to several special cases (regardless of normalisation) in different limits: (i) harmonics, if $T_w\to\infty$ or $m_\mathrm{DK}=0$; (ii) Shannon wavelets, if $0<T_w<\infty$ and $T_b\to\infty$; (iii) impulses, if $T_w\to0$ and $T_b\to\infty$; and (iv) impulse trains (Dirac combs), if $T_w\to0$ and $0<T_b<\infty$.

Assuming the time dependency (\ref{eq:forceDK}), the problem regarding the minimal requirement to sustain periodic bursts with prescribed $T_b$, $T_w$, and $a_{f0}$ can be posed as seeking the force $\hat{\varphi}(y,t)=\hat{\varphi}_\mathrm{sp}(y)\,\gamma(t)$ that maximises the peak intensity of $v$ during a burst period, subject to a unit $|\!|\hat{\varphi}_\mathrm{sp}(y)|\!|_E$.

\begin{problem}[Sustaining periodic bursts]
\label{problem:PeriodicBursts}
    Given $a_{f0}\in\mathbb{R}$,  $T_b>0$ and an integer $m_\mathrm{DK}\geq 0$, solve the optimisation problem
    \begin{subeqnarray}
    \label{eq:PeriodicBursts}
        && I_{v,\,m}^P = \max_{t\in[0,T_b)}\left\{\, \max_{|\!|\hat{\varphi}_\mathrm{sp}(y)|\!|_E\,=\,1}I_v\left[\hat{q}(y,t)\right] \,\right\}, \\ 
        && \textup{where }\ \hat{q}(y,t) = \left\{\frac{1}{T_b}\! \sum_{\ell=-m_\mathrm{DK}}^{m_\mathrm{DK}} \!\!\mathcal{R}\!\left(a_{f0} \!+\! \ell\frac{2\pi}{k_xT_b}\right) \exp\!\left[{-\mathrm{i}k_x \!\left(a_{f0} \!+\! \ell\frac{2\pi}{k_x T_b} \right)t}\right]\right\}\!\hat{\varphi}_\mathrm{sp}(y), \qquad\quad 
    \end{subeqnarray}
    where $\mathcal{R}$ is the resolvent defined as (\ref{eq:defResolvent}). 
    In particular, if $T_w\to\infty$ or $m_\mathrm{DK}=0$, (\ref{eq:PeriodicBursts}) reduces to the classical problem of maximum response to a harmonic force; and if $T_w=T_b/m_\mathrm{DK}\to0$ and $T_b\to\infty$, (\ref{eq:PeriodicBursts}) reduces to the classical problem of maximum unforced transient growth, i.e., (\ref{eq:OPTmaxGrowth}), yielding $I_{v,\,m}^P\to I_{v,\,m}^B$. 
    The problem naturally results in an optimal force $\hat{\varphi}_\mathrm{sp}$ with a zero $x_\bot$-component all the time. The numerical method for this problem is provided in appendix \ref{subsection:AlgorithmPeriodicBursts}.
\end{problem}

\begin{figure}
  \centerline{\includegraphics[width=1.0\textwidth]{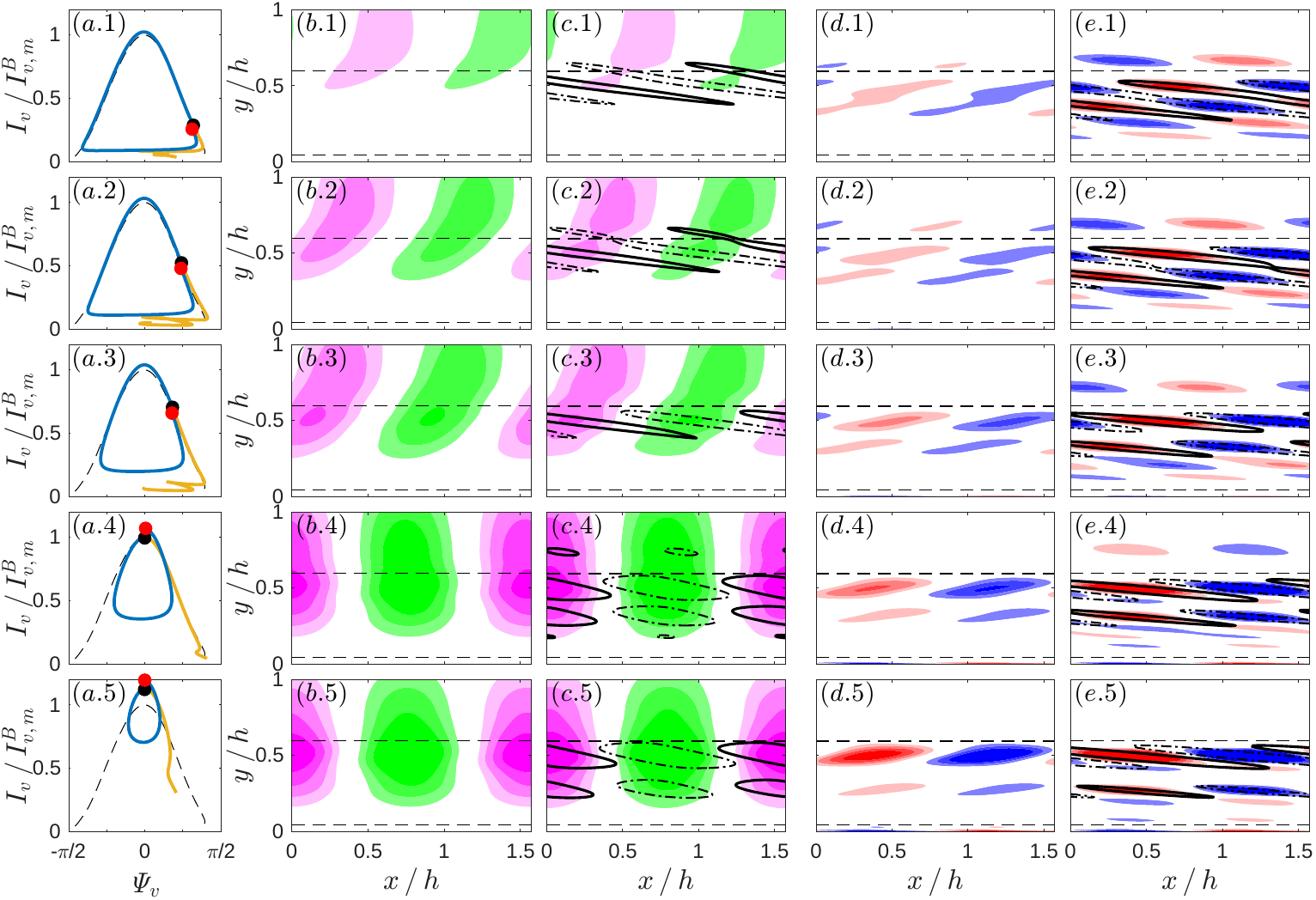}}
  \caption{Optimal periodic-burst solutions under quasi-impulse-train forces for the spatial Fourier mode $(n_x,n_z)=(1,0)$. The forces have identical $a_{f0}\approx U_{bulk}$ and $T_w\approx0.02\,h/u_\tau$ but different $T_b=[1.26,\, 1.12,\, 1.00,\, 0.89,\, 0.71]\,h/u_\tau$ from top to bottom. 
  Column $(a)$: trajectories in the $I_v$-$\mathit{\Psi}_v$ subspace. Blue lines show the periodic solutions, with black and red dots marking the states immediately before and after each quasi-impulse, respectively. Yellow lines show how the states marked by black dots would evolve if the force were withdrawn. The states marked by the black and red dots are shown in columns $(b,d)$ and $(c,e)$, respectively. 
  Columns $(b,c)$: velocity $v$, with magenta and green indicating upward and downward motion, respectively. In $(c)$, solid and dashed contours indicate the upward and downward force $f_y^s$, respectively. 
  Columns $(d,e)$: vorticity $\omega_z$, with red and blue indicating clockwise and counterclockwise rotation, respectively. In $(e)$, solid and dashed contours indicate the clockwise and counterclockwise curl of the force $(\bnabla\times\boldsymbol{f}^s)_z$, respectively. 
  The shading levels given as magnitudes are: $(b,c)$: $[0.71:0.28:1.27]\,I_{v,\,m}^B$; $(d,e)$: $[13.0:5.2:23.4]\,I_{v,\,m}^B/h$. The force-contour levels are: $(c)$: $0.05\,I_{v,\,m}^B$; $(e)$: $11.5\,I_{v,\,m}^B/h$.}
\label{fig:DComb}
\end{figure}

Figure \ref{fig:Amp_Inf}$(b)$ shows the dependency of the maximum peak intensity $I_{v,\,m}^P$ on the parameters $T_b$, $T_w$, and $a_{f0}$. Consider first the $a_{f0}$-$T_w$ plane with the largest period $T_b=16\,h/u_\tau$. For the limiting case with $T_w\gtrsim10\,h/u_\tau$, which essentially recovers the harmonic forcing problem, $I_{v,\,m}^P$ has a maximum at $a_{f0}\approx U_\textit{bulk}$ for fixed $T_w$ and $T_b$. This is broadly consistent with the critical layer framework of \citet{Mckeon2010}, which suggests that the linear response to harmonic forcing with advection speed $a_f$ tends to concentrate around the critical layer at $y=y_c$, where $U(y_c)=a_f$. In this sense, the maximum band-wise intensity is expected when the critical layer lies near the middle of the $y$-band, where $U(y)$ is roughly $U_\textit{bulk}$. 
Such a maximum of $I_{v,\,m}^P$ for fixed $T_w$ and $T_b$ is largely maintained for smaller $T_w$ in the approximate range $1$ - $10\,h/u_\tau$ corresponding to the wavelet forcing cases. As $T_w\lesssim0.5\,h/u_\tau$, $I_{v,\,m}^P$ becomes nearly independent of $a_{f0}$ and is uniformly close to the value of $I_{v,\,m}^B$, which essentially recovers the unforced transient growth problem. As the period $T_b$ decreases from $16$ to $2.5\,h/u_\tau$, the contours of $I_{v,\,m}^P$ on $a_{f0}$-$T_w$ planes change little in the regions with $T_w\leq T_b$. This suggests that for relatively long periods $T_b\gtrsim2\,h/u_\tau$, one burst in the periodic process is largely independent of the others. 

The contours of $I_{v,\,m}^P$ on $a_{f0}$-$T_w$ planes begin to change as $T_b$ further decreases to $1.0\,h/u_\tau$, where $I_{v,\,m}^P$ for some $a_{f0}$ becomes larger than $I_{v,\,m}^B$. This indicates that two successive bursts begin to overlap in their effects and reinforce each other. For several values of $T_b$ around $1.0\,h/u_\tau$, figure \ref{fig:DComb} shows the optimal periodic-burst solutions under quasi-impulse-train forces with $T_w <\!\!< h/u_\tau$. As $T_b$ decreases from $1.26$ to $0.71\,h/u_\tau$, the optimal timing of the quasi-impulsive force gradually shifts from the decaying stage to the moment of peak intensity. Moreover, the state immediately before the quasi-impulsive force increasingly exhibits the characteristics of multilayer vortices as remnants of the previous forcing. Each individual vortex itself, however, is in a highly decayed state after being tilted forward by the local mean shear for an entire period, which precludes a subsequent burst in the absence of forcing. 
In this situation, the role of the quasi-impulsive force is to locally deform an individual vortex into a backward-inclined shape and, meanwhile, to intensify the vorticity. This role of forcing can be regarded as a variant of the breakup effect in the scenario where a multilayer structure is retained in a burst decaying stage.

\subsection[]{Burstiveness of a state:  linearly available energy (LAE)}
\label{subsection:Potential}

We have seen in \S\ref{subsection:weakRestart} and \S\ref{subsection:periodicBursts} that even in an unforced linearised flow, two initial states with similar intensity $I_v$ and inclination angles $\mathit{\Psi}_v$ can exhibit very different behaviour. In particular, an initial state with $\mathit{\Psi}_v>0$ can either decay monotonically or re-burst after a latent period. This indicates that, although $\mathit{\Psi}_v$ well characterises the development of a burst event, it is intrinsically inadequate in characterising its restart. A new quantity is needed to characterise the `burstiveness' of a state, i.e., its potential for future bursting. To this end, we define the linearly available energy (LAE) for a state $\hat{q}(y,t)$ as the integral of $I_v^2$ over its entire future time $[t,+\infty)$ during which the state is assumed to evolve linearly without forcing.

\begin{definition}[Linearly available energy]
    Define the LAE of a state $\hat{q}(y,t)$ as
    \begin{equation}
    \label{eq:DefPotential}
        \mathcal{E}_v\left[\hat{q}(y,t)\right] = \int_{t}^{+\infty} I_v^2\left[e^{\mathcal{L}(\tau-t)}\hat{q}(y,t) \right] \mathrm{d}\tau\,,
    \end{equation}
    in which the integral converges provided that the linearised system is exponentially stable. Using the eigenmode expansion (\ref{eq:EigenExpansion}), $\mathcal{E}_v$ can be expressed explicitly as a quadratic form,
    \begin{subeqnarray}
    \label{eq:DefPotentialMat}
        && \mathcal{E}_v\!\left[\boldsymbol{\kappa}\right] = \boldsymbol{\kappa}_\mathrm{OS}^* \left[\int_{t}^{+\infty}  e^{\mathrm{i}k_x\mathsfbi{c}_\mathrm{OS}^*(\tau-t)} \mathsfbi{M} \, e^{-\mathrm{i}k_x\mathsfbi{c}_\mathrm{OS}^{\vphantom{*}}(\tau-t)} \mathrm{d}\tau\right] \boldsymbol{\kappa}_\mathrm{OS}^{\vphantom{*}} = \boldsymbol{\kappa}_\mathrm{OS}^* \mathsfbi{H} \,\boldsymbol{\kappa}_\mathrm{OS}^{\vphantom{*}}\,, \qquad \\
        && \textup{with }\ \mathsfbi{H}_{[ij]} = \frac{\mathsfbi{M}_{[ij]}}{\mathrm{i}k_x \bigl(c_{\mathrm{OS}j}^{\vphantom{*}}-c_{\mathrm{OS}\,i}^*\bigr) }\,,
    \end{subeqnarray}
    where the subscript $[ij]$ denotes the entry in the $i$-th row and $j$-th column of a matrix. In control theory, $\mathsfbi{H}$ is the observability Gramian expressed in the eigenmode coordinates and under the metric $\mathsfbi{M}$. 
    The LAE $\mathcal{E}_v$ is a measure of a state's linear response integrated over future time and is generally not a conserved quantity. 
    It depends on the linear operator $\mathcal{L}$, the spatial Fourier mode, and the chosen state norm (including the wall-normal band in the definition of $I_v$).
\end{definition}

\begin{figure}
  \centerline{\includegraphics[width=0.75\textwidth]{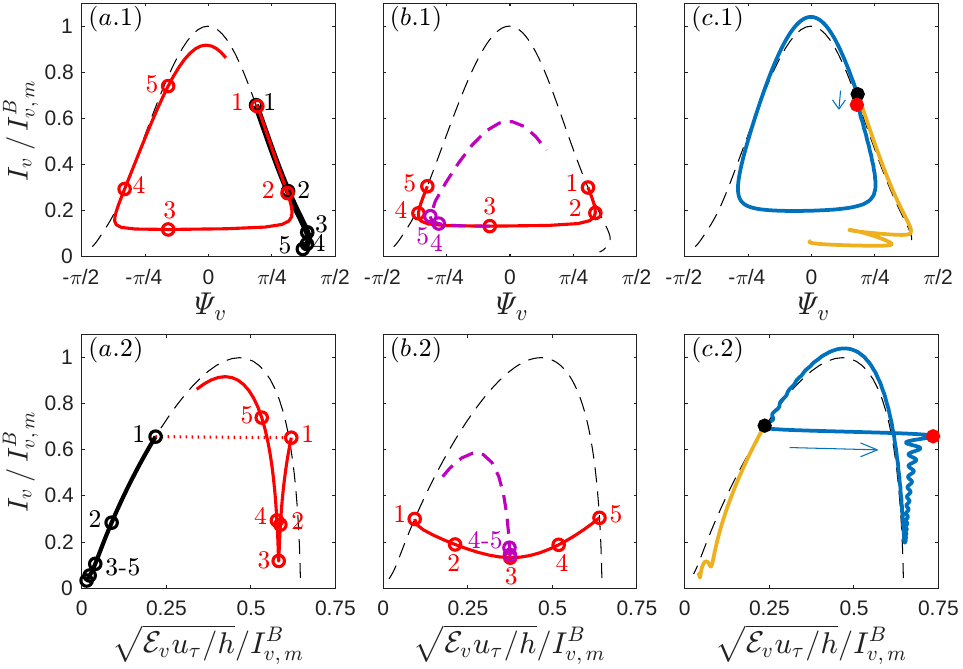}}
  \caption{Trajectories in the $I_v$-$\mathit{\Psi}_v$ subspace (upper row) and the $I_v$-$\sqrt{\mathcal{E}_v}$ subspace (lower row) for: $(a)$ the weak-sense re-burst in figure \ref{fig:RB_Cont0}; $(b)$ the strong-sense re-burst in figure \ref{fig:ShPa_rI0030_rI1030}; and $(c)$ the periodic bursts for the third row in figure \ref{fig:DComb}. 
  Colours and symbols are the same as in the previous corresponding figures. 
  }
\label{fig:threeCases}
\end{figure}

\begin{figure}
  \centerline{\includegraphics[width=1.00\textwidth]{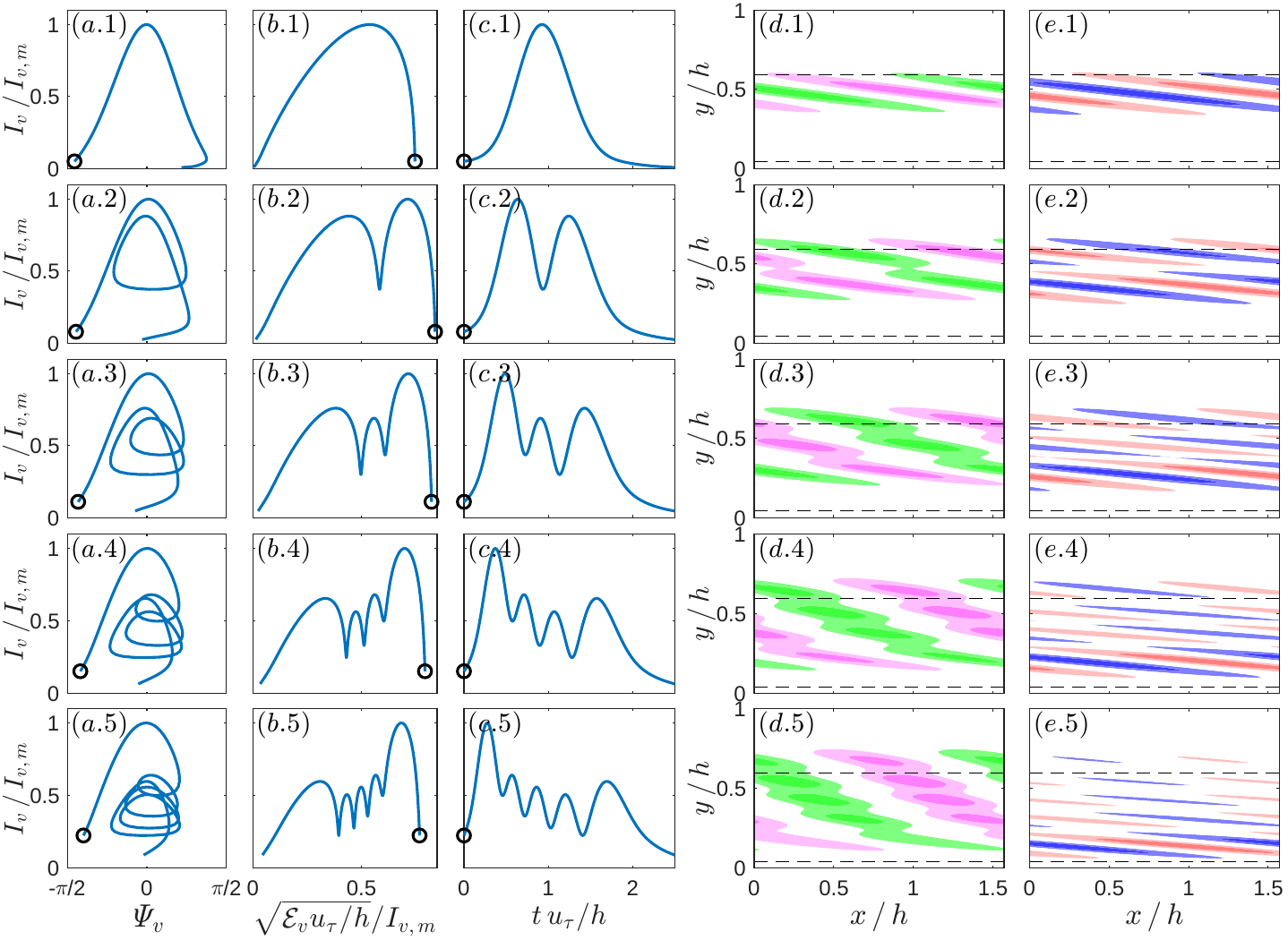}}
  \caption{The first five eigenstates with the largest LAE under a unit two-component energy norm for the spatial Fourier mode $(n_x,n_z)=(1,0)$. The eigenstates $\hat{q}^{Hi}$ are shown in columns $(d,e)$ and marked by black circles in columns $(a,b,c)$. Blue lines in $(a,b,c)$ show their subsequent linear evolution. 
  Columns $(a,b)$: trajectories in the $I_v$-$\mathit{\Psi}_v$ and $I_v$-$\sqrt{\mathcal{E}_v}$ subspaces, respectively. 
  Column $(c)$: time dependence of $I_v$. 
  Column $(d)$: velocity $v$, with magenta and green indicating upward and downward motion, respectively. 
  Column $(e)$: vorticity $\omega_z$, with red and blue indicating clockwise and counterclockwise rotation, respectively. 
  In $(a,b,c)$, $I_v$ is normalised by the peak intensity of individual evolution, $I_{v,\,m}$, whose values are $[0.92,\,0.63,\,0.49,\,0.40,\,0.32]\,I_{v,\,m}^B$ from top to bottom. 
  In $(d,e)$, the shading levels given as magnitudes are $[0.56, 0.78]$ times by the maximum value in the field of each panel.}
\label{fig:LLT_Eig}
\end{figure}

\begin{figure}
  \centerline{\includegraphics[width=1.00\textwidth]{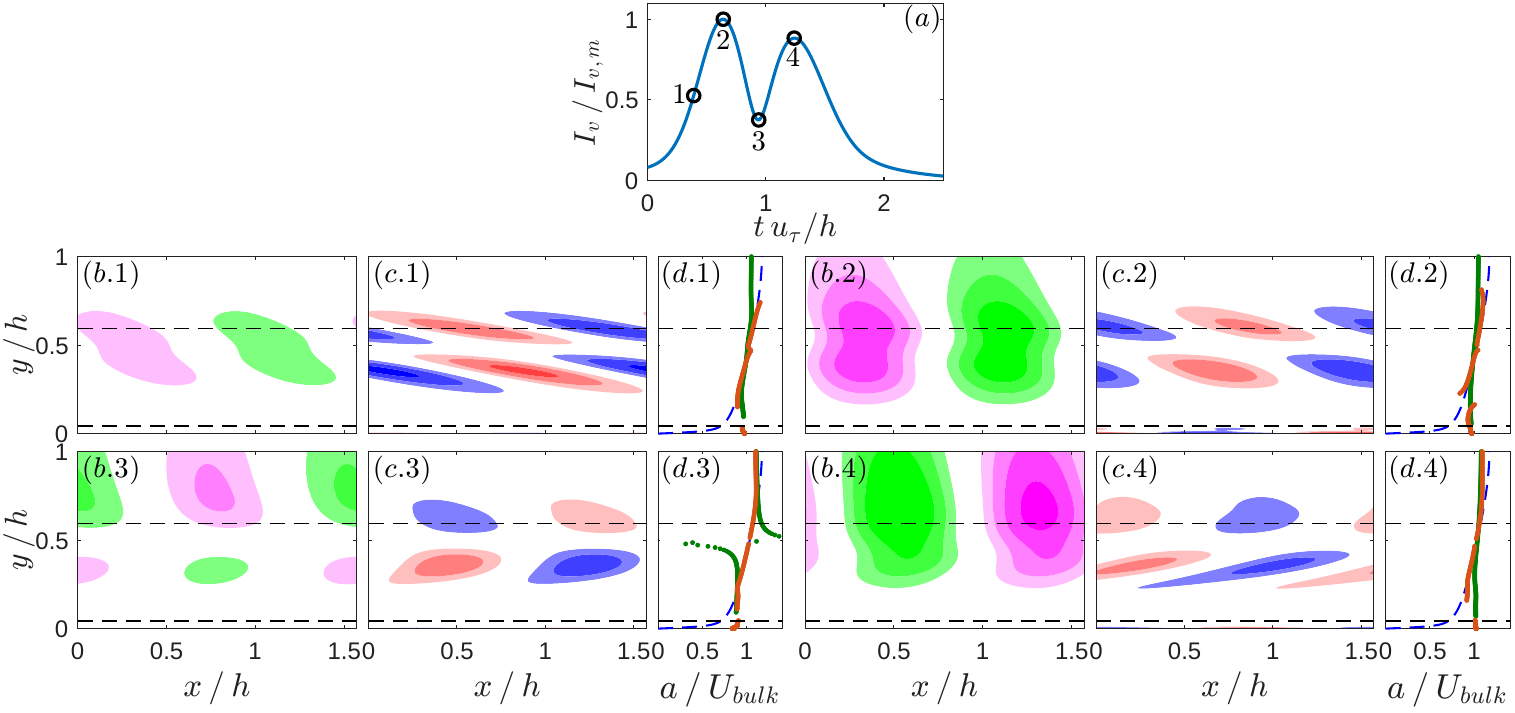}}
  \caption{Linear evolution of the eigenstate with the second largest LAE for the spatial Fourier mode $(n_x,n_z)=(1,0)$. The flow fields in $(b,c)$ and advection speed profiles in $(d)$ correspond to the four time instants marked 1-4 in $(a)$. All flow fields are viewed in a reference frame moving at the constant streamwise speed $U_\textit{bulk}$. 
  Panels $(b)$: velocity $v$, with magenta and green indicating upward and downward motion, respectively. 
  Panels $(c)$: vorticity $\omega_z$, with red and blue indicating clockwise and counterclockwise rotation, respectively. 
  Panels $(d)$: $y$-profiles of the advection speeds $a$ of $v$ (green) and $\omega_z$ (brown). Blue dashed lines indicate the mean velocity $U$. The shading levels given as magnitudes are: $(b)$: $[0.75:0.30:1.65]\,I_{v,\,m}$; $(c)$: $[12.5:5.0:22.5]\,I_{v,\,m}/h$.}
\label{fig:LLT_Eig2Evol}
\end{figure}

Substituting the definition into (\ref{eq:governEqnCoef}), we obtain the time derivative of $\mathcal{E}_v$ as
\begin{equation}
\label{eq:dtPotential}
    \partial_t\mathcal{E}_v = - I_v^2 + 2\mathrm{Re}\!\left(\boldsymbol{\kappa}_\mathrm{OS}^*\mathsfbi{H}\boldsymbol{\chi}_\mathrm{OS}^{\vphantom{*}}\right),
\end{equation}
where `$\mathrm{Re}$' stands for the real part. It follows that in an unforced linearised flow, we have $\partial_t\mathcal{E}_v = - I_v^2$, indicating that $\mathcal{E}_v$ always decreases with time. Therefore, to restart a burst, one indispensable role of the external forcing is to increase $\mathcal{E}_v$. Figure \ref{fig:threeCases} shows the trajectories of some solutions presented in \S\ref{subsection:weakRestart}-\S\ref{subsection:periodicBursts} in both the $I_v$-$\mathit{\Psi}_v$ and $I_v$-$\sqrt{\mathcal{E}_v}$ subspaces (in constructing the latter, we use the intensity-like quantity $\sqrt{\mathcal{E}_v}$ rather than the energy-like quantity $\mathcal{E}_v$, because $\sqrt{\mathcal{E}_v}$, like $I_v$, scales linearly with the state amplitude). For a typical burst development, the forward-tilting process in the $I_v$-$\mathit{\Psi}_v$ subspace manifests as a continuous decrease in $\mathcal{E}_v$ in the $I_v$-$\sqrt{\mathcal{E}_v}$ subspace. The burst-restarting forces do not necessarily alter $I_v$ or $\mathit{\Psi}_v$, but they need to significantly increase $\mathcal{E}_v$ to a sufficiently high level. 

The LAE $\mathcal{E}_v$ will be further used in the analysis of real turbulence in section \ref{section:RealTurb}. Before proceeding, let us first develop some intuition for the states that have relatively large $\mathcal{E}_v$. Such intuition can be obtained by observing the eigenstates corresponding to the first few largest $\mathcal{E}_v$ under a unit energy norm of a state.

\begin{problem}[States with high linearly available energy]
\label{problem:EigenPotential}
For a state $\hat{q}$ with a unit two-component energy norm, i.e.,
\begin{equation}
\label{eq:unitEngeryNorm}
    |\!|\,\hat{q}\,|\!|_{E_\mathrm{2C}} = |\!|\,\boldsymbol{\theta}\,|\!|_2 = 1, \quad \textup{where }\: \boldsymbol{\theta}=\mathsfbi{E}^{1/2}\boldsymbol{\kappa}_\mathrm{OS}^{\vphantom{*}}\,,
\end{equation}
its LAE is
\begin{equation}
\label{eq:EigenPotential}
    \mathcal{E}_v\!\left[\hat{q}\right] = \boldsymbol{\kappa}_\mathrm{OS}^* \mathsfbi{H} \,\boldsymbol{\kappa}_\mathrm{OS}^{\vphantom{*}} = \boldsymbol{\theta}^*\mathsfbi{\Tilde{H}}\,\boldsymbol{\theta}\,,\quad\textup{where }\: \mathsfbi{\Tilde{H}} = \mathsfbi{E}^{-1/2*}\mathsfbi{H}\,\mathsfbi{E}^{-1/2}\,.
    \end{equation}
    Calculate the eigenstates $\hat{q}^{Hi},\,i=1,2,\cdots$ that correspond to the first few largest eigenvalues of $\mathsfbi{\Tilde{H}}$.
\end{problem}

Figure \ref{fig:LLT_Eig} shows the first five eigenstates obtained from problem \ref{problem:EigenPotential} for the mode $(n_x,n_z)=(1,0)$. The first one $\hat{q}^{H1}$ is similar to the baseline initial state $\hat{q}_0^B$, both of which are characterised by backward-inclined vortices and experience typical burst developments in their subsequent linear evolutions. The second eigenstate $\hat{q}^{H2}$ is characterised by two layers of backward-inclined vortices. This structure leads to two instances of co-rotating catch-up in its subsequent linear evolution, as shown in figure \ref{fig:LLT_Eig2Evol}, thereby giving rise to two successive bursts for $I_v$ at different times before the final monotonic decay. The evolution between the two burst events is similar to the weak-sense re-burst in \S\ref{subsection:weakRestart}, but the `re-burst' for $\hat{q}^{H2}$ originates from the vortex breakup that occurs even before the earlier burst event. 

As the eigenstate order increases, the eigenstates in figure \ref{fig:LLT_Eig} exhibit an increasing number of vortex layers. Correspondingly, their subsequent linear evolutions involve an increasing number of $I_v$-peak events, although the peak intensity $I_{v,\,m}$ tends to decrease relative to $I_{v,\,m}^B$. A more detailed observation of the linear evolution of a higher-order eigenstate in appendix \ref{section:HighEigenState} indicates that an $I_v$-peak event is the combined effect of two mechanisms: the forward-tilting of vortices within each vortex layer; and the co-rotating and counter-rotating catch-up of vortices across layers.

The eigenstate analysis above suggests that states with relatively high LAE generally have backward-inclined and/or multilayer structures of vortices, corroborating the essential role of breakup in burst restarts, as suggested in \S\ref{subsection:weakRestart}-\S\ref{subsection:periodicBursts}. In addition, the evolution of these eigenstates indicates that even in an unforced linearised flow, an initial state with a multilayer structure of broken-up vortices can lead to multiple successive burst events in its subsequent evolution. In real turbulence, this means that the nonlinear origin of a detected burst event may be seeded even before several preceding bursts.

\section{Burst restarts in real turbulence}
\label{section:RealTurb}

We now analyse real turbulence in the channel from the database introduced in \S\ref{subsection:Database}. The nonlinear terms, which are treated as external forces computed \textit{a posteriori} in \S\ref{section:Requirements}, are available in the database. 
Whereas the model solutions in \S\ref{section:Requirements} can be analysed by tracking individual burst events, applying such an event-tracking approach to real turbulence is significantly more complicated because, unlike the model solutions, states in real turbulence are widely dispersed across the space of realisable flow states and evolve chaotically in time. Accordingly, this section projects turbulence in two-dimensional subspaces and focuses on its statistical behaviour conditioned on different regions of these subspaces.

\subsection[]{Linear and nonlinear effects in the burst cycle}
\label{subsection:Arrows}

We begin the analysis of turbulence by distinguishing between the linear and nonlinear contributions to the change in a turbulence state based on equation (\ref{eq:OSSQconciseFourier}). For a state $\hat{q}$, the linear and nonlinear contributions to its instantaneous rate of change $\partial_t\hat{q}$ are $\mathcal{L}\hat{q}$ and $\hat{\varphi}$, respectively. We accordingly calculate the averages of the three rates conditioned on each cell in the coarse-grained $I_v$-$\mathit{\Psi}_v$ subspace and represent the results as instantaneous orbit velocity $(\partial_t\mathit{\Psi}_v,\partial_tI_v)$. 

\begin{figure}
  \centerline{\includegraphics[width=1.0\textwidth]{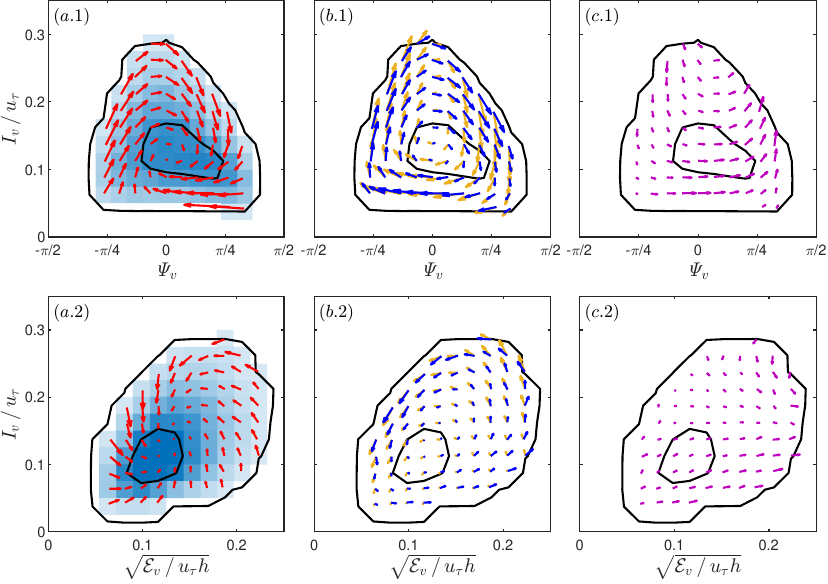}}
  \caption{Short-time displacement $(a)$ and instantaneous orbit velocity $(b,c)$ of turbulence states projected onto the $I_v$-$\mathit{\Psi}_v$ subspace (upper row) and the $I_v$-$\sqrt{\mathcal{E}_v}$ subspace (lower row) for the spatial Fourier mode $(n_x,n_z)=(1,0)$. 
  In $(a)$, the blue shading represents the probability density, and the red arrows indicate the average displacement over a time interval of approximately $0.05\,h/u_\tau$ (note that panel $(a.1)$ is a duplicate of panel $(a)$ from figure \ref{fig:PFO}). 
  In $(b)$, the arrows indicate the average instantaneous orbit velocity, with the blue ones for real turbulence (i.e., including both the linear and nonlinear contributions) and the yellow ones for the linear contribution. 
  In $(c)$, the purple arrows indicate the nonlinear contributions to the instantaneous orbit velocity. 
  The length of an arrow in $(b,c)$ indicates the orbit velocity multiplied by $0.05\,h/u_\tau$. 
  In each panel, the black contours contain 30\% and 95\% of the probability mass, and each cell within the 95\% contour contains at least 100 samples.}
\label{fig:arrows}
\end{figure}

The resulting orbit velocity $(\partial_t\mathit{\Psi}_v,\partial_tI_v)$ for the straight Fourier mode $(n_x,n_z)=(1,0)$ is shown in panels $(b.1)$ and $(c.1)$ of figure \ref{fig:arrows}. In $(b.1)$, a clockwise trend of orbit velocity is observed not only for $\partial_t\hat{q}$ but also for $\mathcal{L}\hat{q}$. In particular, the orbit velocity for $\mathcal{L}\hat{q}$ also exhibits a backward-tilting trend in the restart stage. This is consistent with the results of the linearised models in \S\ref{subsection:weakRestart}, \S\ref{subsection:periodicBursts}, and \S\ref{subsection:Potential}, suggesting that the model-inferred counter-rotating catch-up process probably characterises the restart stage in real turbulence. In $(c.1)$, the orbit velocity for $\hat{\varphi}$, which represents the nonlinear effect, exhibits a counterclockwise trend centred outside of the probability mass. In the amplification and decaying stages of bursts, the nonlinearity tends to decrease and increase $I_v$, respectively, suggesting that, on average, it has a moderating effect that counteracts burst development. 
In the restart stage, the nonlinearity tends to increase $\mathit{\Psi}_v$. This trend is somewhat counter-intuitive considering that previous studies \citep{Jimenez2013,Jimenez2023,Encinar2020} suggested that nonlinearity causes burst restart, which is an apparent backward-tilting process. 
However, the results in $(c.1)$ show that, although nonlinearity is the ultimate cause of any finite-amplitude bursts, it does not manifest itself as a backward-tilting mechanism on average. 
That is, the inclination angle $\mathit{\Psi}_v$ is inadequate for revealing the essential role of nonlinearity in burst restarts in real turbulence.

Besides the inclination angle $\mathit{\Psi}_v$, the LAE $\mathcal{E}_v$, defined in \S\ref{subsection:Potential}, provides another perspective for understanding the nonlinear effects in real turbulence. Panel $(a.2)$ of figure \ref{fig:arrows} shows the short-time mean displacement of states in the $I_v$-$\sqrt{\mathcal{E}_v}$ subspace, where the burst cycle is reflected by the counterclockwise trend. In general, the upper-right edge of the probability distribution, which is characterised by increases in $I_v$ with decreasing $\mathcal{E}_v$, represents the amplification stage of bursts, while the upper-left edge, where $I_v$ decreases with largely decreasing $\mathcal{E}_v$, represents the decaying stage. The lower part of the probability distribution with $I_v\lesssim0.1 \,u_\tau$, which approximately corresponds to the restart stage, features both trends of increases in $\mathcal{E}_v$ and a spiral towards the centre.

Representing the burst cycle in the $I_v$-$\sqrt{\mathcal{E}_v}$ subspace helps reveal the nonlinear effects more clearly, especially in the restart stage. We have shown from equation (\ref{eq:dtPotential}) that $\mathcal{E}_v$ always decreases in the absence of nonlinear terms. Therefore, to close the burst cycle in real turbulence, one indispensable effect of the nonlinearity is to compensate for the consumption of $\mathcal{E}_v$ by the linear term. This effect, as illustrated by the nonlinear contribution to the orbit velocity in panel $(c.2)$ of figure \ref{fig:arrows}, is relatively weak in the amplification and decaying stages, but becomes particularly significant in the restart stage. It is this `recharge' of $\mathcal{E}_v$ by nonlinearity in the restart stage that `powers' the subsequent burst of $I_v$.

\begin{figure}
  \centerline{\includegraphics[width=1.0\textwidth]{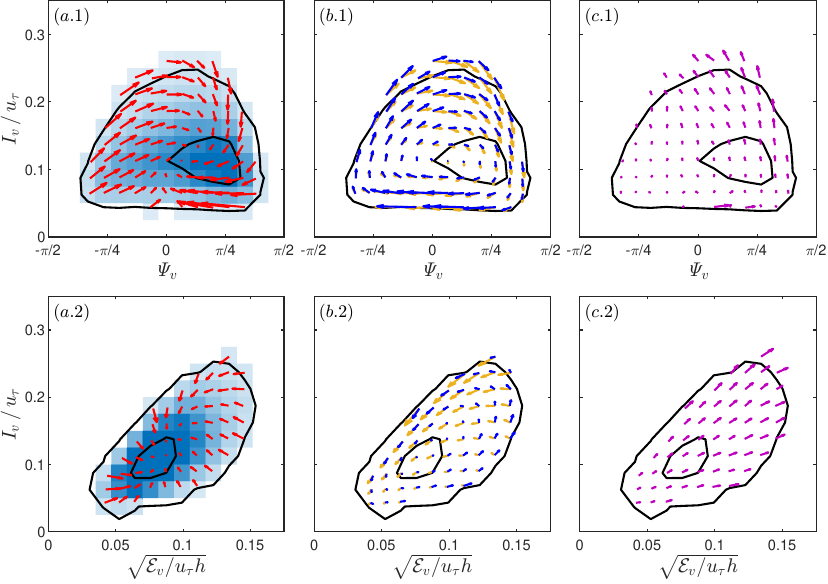}}
  \caption{Short-time displacement $(a)$ and instantaneous orbit velocity $(b,c)$ of turbulence states projected onto the $I_v$-$\mathit{\Psi}_v$ subspace (upper row) and the $I_v$-$\sqrt{\mathcal{E}_v}$ subspace (lower row) for the spatial Fourier mode $(n_x,n_z)=(1,1)$. 
  All graphical conventions are the same as in figure \ref{fig:arrows}.}
\label{fig:arrows_oblq}
\end{figure}

The nonlinear effects for the oblique mode $(n_x,n_z)=(1,1)$ differ from those for the straight mode in some respects, as shown in figure \ref{fig:arrows_oblq}. In panels $(a.1)$ and $(a.2)$, the short-time displacement of states in both subspaces exhibits more apparent trends of inward spirals. In the $I_v$-$\mathit{\Psi}_v$ subspace, panel $(c.1)$ shows that the main effect of nonlinearity is to increase $I_v$ around the upper edge of the probability distribution, i.e., the nonlinear terms directly inject energy into $I_v$ to intensify an existing burst event. Such a role of nonlinearity is distinct from that in the straight mode case, where nonlinearity acts to moderate bursts. In the $I_v$-$\sqrt{\mathcal{E}_v}$ subspace, panel $(c.2)$ shows that the nonlinearity acts to increase $\mathcal{E}_v$ not only in the restart stage but also throughout the entire subspace. Moreover, for each cell, the nonlinear contribution to the orbit velocity $(\partial_t\sqrt{\mathcal{E}_v},\partial_t I_v)$ roughly satisfies $\partial_t\sqrt{\mathcal{E}_v}/\partial_t I_v\approx\sqrt{\mathcal{E}_v}/I_v$, i.e., $\partial_t(\mathcal{E}_v/I_v^2)\approx0$. This suggests that, on average, the nonlinearity increases a state's band-wise intensity but does not bias the $y$-profile changes in any particular direction. All of the above characteristics may be associated with the fact that the burst cycle for the oblique mode is more affected by nonlinearity and is less coherent than for the straight mode.

In addition to the short-time tendencies illustrated in figures \ref{fig:arrows} and \ref{fig:arrows_oblq}, appendix \ref{section:lossPred} examines finite-time predictability within the $I_v$-$\mathit{\Psi}_v$ and $I_v$-$\sqrt{\mathcal{E}_v}$ subspaces. It shows that low-$I_v$ initial states with smaller $\mathit{\Psi}_v$ or larger $\mathcal{E}_v$ are more likely to evolve into higher-$I_v$ states over $0.2$-$0.3\,h/u_\tau$. This tendency is observed not only under the assumed linear evolution but also under the real-turbulence evolution. It suggests that $\mathit{\Psi}_v$ and $\mathcal{E}_v$ have some predictive power for subsequent bursting in real turbulence, although their relative predictive power cannot be determined from the available data.

\subsection[]{Flow patterns in the burst cycle}
\label{subsection:Patterns}

Building on the understanding of the linear and nonlinear effects from \S\ref{subsection:Arrows}, we now visualise how these effects manifest themselves in the flow patterns in different stages of the burst cycle. Specifically, with the burst cycle represented in the coarse-grained $I_v$-$\sqrt{\mathcal{E}_v}$ subspace, our focus is to extract relevant information from the turbulence states $\hat{q}(t)$ conditioned on each individual cell. Such cell-conditioned information should capture both representative flow patterns and their short-time dynamical evolution, with the latter providing a basis for separating the linear and nonlinear effects on changes in flow states. To this end, we introduce a procedure for extracting the dominant joint mode in \S\ref{subsubsection:AdaptedPOD} and present the results in \S\ref{subsubsection:FlowPatterns}.

\subsubsection{Extraction of the dominant joint mode (DJM)}
\label{subsubsection:AdaptedPOD}

We concentrate on the component $\hat{v}$ of a state $\hat{q}=[\hat{v}\quad\hat{\omega}_y]^\mathrm{T}$. For a long time sequence of turbulence states $\hat{v}(t)$, suppose that $t_1$, $t_2$, $\cdots$, $t_{n_c}$ are the times at which the states belong to a given cell in the coarse-grained $I_v$-$\sqrt{\mathcal{E}_v}$ subspace. Regard $\hat{v}(t_1)$, $\hat{v}(t_2)$, $\cdots$, $\hat{v}(t_{n_c})$ as $n_c$ sample states. Following the concept of standard pattern extraction techniques, such as proper orthogonal decomposition \citep[POD,][]{Berkooz1993}, a representative flow pattern can be defined as the dominant mode $\hat{v}^\mathrm{D}$ capturing the highest intensity $I_v$ in these samples:
\begin{equation}
\label{eq:DominantSingleMode}
    \hat{v}^\mathrm{D} = \arg\max_{I_v[\hat{\upsilon}]\,=\,1} \mathbb{E}\left[\,\left|\,\langle\,\hat{v}(t_i),\,\hat{\upsilon}\,\rangle\,\right|^2\,\right],
\end{equation}
where $\langle\,\cdot\,,\,\cdot\,\rangle$ denotes the inner product associated with (\ref{eq:defIntensity}) and is defined as
\begin{equation}
\label{eq:originalInnerProduct}
    \langle\,\hat{v}_1,\,\hat{v}_2\,\rangle = \frac{1}{y_\mathrm{upp}-y_\mathrm{low}}\int_{y_\mathrm{low}}^{y_\mathrm{upp}} \,\hat{v}_1^*\,\hat{v}_2 \, \mathrm{d}y\,.
\end{equation}
The representativeness of $\hat{v}^\mathrm{D}$ can be quantified by the fraction of the total energy captured by $\hat{v}^\mathrm{D}$:
\begin{equation}
\label{eq:energyFractionSingle}
    \rho\!\left[\hat{v}^\mathrm{D}\right] = \frac{\mathbb{E}\left[\,\left|\,\langle\,\hat{v}(t_i),\,\hat{v}^\mathrm{D}\,\rangle\,\right|^2\,\right]}{\mathbb{E}\left[\,\langle\,\hat{v}(t_i),\,\hat{v}(t_i)\,\rangle\,\right]}.
\end{equation}

Instead of directly solving (\ref{eq:DominantSingleMode}), we first pair each sample state $\hat{v}(t_i)$ with the one from a short time $\Delta t$ later in the time sequence, $\hat{v}(t_i\!+\!\Delta t)$ (which may or may not belong to the same cell), where $0<\Delta t<\!<h/u_\tau$, thereby constructing $n_c$ joint samples $\hat{\eta}(t_i)$:
\begin{equation}
\label{eq:augmentedStates}
    \hat{\eta}(t_1)\!=\! 
    \left[\begin{array}{l}
    \hat{v}(t_1) \\
    \hat{v}(t_1\!+\!\Delta t)
    \end{array}\right],\ \ 
    \hat{\eta}(t_2)\!=\! 
    \left[\begin{array}{l}
    \hat{v}(t_2) \\
    \hat{v}(t_2\!+\!\Delta t)
    \end{array}\right],\ \ 
    \cdots,\ \ 
    \hat{\eta}(t_{n_c})\!=\! 
    \left[\begin{array}{l}
    \hat{v}(t_{n_c}) \\
    \hat{v}(t_{n_c}\!+\!\Delta t)
    \end{array}\right].
\end{equation}
We then search for the dominant joint mode (DJM) $\hat{\eta}^\mathrm{D}$ defined as
\begin{equation}
\label{eq:DominantJointMode}
    \hat{\eta}^\mathrm{D} = 
    \left[\begin{array}{c}
        \hat{v}_\alpha^\mathrm{D}  \\
        \hat{v}_\beta^\mathrm{D}
    \end{array}\right] = \arg\max_{I_\eta[\hat{\zeta}]\,=\,1} \mathbb{E}\left[\,\left|\,\langle\,\hat{\eta}(t_i),\,\hat{\zeta}\,\rangle\,\right|^2\,\right],
\end{equation}
where the inner product is defined as a natural extension of (\ref{eq:originalInnerProduct}): 
\begin{equation}
\label{eq:augmentedInnerProduct}
    \langle\,\hat{\eta}_1,\,\hat{\eta}_2\,\rangle = \frac{1}{2(y_\mathrm{upp}-y_\mathrm{low})}\int_{y_\mathrm{low}}^{y_\mathrm{upp}} \,\hat{\eta}_1^*\,\hat{\eta}_2 \, \mathrm{d}y\,,
\end{equation}
and $I_\eta$ is the associated intensity, which satisfies $I_\eta^2\left[\hat{\eta}(t_i)\right] = \langle\,\hat{\eta}(t_i),\,\hat{\eta}(t_i)\,\rangle = (I_v^2\left[\hat{v}(t_i)\right]+I_v^2\left[\hat{v}(t_i\!+\!\Delta t)\right])\,/\,2$. The corresponding representativeness of $\hat{\eta}^\mathrm{D}$ is defined as
\begin{equation}
\label{eq:energyFractionJoint}
    \rho\!\left[\hat{\eta}^\mathrm{D}\right] = \frac{\mathbb{E}\left[\,\left|\,\langle\,\hat{\eta}(t_i),\,\hat{\eta}^\mathrm{D}\,\rangle\,\right|^2\,\right]}{\mathbb{E}\left[\,\langle\,\hat{\eta}(t_i),\,\hat{\eta}(t_i)\,\rangle\,\right]}.
\end{equation}
It follows that $\hat{\eta}^\mathrm{D}$ resulting from (\ref{eq:DominantJointMode}) captures the highest intensity $I_\eta^2$ -- the average of $I_v^2$ over two $\hat{v}$ states separated by $\Delta t$ -- in the joint samples. Appendix \ref{section:dominantJointMode} shows that the two states $\hat{v}_\alpha^\mathrm{D}$ and $\hat{v}_\beta^\mathrm{D}$ constituting $\hat{\eta}^\mathrm{D}$ deviate from the standard dominant mode $\hat{v}^\mathrm{D}$ by terms of order $O(\Delta t)$, i.e.,
\begin{subeqnarray}
\label{eq:JointModeOrder}
    \hat{v}_\alpha^\mathrm{D} &=& \hat{v}^\mathrm{D} + O(\Delta t)\,, \\
    \hat{v}_\beta^\mathrm{D} &=& \hat{v}^\mathrm{D} + O(\Delta t)\,, \\
    \rho\!\left[\hat{\eta}^\mathrm{D}\right] &=&  \rho\!\left[\hat{v}^\mathrm{D}\right] + O(\Delta t)\,. 
\end{subeqnarray}

The two states in  $\hat{\eta}^\mathrm{D} = [\hat{v}_\alpha^\mathrm{D}\quad\hat{v}_\beta^\mathrm{D}]^\mathrm{T}$ are coupled, retaining information on the temporal correlation between the two states in individual joint samples $\hat{\eta}(t_i)=[\hat{v}(t_i)\quad\hat{v}(t_i\!+\!\Delta t)]^\mathrm{T}$. The two states in $\hat{\eta}(t_i)$ are the initial and final states of a short-time evolution and are thus correlated via equation (\ref{eq:GenSolForm}) as
\begin{equation}
\label{eq:individualCorrelation}
    \hat{v}(t_i\!+\!\Delta t) - e^{\mathcal{L}_\mathrm{OS}\Delta t} \hat{v}(t_i) = \hat{r}(t_i) \Delta t\,,
\end{equation}
where
\begin{equation}
\label{eq:defNonlinearEff}
    \hat{r}(t_i) = \frac{1}{\Delta t}\int_{t_i}^{t_i+\Delta t} e^{\mathcal{L}_\mathrm{OS}(t_i+\Delta t -\tau)} \hat{f}^s_y(\tau) \,\mathrm{d}\tau
\end{equation}
represents the short-time-averaged nonlinear effect over $[t_i,\, t_i\!+\!\Delta t]$. Correspondingly, appendix \ref{section:dominantJointMode} shows that the two states in $\hat{\eta}^\mathrm{D}$ satisfy
\begin{equation}
\label{eq:CorrelationDominantJoint}
    \hat{v}_\beta^\mathrm{D} - e^{\mathcal{L}_\mathrm{OS}\Delta t}\hat{v}_\alpha^\mathrm{D} = \frac{\mathbb{E}\left[\,\hat{r}(t_i)\,\langle\,\hat{v}(t_i),\,\hat{v}^\mathrm{D}\,\rangle\,\right]}{\mathbb{E}\left[\,\left|\,\langle\,\hat{v}(t_i),\,\hat{v}^\mathrm{D}\,\rangle\,\right|^2\,\right]}\Delta t + O(\Delta t^2)\,.
\end{equation}
On the right-hand side of (\ref{eq:CorrelationDominantJoint}), the cross-correlation between $\hat{r}(t_i)$ and $\langle\,\hat{v}(t_i),\,\hat{v}^\mathrm{D}\,\rangle$ indicates the mean tendency of the  nonlinearity associated with the state component along the dominant mode $\hat{v}^\mathrm{D}$. This suggests that $\hat{v}_\alpha^\mathrm{D}$ and $\hat{v}_\beta^\mathrm{D}$ can be interpreted as the initial and final states of a notional, representative short-time evolution, over which the cumulative nonlinear effect $\hat{v}_\beta^\mathrm{D} - e^{\mathcal{L}_\mathrm{OS}\Delta t}\hat{v}_\alpha^\mathrm{D}$ captures the mean tendency of the nonlinearity in the joint samples.

The above discussion suggests that the DJM $\hat{\eta}^\mathrm{D} = [\hat{v}_\alpha^\mathrm{D}\quad\hat{v}_\beta^\mathrm{D}]^\mathrm{T}$ obtained from (\ref{eq:DominantJointMode}) contains key information on both the flow pattern and its short-time dynamics for the turbulence states conditioned on a given cell in the $I_v$-$\sqrt{\mathcal{E}_v}$ subspace. The state $\hat{v}_\alpha^\mathrm{D}$ or $\hat{v}_\beta^\mathrm{D}$ captures the dominant flow pattern with the highest intensity, as indicated by (\ref{eq:JointModeOrder}), and the difference between $\hat{v}_\beta^\mathrm{D}$ and $e^{\mathcal{L}_\mathrm{OS}\Delta t}\hat{v}_\alpha^\mathrm{D}$ captures the mean nonlinear effect associated with the pattern, as indicated by (\ref{eq:CorrelationDominantJoint}).

\subsubsection{Results of flow patterns}
\label{subsubsection:FlowPatterns}

\begin{figure}
  \centerline{\includegraphics[width=1.00\textwidth]{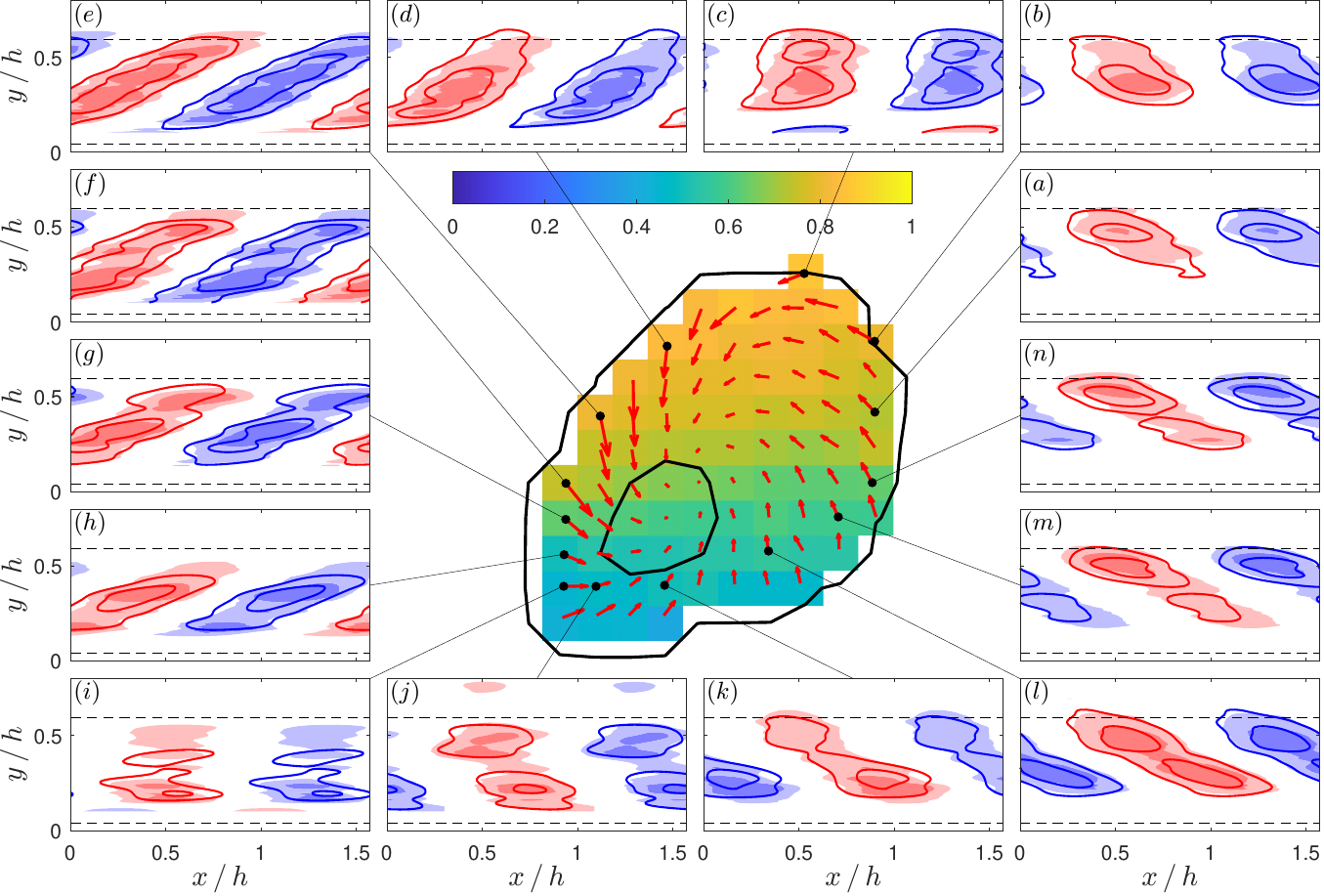}}
  \caption{Flow patterns represented by the dominant joint modes conditioned on different stages of the burst cycle for the spatial Fourier mode $(n_x,n_z)=(1,0)$. 
  The central panel is a duplicate of panel $(a.2)$ from figure \ref{fig:arrows}, except that the colour indicates the representativeness $\rho\!\left[\hat{\eta}^\mathrm{D}\right]$. 
  Panels $(a)$-$(n)$, arranged in counterclockwise order, are flow fields corresponding to 14 cells in the central panel. 
  In each of $(a)$-$(n)$, the contour lines and contour shading represent the linearly and nonlinearly evolved states, $e^{\mathcal{L}_\mathrm{OS}\Delta t}\hat{v}_\alpha^\mathrm{D}$ and $\hat{v}_\beta^\mathrm{D}$, respectively, with $\Delta t\approx0.075\,h/u_\tau$. 
  The red and blue colours represent the clockwise and counterclockwise vorticity $\omega_z$, respectively, at the levels $[0.5, 0.8]$ multiplied by the maximum value in the region $y/h\in[0.1,0.6]$ for the state $\hat{v}_\beta^\mathrm{D}$. 
  Note that the near-wall region with $y/h<0.1$, where the vorticity has large magnitudes but is essentially irrelevant to the logarithmic-layer bursts, is not shown.}
\label{fig:CondPOD1}
\end{figure}

We apply the method above to the burst cycle for the straight Fourier mode $(n_x,n_z)=(1,0)$. The time interval $\Delta t$ between the two states in each joint sample $\hat{\eta}(t_i)=[\hat{v}(t_i)\quad\hat{v}(t_i\!+\!\Delta t)]^\mathrm{T}$ is set to approximately $0.075\,h/u_\tau$, which is large enough to capture the nonlinear effect accumulated over this interval, yet small enough to preserve the temporal correlation between the two states. With the DJM $\hat{\eta}^\mathrm{D} = [\hat{v}_\alpha^\mathrm{D}\quad\hat{v}_\beta^\mathrm{D}]^\mathrm{T}$ obtained from (\ref{eq:DominantJointMode}), we focus on the comparison between the flow patterns of $e^{\mathcal{L}_\mathrm{OS}\Delta t}\hat{v}_\alpha^\mathrm{D}$ and $\hat{v}_\beta^\mathrm{D}$, which conceptually represent the linearly and nonlinearly evolved states, respectively, starting from the same state $\hat{v}_\alpha^\mathrm{D}$ at a time $\Delta t$ earlier. Figure \ref{fig:CondPOD1} shows the results in different stages of the burst cycle in the $I_v$-$\sqrt{\mathcal{E}_v}$ subspace. For each cell, the representativeness $\rho\!\left[\eta^\mathrm{D}\right]$ is at least 0.4, suggesting that $\eta^\mathrm{D}$ provides a reasonable representation of the flow for the cell. Although the flow patterns across different cells cannot be interpreted as time evolution, they still shed light on how vortex structures are associated with different stages of the burst cycle.

In figure \ref{fig:CondPOD1}, panels $(a)$-$(e)$ essentially demonstrate the typical development of a burst as an Orr process, during which a roughly intact vortex is gradually tilted forward. Panels $(e)$-$(i)$ represent the late decaying stage with highly forward-inclined vortices. In each panel, the nonlinearly evolved state $\hat{v}_\beta^\mathrm{D}$ exhibits a more evident characteristic of the multilayer structure than the linearly evolved one $e^{\mathcal{L}_\mathrm{OS}\Delta t}\hat{v}_\alpha^\mathrm{D}$. This suggests that the nonlinearity tends to statistically facilitate the breakup of forward-inclined vortices during the late decaying stage, which is consistent with the suggestion from the weak-sense burst-restart model in \S\ref{subsection:weakRestart}. Panels $(j)$-$(n)$ represent a stage during which the LAE gradually accumulates towards forming a state that is ready to enter the next Orr process. In each panel, the comparison between $e^{\mathcal{L}_\mathrm{OS}\Delta t}\hat{v}_\alpha^\mathrm{D}$ and $\hat{v}_\beta^\mathrm{D}$ suggests that the nonlinearity tends to statistically facilitate the merging of co-rotating, broken-up vortices across different layers to form merged, backward-inclined vortices. This role of the nonlinearity is consistent with the suggestion from the strong-sense burst-restart model in \S\ref{subsection:strongRestart}. It is worth reiterating that all the cell-conditioned flow patterns in figure \ref{fig:CondPOD1} reveal the vortex structures in the burst cycle only in a statistical sense and should not be interpreted as the time evolution of a burst event.

\section{Conclusions}
\label{section:Conclusions}

This paper discusses the restarts of bursts of the wall-normal velocity in a log-minimal channel. The work is motivated by two aspects. The first is the increasing evidence in recent years \citep{Jimenez2022,Martinez2023,Martinez2025,Maeyama2023} suggesting that the burst cycle, including development and restart, does not directly depend on the presence of larger-scale structures, typically streaks. This suggests that the burst cycle mechanism should be approached without relying on specific \textit{a priori} knowledge and bias regarding other structures. The second is the fact that the development of a burst, which is `one half' of a burst cycle, can essentially be interpreted as a linearised Orr process \citep{Jimenez2013,Jimenez2015,Encinar2020}. This fact provides a starting point to approach `the other half' of the cycle, i.e., the restart. Accordingly, we discuss the problem of burst restarts within the framework of a linearised Navier-Stokes system with forcing terms that encapsulate the nonlinear effects of all other turbulence structures. Within this framework, the forcing is an immediate cause of burst restarts, and an understanding of it provides a basis for inferring the underlying nonlinear mechanism in the full Navier-Stokes system from which this immediate cause arises. Two of the three generic questions introduced in \S\ref{section:Introduction} are addressed.

The first one is about the conditions for burst-restart-like solutions for the forced linearised system itself. 
We pose three sets of optimisation problems to obtain the `minimal requirements' for, respectively: a weak-sense burst restart, a strong-sense burst restart, and periodic bursting. 
A summary of the optimal solutions to all these problems illustrates that a typical burst restart process involves three conceptual periods: a breakup period in which spanwise vortices in the decaying stage of the preceding burst break up and form multiple layers of vortices; a latent period when counter-rotating vortices across different layers catch up to one another due to mean advection; and a re-burst period characterised by the catch-up of co-rotating vortices from different layers. During this process, external forces play an important role in breaking up vortices; in addition, they can also contribute to the burst restart by merging the co-rotating, catching-up vortices before the re-burst. 
We therefore conjecture that the modeled burst restart process represents a plausible pathway for burst restarts in real turbulence and that the roles of the modeled forces in vortex breakup and merging may correspond to key effects of the nonlinear terms in the Navier-Stokes equations.


The discussion of the linearised system also motivates us to define a new state quantity -- the linearly available energy (LAE), $\mathcal{E}_v$ -- that is more relevant than the inclination angle $\mathit{\Psi}_v$ for parameterising a burst restart process. A burst-restarting force does not necessarily alter either $\mathit{\Psi}_v$ or the intensity $I_v$, but it needs to significantly increase $\mathcal{E}_v$ to a sufficiently high level. An eigenstate analysis shows that states with relatively high LAE generally have backward-inclined and/or multilayer structures of vortices, corroborating the essential role of the breakup effect in burst restarts. The analysis also indicates that multiple successive burst events in a linear evolution are possible. In real turbulence, this means that the nonlinear origin of a detected burst event may be seeded even before several preceding bursts.

The second question is about the contributory features to the observed burst restarts in real turbulence. We first project the real turbulence data onto both the intensity-inclination ($I_v$-$\mathit{\Psi}_v$) subspace and the intensity-LAE ($I_v$-$\sqrt{\mathcal{E}_v}$) subspace to analyse the linear and nonlinear effects in different stages of the burst cycle. We find that during the burst restart, the linear effect acts as a backward-tilting mechanism, whereas the nonlinear effect primarily induces forward tilting or no tilting. This somewhat counter-intuitive finding leads to two suggestions: 
the backward-tilting process in real turbulence is probably consistent with the latent period featuring the catch-up of counter-rotating vortices, as illustrated by the linearised models; 
and the inclination angle $\mathit{\Psi}_v$ is inadequate for revealing the essential role of nonlinearity in burst restarts. 
Instead of altering $\mathit{\Psi}_v$, a more essential role of nonlinearity is to replenish $\mathcal{E}_v$, which occurs primarily in the restart stage for the highly coherent straight Fourier mode and throughout the burst cycle for the less coherent oblique mode. 
We then extract flow patterns that statistically represent different stages in the burst cycle to visualise how the linear and nonlinear effects manifest themselves in the flow fields. For each cell in the coarse-grained $I_v$-$\sqrt{\mathcal{E}_v}$ subspace, we extract the dominant joint mode (DJM) to capture both the flow pattern and its short-time dynamics. The resulting representative flow patterns suggest that the nonlinearity tends to facilitate both the breakup of forward-inclined vortices in the late decaying stage and the merging of co-rotating, catching-up vortices before the subsequent burst. 
These statistical contributory features -- breakup and merging -- of the nonlinearity in real turbulence are consistent with our conjectures based on the linearised models.

The third question, which is to be addressed in the future, concerns identifying the real causal structures manifested as the above contributory features. Some general methodologies based on dyadic interactions of waves have been proposed to identify flow structures indicated by nonlinear terms \citep{Bae2021,Young2024}, which can contribute to addressing this question. Based on these methodologies, non-interventional  experiments can be designed to diagnose possible causal structures, which can be followed by interventional experiments with opposition control for verification.

\begin{bmhead}[Acknowledgements]
This work is supported by the European Research Council under the Caust grant ERC-AdG-101018287. 
\end{bmhead}

\begin{bmhead}[Declaration of interests]
The authors report no conflict of interest.
\end{bmhead}

\begin{bmhead}[Author ORCIDs]
Z. Hao, https://orcid.org/0000-0003-2199-6938; 
J. Jim\'enez, https://orcid.org/0000-0003-0755-843X
\end{bmhead}

\begin{appen}

\section{Numerical methods for the optimisation problems}
\label{section:AppNumericalMethods}

\subsection{Baseline solution}
\label{subsection:AlgorithmBaseline}

Following \citet{Schmid2001}, we use the eigenmode expansion (\ref{eq:EigenExpansion}) and reformulate equation (\ref{eq:OPTmaxGrowth}) for problem \ref{problem:Baseline} as 
\begin{subeqnarray}
\label{eq:OPTmaxGrowthMatrix}
    & & I_{v,\,m}^B = \max_{t\,>\,0} \left\{\, \max_{|\!|\,\boldsymbol{\theta}_0\,|\!|_2\,=\,1} |\!|\,\mathsfbi{N}(t)\,\boldsymbol{\theta}_0\,|\!|_2\,\right\}, \\
    & & \textup{where }\ \mathsfbi{N}(t)=\mathsfbi{M}^{1/2} e^{-\mathrm{i}k_x\mathsfbi{c}_\mathrm{OS}^{\vphantom{*}}t} \mathsfbi{E}^{-1/2}, \ \ \boldsymbol{\theta}_0=\mathsfbi{E}^{1/2}\boldsymbol{\kappa}_{\mathrm{OS}\,0}^{\vphantom{*}}\,. \qquad
\end{subeqnarray}
The inner maximisation of (\ref{eq:OPTmaxGrowthMatrix}$a$) can therefore be solved by calculating the singular vector of the matrix $\mathsfbi{N}(t)$ corresponding to its largest singular value. 

\subsection{Weak-sense burst restart}
\label{subsection:AlgorithmWeakRestart}

Equation (\ref{eq:OPTreburst}) for problem \ref{problem:weakRestart} can be reformulated in matrix form as 
\begin{subeqnarray}
\label{eq:OPTreburstMatrix}
    A(t,m_f;\boldsymbol{\kappa}_\mathrm{I}) & = & \max_{|\!|\,\boldsymbol{\zeta}\,|\!|_2\,=\,m_f} |\!|\,\mathsfbi{N}(t)\, (\boldsymbol{\theta}_\mathrm{I}+\boldsymbol{\zeta})\,|\!|_2 \nonumber \\
    & = & \max_{|\!|\,\boldsymbol{\zeta}\,|\!|_2\,=\,m_f} \sqrt{\, |\!|\,\mathsfbi{N}(t)\,\boldsymbol{\theta}_\mathrm{I}\,|\!|_2^2 +2\mathrm{Re}\!\left[\boldsymbol{\theta}_\mathrm{I}^*\,\mathsfbi{N}(t)^*\mathsfbi{N}(t)\,\boldsymbol{\zeta}\right] + |\!|\,\mathsfbi{N}(t)\,\boldsymbol{\zeta}\,|\!|_2^2 \,}\,,\qquad \\
    \textup{where } \ \mathsfbi{N}(t) & = & \mathsfbi{M}^{1/2} e^{-\mathrm{i}k_x\mathsfbi{c}_\mathrm{OS}^{\vphantom{*}}t} \mathsfbi{E}^{-1/2}, \ \ \boldsymbol{\theta}_\mathrm{I}=\mathsfbi{E}^{1/2}\boldsymbol{\kappa}_\mathrm{OS\,I}^{\vphantom{*}}\,, \ \ \boldsymbol{\zeta}=\mathsfbi{E}^{1/2}\boldsymbol{\chi}_\mathrm{OS}^{\vphantom{*}}\,.
\end{subeqnarray}
Equation (\ref{eq:OPTreburstMatrix}$a$) is the maximisation of an inhomogeneous quadratic form with a constraint and can be solved using the Lagrange-Newton method.

\subsection{Strong-sense burst restart}
\label{subsection:AlgorithmStrongRestart}

The inner minimisation in equation (\ref{eq:OPTrecover}) for problem \ref{problem:strongRestart} 
can be rewritten as
\begin{equation}
\label{eq:OPTrecoverConcise}
    \min\left\langle\,\boldsymbol{\chi}_\mathrm{OS}^{\vphantom{*}},\,\boldsymbol{\chi}_\mathrm{OS}^{\vphantom{*}}\,\right\rangle_{L_\mathsfbi{E}^2}, \quad\textup{subject to }\ \mathcal{K}\boldsymbol{\chi}_\mathrm{OS}^{\vphantom{*}} = \boldsymbol{r},
\end{equation}
where $\mathcal{K}$ is the linear operator defined as 
\begin{equation}
\label{eq:linOprK}
    \mathcal{K}: \: L_\mathsfbi{E}^2\!\left([0,T_R];\mathbb{C}^{N_\mathrm{OS}}\right) \to \mathbb{C}^{N_\mathrm{OS}}, \quad \mathcal{K}\,\boldsymbol{\xi} = \int_0^{T_R} e^{-\mathrm{i}k_x \mathsfbi{c}_\mathrm{OS}^{\vphantom{*}} (T_R-t)}\boldsymbol{\xi}(t)\,\mathrm{d}t\,,
\end{equation}
the inner product $\left\langle\,\cdot\,,\,\cdot\,\right\rangle_{L_\mathsfbi{E}^2}$ on the space $L_\mathsfbi{E}^2\!\left([0,T_R];\mathbb{C}^{N_\mathrm{OS}}\right)$ is defined as
\begin{equation}
\label{eq:innerProductLE2}
    \left\langle\,\boldsymbol{\xi},\,\boldsymbol{\eta}\,\right\rangle_{L_\mathsfbi{E}^2} = \int_0^{T_R} \boldsymbol{\xi}^*\!(t)\,\mathsfbi{E}\,\boldsymbol{\eta}(t)\,\mathrm{d}t\,,
\end{equation}
and the vector $\boldsymbol{r}\in\mathbb{C}^{N_\mathrm{OS}}$ is
\begin{equation}
\label{eq:residualvector}
    \boldsymbol{r} = e^{\mathrm{i}\theta_\mathrm{off}}\boldsymbol{\kappa}_{\mathrm{OS\,II}}^{\vphantom{*}} - e^{-\mathrm{i}k_x \mathsfbi{c}_\mathrm{OS}^{\vphantom{*}} T_R}\boldsymbol{\kappa}_{\mathrm{OS\,I}}^{\vphantom{*}}\,.
\end{equation}
The optimal solution to (\ref{eq:OPTrecoverConcise}) is thus 
\begin{equation}
\label{eq:formalSolStrongRestart}
    \boldsymbol{\chi}_\mathrm{OS}^{\vphantom{*}} = \mathcal{K}^\dagger(\mathcal{K}\mathcal{K}^\dagger)^{-1} \boldsymbol{r}\,,
\end{equation}
where $\mathcal{K}^\dagger$ is the adjoint operator of $\mathcal{K}$ defined by 
\begin{equation}
\label{eq:DefAdjointK}
    \left(\mathcal{K}\,\boldsymbol{\xi}\right)^*\boldsymbol{\beta} = \left\langle\,\boldsymbol{\xi},\,\mathcal{K}^\dagger\boldsymbol{\beta}\,\right\rangle_{L_\mathsfbi{E}^2},\quad\forall\,\boldsymbol{\xi}\in L_\mathsfbi{E}^2,\,\boldsymbol{\beta}\in\mathbb{C}^{N_\mathrm{OS}}.
\end{equation}
To obtain $\mathcal{K}^\dagger$, we substitute (\ref{eq:linOprK}) into the left-hand side of (\ref{eq:DefAdjointK}), yielding
\begin{eqnarray}
\label{eq:deriveAdjont}
    \left(\mathcal{K}\,\boldsymbol{\xi}\right)^*\boldsymbol{\beta} \!\!\!\! & = & \!\!\!\! \int_0^{T_R} \boldsymbol{\xi}^*\!(t) e^{\mathrm{i} k_x \mathsfbi{c}_\mathrm{OS}^* (T_R-t)}\boldsymbol{\beta}\,\mathrm{d}t \nonumber \\ 
    & = & \!\!\!\! \int_0^{T_R} \boldsymbol{\xi}^*\!(t) \, \mathsfbi{E} \left[\mathsfbi{E}^{-1}e^{\mathrm{i} k_x \mathsfbi{c}_\mathrm{OS}^* (T_R-t)}\boldsymbol{\beta}\right]\mathrm{d}t 
    = \left\langle\,\boldsymbol{\xi},\,\mathsfbi{E}^{-1}e^{\mathrm{i} k_x \mathsfbi{c}_\mathrm{OS}^* (T_R-t)}\boldsymbol{\beta}\,\right\rangle_{L_\mathsfbi{E}^2}.\qquad
\end{eqnarray}
Comparing the right-hand sides of (\ref{eq:DefAdjointK}) and (\ref{eq:deriveAdjont}) leads to the form of $\mathcal{K}^\dagger$ as
\begin{equation}
\label{eq:KAdjointDerived}
    \left(\mathcal{K}^\dagger\boldsymbol{\beta}\right)(t) = \mathsfbi{E}^{-1}e^{\mathrm{i} k_x \mathsfbi{c}_\mathrm{OS}^* (T_R-t)}\boldsymbol{\beta}\,.
\end{equation}
Substituting (\ref{eq:linOprK}) and (\ref{eq:KAdjointDerived}) into (\ref{eq:formalSolStrongRestart}), we finally obtain the optimal solution
\begin{equation}
\label{eq:SolStrongRestart}
    \boldsymbol{\chi}_\mathrm{OS}^{\vphantom{*}}(t) = \mathsfbi{E}^{-1} e^{\mathrm{i}k_x \mathsfbi{c}_\mathrm{OS}^*(T_R-t)} \mathsfbi{W}(T_R)^{-1}\boldsymbol{r}\,,
\end{equation}
where $\mathsfbi{W}\in\mathbb{C}^{N_\mathrm{OS}\times N_\mathrm{OS}}$ is the controllability Gramian defined as
\begin{equation}
\label{eq:DefGramian}
    \mathsfbi{W}(s) = \int_0^{s} e^{-\mathrm{i}k_x \mathsfbi{c}_\mathrm{OS}^{\vphantom{*}} (s-t)}\mathsfbi{E}^{-1} e^{\mathrm{i}k_x \mathsfbi{c}_\mathrm{OS}^* (s-t)}\,\mathrm{d}t\,,
\end{equation}
whose entry in the $i$-th row and $j$-th column, $\mathsfbi{W}(s)_{\left[ij\right]}$, can be explicitly expressed as
\begin{equation}
\label{eq:DefGramianElement}
    \mathsfbi{W}(s)_{\left[ij\right]} = \frac{1-e^{-\mathrm{i}k_x(c_{\mathrm{OS}i}^{\vphantom{*}}-c_{\mathrm{OS}j}^*)s}}{\mathrm{i}k_x \bigl(c_{\mathrm{OS}\,i}^{\vphantom{*}}-c_{\mathrm{OS}j}^*\bigr) }(\mathsfbi{E}^{-1})_{\left[ij\right]}\,.
\end{equation}
Corresponding to the optimal $\boldsymbol{\chi}_\mathrm{OS}^{\vphantom{*}}(t)$ in (\ref{eq:SolStrongRestart}), the minimum time-integral forcing energy is
\begin{equation}
\label{eq:minTimeIntForceEnergy}
    \left\langle\,\boldsymbol{\chi}_\mathrm{OS}^{\vphantom{*}},\,\boldsymbol{\chi}_\mathrm{OS}^{\vphantom{*}}\,\right\rangle_{L_\mathsfbi{E}^2} = \int_0^{T_R} \boldsymbol{\chi}_\mathrm{OS}^*(t)\,\mathsfbi{E}\,\boldsymbol{\chi}_\mathrm{OS}^{\vphantom{*}}(t)\,\mathrm{d}t = \boldsymbol{r}^*\mathsfbi{W}(T_R)^{-1}\boldsymbol{r}\,,
\end{equation}
and the optimal coefficient vector for the velocity $\hat{v}(t)$ is 
\begin{eqnarray}
\label{eq:SolStrongRestartKappa}
    \boldsymbol{\kappa}_\mathrm{OS}^{\vphantom{*}}(t) \!\!\!\! & = & \!\!\!\! e^{-\mathrm{i}k_x \mathsfbi{c}_\mathrm{OS}^{\vphantom{*}}t} \boldsymbol{\kappa}_\mathrm{OS\,I}^{\vphantom{*}} + \int_0^t e^{-\mathrm{i}k_x \mathsfbi{c}_\mathrm{OS}^{\vphantom{*}}(t-\tau)}\boldsymbol{\chi}_\mathrm{OS}^{\vphantom{*}}(\tau) \,\mathrm{d}\tau \nonumber \\
    \!\!\!\! & = & \!\!\!\! e^{-\mathrm{i}k_x \mathsfbi{c}_\mathrm{OS}^{\vphantom{*}}t} \boldsymbol{\kappa}_\mathrm{OS\,I}^{\vphantom{*}} + e^{-\mathrm{i}k_x \mathsfbi{c}_\mathrm{OS}^{\vphantom{*}}(t-T_R)}\left[\mathsfbi{W}(T_R)-\mathsfbi{W}(T_R-t)\right]\mathsfbi{W}(T_R)^{-1}\boldsymbol{r}\,.\qquad
\end{eqnarray}

\subsection{Sustaining periodic bursts}
\label{subsection:AlgorithmPeriodicBursts}

Equation (\ref{eq:PeriodicBursts}) for problem \ref{problem:PeriodicBursts} can be reformulated as
\begin{subeqnarray}
\label{eq:PeriodicBurstsMatrix}
    && I_{v,\,m}^P = \max_{t\in[0,T_b)}\left\{\, \max_{|\!|\,\boldsymbol{\zeta}\,|\!|_2\,=\,1} |\!|\,\mathsfbi{G}(t)\,\boldsymbol{\zeta}\,|\!|_2 \,\right\}, \\ 
    && \textup{where }\ \mathsfbi{G}(t) = \mathsfbi{M}^{1/2}\left\{\frac{1}{T_b}\! \sum_{\ell=-m_\mathrm{DK}}^{m_\mathrm{DK}}\mathsfbi{R}\!\left(a_{f\ell}\right) \exp\!\left({-\mathrm{i}k_x a_{f\ell}\, t}\right)\right\}\mathsfbi{E}^{-1/2}\,, \nonumber \\ 
    && \qquad \quad \mathsfbi{R}\!\left(a_{f\ell}\right) = \left[\,\mathrm{i}k_x\!\left(\mathsfbi{c}_\mathrm{OS}^{\vphantom{*}}-a_{f\ell}\mathsfbi{I}\,\right)\,\right]^{-1}\!, \ a_{f\ell} = a_{f0} + \ell\frac{2\pi}{k_x T_b}, \ \boldsymbol{\zeta}=\mathsfbi{E}^{1/2}\boldsymbol{\chi}_\mathrm{OS}^{\vphantom{*}}. \qquad \quad
\end{subeqnarray}
Similar to \S\ref{subsection:AlgorithmBaseline}, the inner maximisation of (\ref{eq:PeriodicBurstsMatrix}$a$) can be solved by calculating the singular vector of the matrix $\mathsfbi{G}(t)$ corresponding to its largest singular value.

\section{Robustness of the model-inferred breakup mechanism}
\label{section:Robustness}

To assess the robustness of the vortex breakup mechanism inferred in \S\ref{subsection:weakRestart}, we consider optimal solutions to problem \ref{problem:weakRestart} under six modified conditions: (i, ii) the $E$-norm in the constraint on $\hat{\varphi}_\mathrm{sp}$ in (\ref{eq:OPTreburst}$a$) is replaced with the one-component $E_{\hat{v}}$-norm and $E_{\hat{u}_{/\!/}}$-norm, respectively, each defined by retaining only the corresponding velocity component in (\ref{eq:defEnergyNorm}); (iii, iv) the $y$-band over which the intensity $I_v$ is defined is restricted to the lower and upper halves of the original band, respectively; and (v, vi) the spatial Fourier mode of interest, $(n_x, n_z)$, is changed to $(2,0)$ and $(2,1)$, respectively. Figure \ref{fig:RB_Cont0_rT125_assembled} shows the flow fields at $t=0$ and $0.18\,h/u_\tau$ for each of these six cases. Although the solutions differ in detail from those in figures \ref{fig:RB_Cont0} and \ref{fig:RBoblq_Cont0}, all six exhibit vortex breakup as a prominent feature.

\begin{figure}
  \centerline{\includegraphics[width=1.00\textwidth]{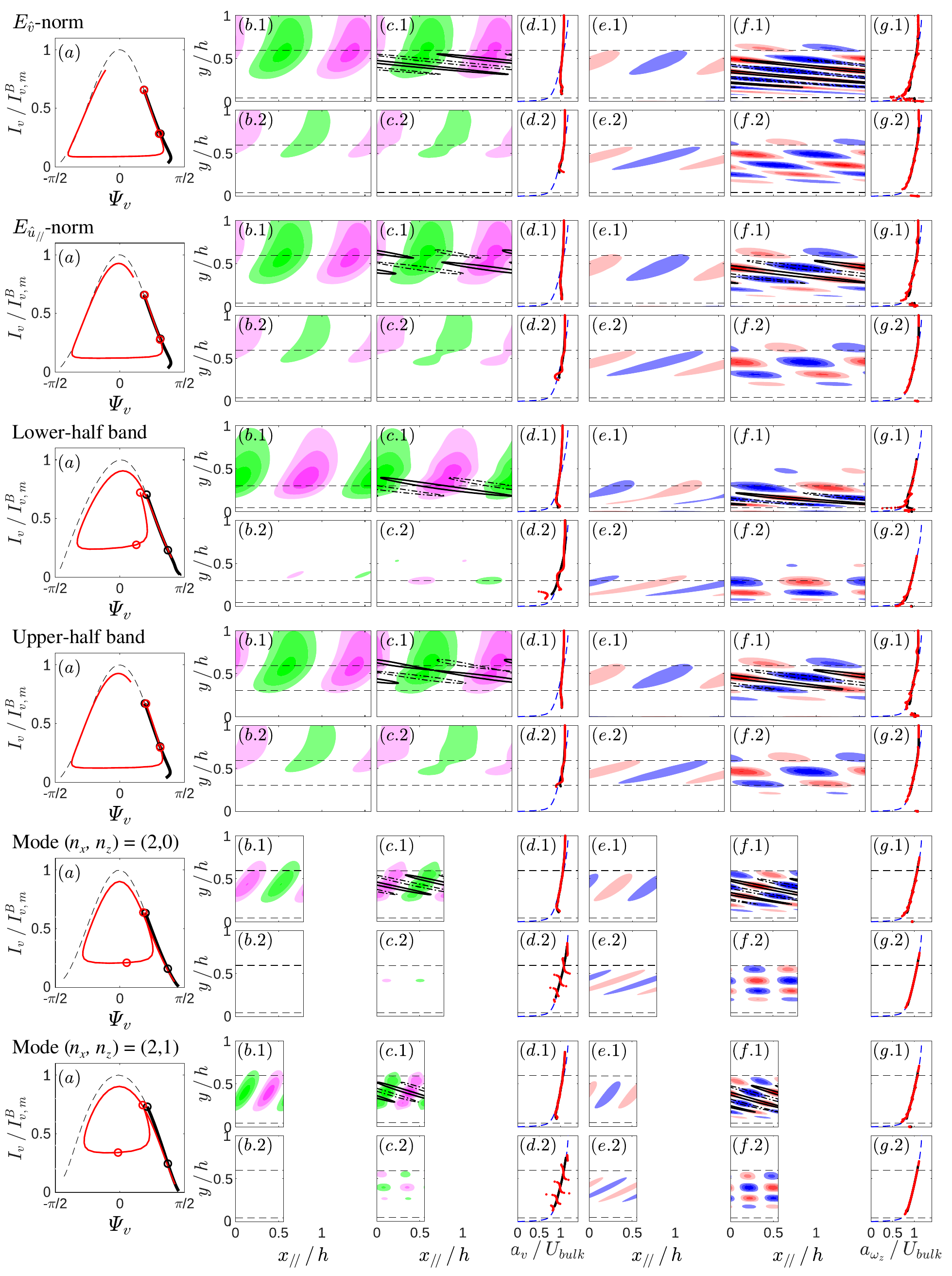}}
  \caption{Optimal weak-sense re-burst solutions to problem \ref{problem:weakRestart} under six modified conditions. 
  The six conditions are described in the main text of appendix \ref{section:Robustness}. 
  Each case includes one trajectory panel $(a)$ and two rows of panels $(b)$-$(g)$ representing flows at two times $t=0$ and $0.18\,h/u_\tau$, respectively. 
  Plotting conventions follow figure \ref{fig:RB_Cont0}.}
\label{fig:RB_Cont0_rT125_assembled}
\end{figure}

These tests support the robustness of vortex breakup to the modifications considered here, but do not establish that it is a general property of burst-restart-like solutions under all realisable conditions. Establishing such robustness would require a systematic exploration of the space of realisable conditions. Such a task goes beyond the intended purpose of our conceptual, interpretable models, which aim to infer physical mechanisms rather than establish their validity under all realisable conditions, and thus lies outside the scope of the present study.

\section{Dependence of optimal recovery processes on recovery times}
\label{section:RecoveryTime}

Figure \ref{fig:ShPa_rI0030_rI1030_traj}$(a)$ shows the trajectories of the optimal recovery processes obtained from problem \ref{problem:strongRestart} in the $I_v$-$\mathit{\Psi}_v$ subspace for different prescribed recovery time $T_R$, given fixed initial and final states $\hat{q}_\mathrm{I}$ and $\hat{q}_\mathrm{II}$ with $I_v[\hat{q}_{\mathrm{I}}]=I_v[\hat{q}_{\mathrm{II}}]=0.3\,I_{v,\,m}^B$. There exist two limiting trajectories. One occurs as $T_R\gtrsim2\,h/u_\tau$, for which the recovery process includes two decoupled stages: the unforced decay of $\hat{q}_\mathrm{I}$ to zero and the production of $\hat{q}_\mathrm{II}$ from zero, both independent of $T_R$. The latter stage's independence from $T_R$ is illustrated by the collapse of the recovery force magnitudes $|\!|\hat{\varphi}|\!|_E$ in figure \ref{fig:ShPa_rI0030_rI1030_traj}$(b)$ for the cases $T_R=2\,h/u_\tau$ and $6\,h/u_\tau$, provided that the time is normalised by the same scale. In figure \ref{fig:ShPa_rI0030_rI1030_traj}$(c)$, the time-integral forcing energy $J_v$ for $T_R\gtrsim2\,h/u_\tau$ is asymptotic to the minimum value, which is an intrinsic property of state $\hat{q}_\mathrm{II}$ alone. For this long-time limiting case, the optimal process producing $\hat{q}_\mathrm{II}$ from zero is shown in figure \ref{fig:ShPa_rI0030_rI1030_limits}$(a$-$d)$. During the process, the spanwise curl of the force first induces backward-inclined vortices and is then gradually withdrawn to let these vortices develop under the Orr mechanism towards the state $\hat{q}_\mathrm{II}$.

\begin{figure}
  \centerline{\includegraphics[width=1.00\textwidth]{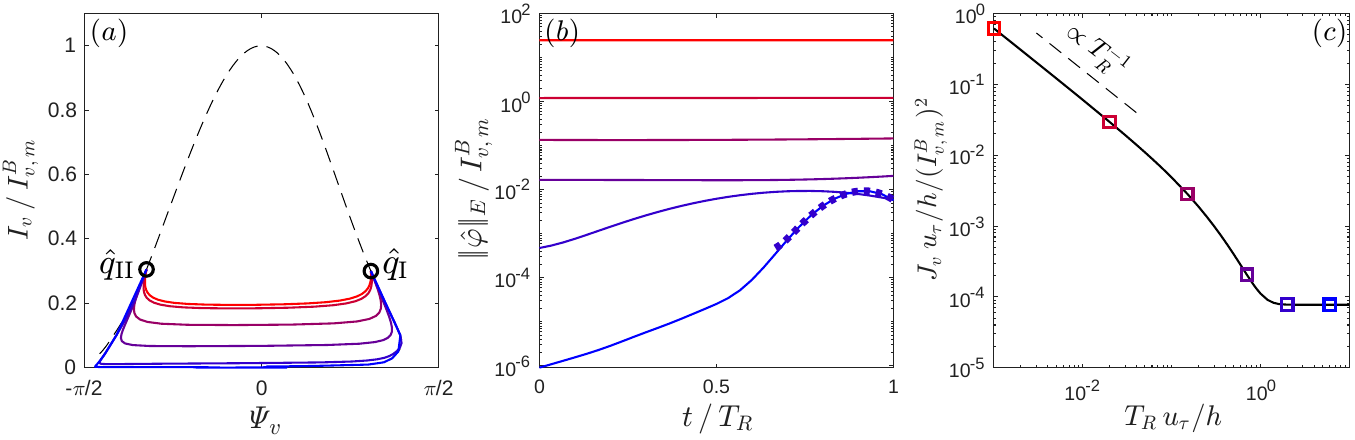}}
  \caption{Optimal recovery processes with different recovery times $T_R$ for the Fourier mode $(n_x,n_z)=(1,0)$. Colours from red to blue represent the six cases with $T_R=[0.001,\,0.02,\,0.15,\,0.7,\,2,\,6]\,h/u_\tau$. All the cases share the same initial state $\hat{q}_\mathrm{I}$ and final state $\hat{q}_\mathrm{II}$ with $I_v[\hat{q}_{\mathrm{I}}]=I_v[\hat{q}_{\mathrm{II}}]=0.3\,I_{v,\,m}^B$. Panel $(a)$ shows the trajectories in the $I_v$-$\mathit{\Psi}_v$ subspace, in which the dashed black line represents the baseline solution. Panel $(b)$ presents the energy norms of the optimal recovering forces, $|\!|\hat{\varphi}|\!|_E$, changed over time; the dotted line is obtained from the solid line for the case $T_R=2\,h/u_\tau$ by applying the following transformation: $t$ is rescaled by $6\,h/u_\tau$ and shifted to align the final time instant of recovery. Panel $(c)$ shows the time-integral forcing energy $J_v$ changed with $T_R$, with the square markers corresponding to the six cases in $(a,b)$.}
\label{fig:ShPa_rI0030_rI1030_traj}
\end{figure}

\begin{figure}
  \centerline{\includegraphics[width=1.00\textwidth]{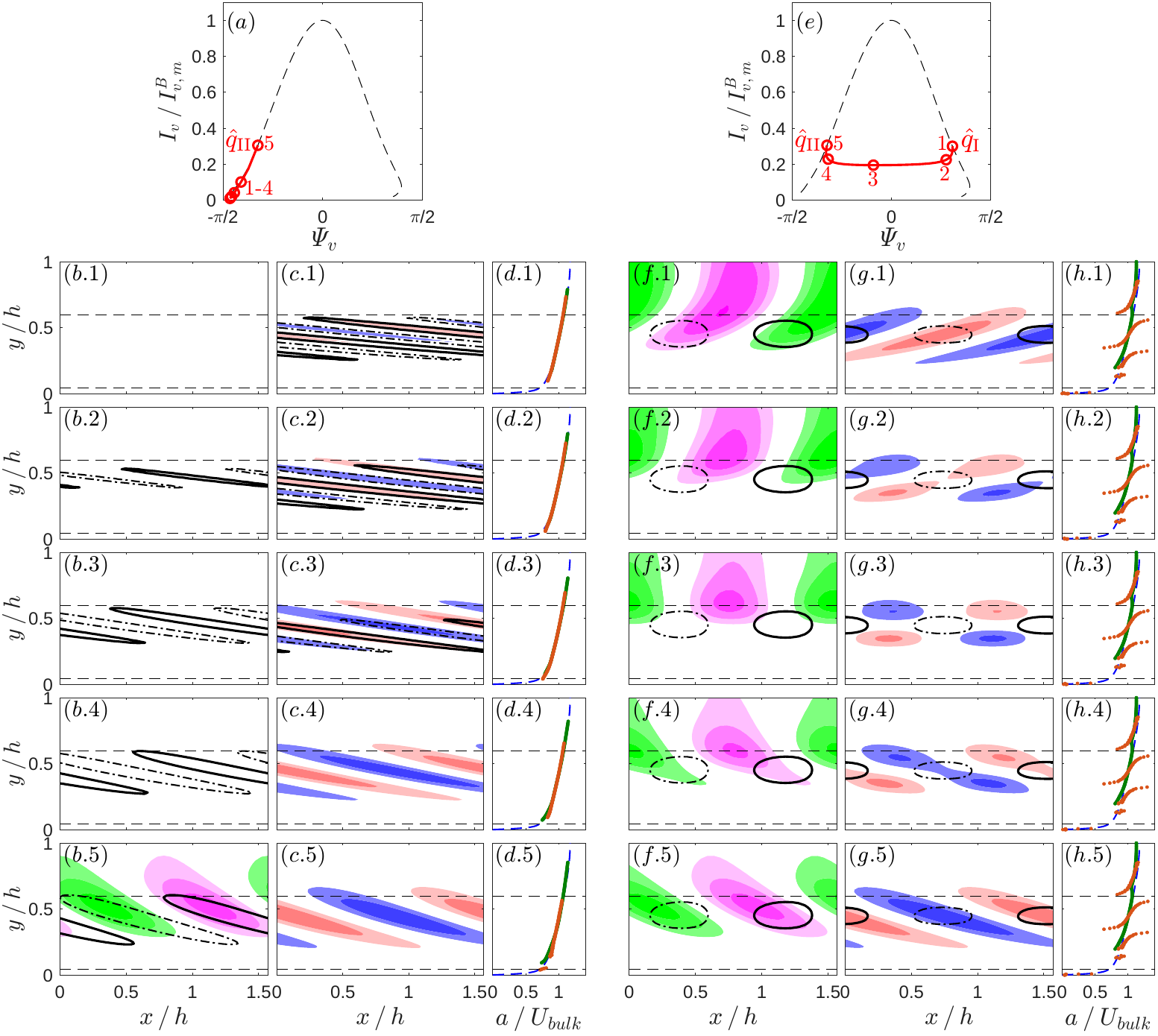}}
  \caption{Limiting cases of optimal recovery processes for the spatial Fourier mode $(n_x,n_z)=(1,0)$, with the recovery time $T_R\to\infty$ in $(a)$-$(d)$ and $T_R\to0$ in $(e)$-$(h)$. In both cases, the initial and final states, $\hat{q}_{\mathrm{I}}$ and $\hat{q}_{\mathrm{II}}$, have the intensity $I_v[\hat{q}_{\mathrm{I}}]=I_v[\hat{q}_{\mathrm{II}}]=0.3\,I_{v,\,m}^B$. All flow and force fields are viewed in a reference frame moving at the constant streamwise speed $U_\textit{bulk}$. 
  Panels $(a,e)$: trajectories in the $I_v$-$\mathit{\Psi}_v$ subspace. Black dashed and red solid lines indicate the baseline solution and optimal recovery process, respectively. Markers 1-5 in $(a)$ denote the states at $[0.8:-0.2:0]\,h/u_\tau$ before the final state $\hat{q}_\mathrm{II}$ is reached, and markers 1-5 in $(e)$ denote the states at $[0:0.25:1]\,T_R$ after the recovery force is applied. The marked states correspond to the five rows of panels in $(b)$-$(d)$ and $(f)$-$(h)$. 
  Columns $(b,f)$: velocity $v$, with magenta and green indicating upward and downward motion, respectively. Solid and dashed contours indicate the upward and downward force $f^s_y$, respectively.
  Columns $(c,g)$: vorticity $\omega_z$, with red and blue indicating clockwise and counterclockwise rotation, respectively. Solid and dashed contours indicate the clockwise and counterclockwise curl of the force $(\bnabla\times\boldsymbol{f}^s)_z$, respectively.
  Columns $(d,h)$: $y$-profiles of the advection speeds $a$ of $f^s_y$ (green) and $(\bnabla\times\boldsymbol{f}^s)_z$ (brown). Blue dashed lines indicate the mean velocity $U$. 
  The shading levels given as magnitudes are: $(b,f)$: $[0.29:0.12:0.53]\,I_{v,\,m}^B$; $(c,g)$: $[8.1:3.2:14.5]\,I_{v,\,m}^B/h$. 
  The force-contour levels are: $(b)$: $0.056\,I_{v,\,m}^B u_\tau/h$; $(c)$: $12.6\,I_{v,\,m}^B u_\tau/h^2$; $(f)$: $0.51\,I_{v,\,m}^B/T_R$; $(g)$: $13.3\,I_{v,\,m}^B/T_Rh$.}
\label{fig:ShPa_rI0030_rI1030_limits}
\end{figure}

The other limiting trajectory in figure \ref{fig:ShPa_rI0030_rI1030_traj}$(a)$ occurs as $T_R\lesssim10^{-2}h/u_\tau$, for which the recovery process takes place mostly at the intensity level $I_v\approx0.2\,I_{v,\,m}^B$, moderately lower than $I_v[\hat{q}_{\mathrm{I}}]$ and $I_v[\hat{q}_{\mathrm{II}}]$. During such a short time, the recovery force is almost constant, i.e., $\hat{\varphi}(y,t) \simeq \left(\hat{q}_\mathrm{II}(y)-\hat{q}_\mathrm{I}(y)\right)/\,T_R$, as shown by its magnitudes in figure \ref{fig:ShPa_rI0030_rI1030_traj}$(b)$. The time-integral forcing energy $J_v$ is thus proportional to $T_R^{-1}$, as shown in figure \ref{fig:ShPa_rI0030_rI1030_traj}$(c)$. For this short-time limiting case, figure \ref{fig:ShPa_rI0030_rI1030_limits}$(e$-$h)$ shows that the temporally constant force acts first to break up the forward-inclined vortices and then to reconnect co-rotating, catching-up vortices to form new merged, backward-inclined vortices. Additional tests indicate that such breakup and merging effects of the recovery forces can be observed for any $T_R\lesssim0.5\,h/u_\tau$ without qualitative differences.

\section{Linear evolution of a high-order high-LAE eigenstate}
\label{section:HighEigenState}

Figure \ref{fig:LLT_Eig4Evol} visualises the linear evolution of the eigenstate $\hat{q}^{H4}$ obtained from problem \ref{problem:EigenPotential}. The evolution involves four $I_v$-peak events, as indicated by markers 2, 4, 6, and 8. Although the eigenstate has four layers of broken-up vortices, the first three $I_v$-peak events are mainly caused by the vortices farthest from the wall at $y\approx0.7\,h$ catching up to the co-rotating vortices nearest to the wall at $y\approx0.2\,h$. The last $I_v$-peak event is mainly caused by the `sub-Orr' process for the farthest-wall vortices themselves, during which they are tilted forward by the local mean shear with a much slower rate than the band-wise averaged one. The three $I_v$-trough moments, as indicated by markers 3, 5, and 7, are characterised by the counter-rotating catch-up between the farthest- and nearest-wall vortices, between which the wall-normal velocity $v$ is almost zero. 

\begin{figure}
  \centerline{\includegraphics[width=1.00\textwidth]{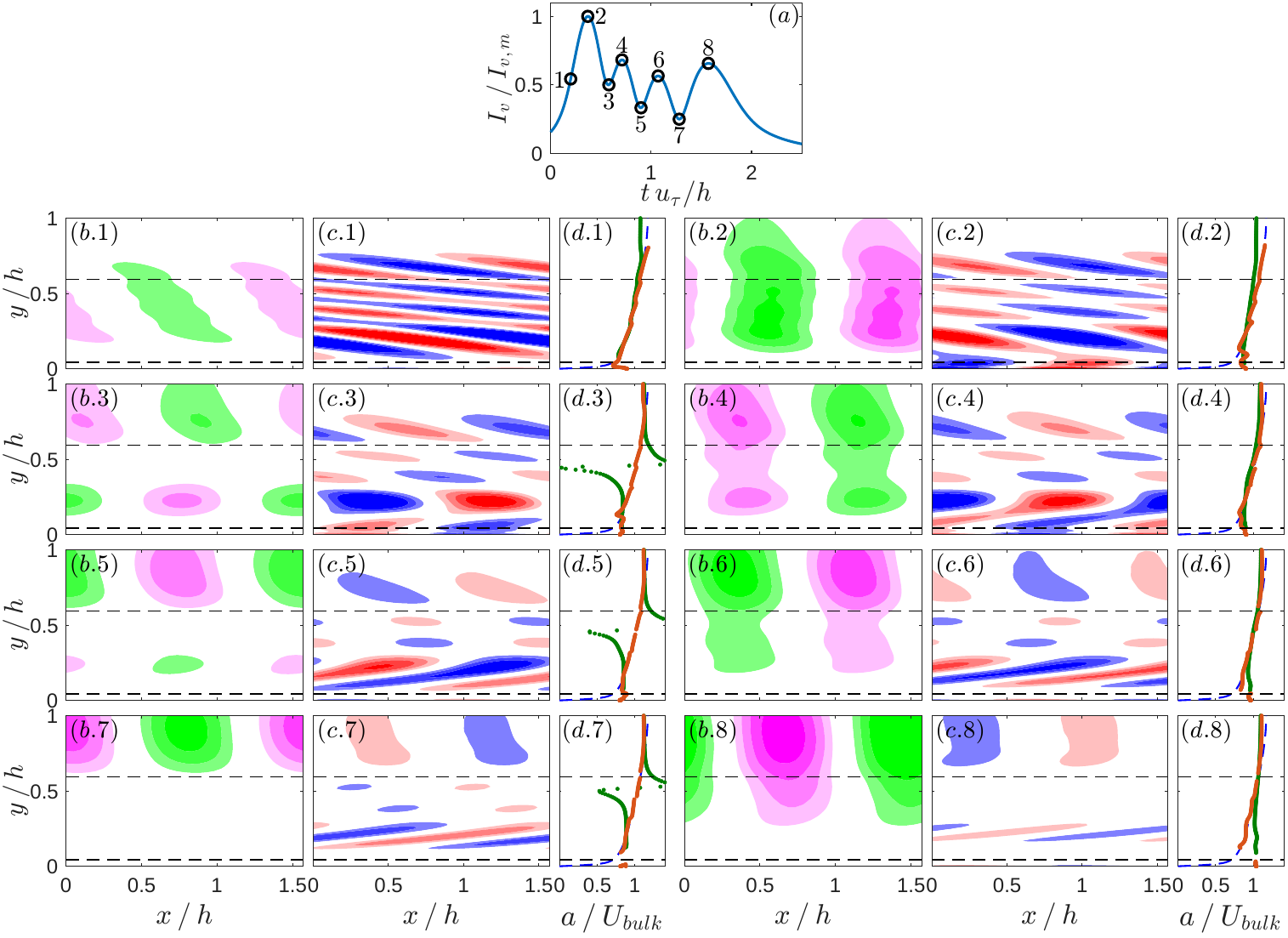}}
  \caption{Linear evolution of the eigenstate with the fourth largest LAE for the spatial Fourier mode $(n_x,n_z)=(1,0)$. The flow fields in $(b,c)$ and advection speed profiles in $(d)$ correspond to the eight time instants marked 1-8 in $(a)$. All flow fields are viewed in a reference frame moving at the constant streamwise speed $U_\textit{bulk}$. 
  Panels $(b)$: velocity $v$, with magenta and green indicating upward and downward motion, respectively. 
  Panels $(c)$: vorticity $\omega_z$, with red and blue indicating clockwise and counterclockwise rotation, respectively. 
  Panels $(d)$: $y$-profiles of the advection speeds $a$ of $v$ (green) and $\omega_z$ (brown). Blue dashed lines indicate the mean velocity $U$. The shading levels given as magnitudes are: $(b)$: $[0.75:0.30:1.65]\,I_{v,\,m}$; $(c)$: $[12.5:5.0:22.5]\,I_{v,\,m}/h$.}
\label{fig:LLT_Eig4Evol}
\end{figure}

\begin{figure}
  \centerline{\includegraphics[width=1.00\textwidth]{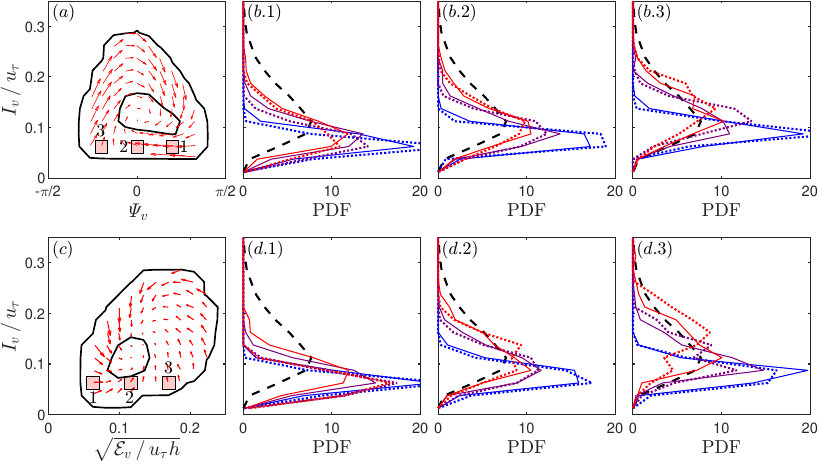}}
  \caption{Probability densities of the future intensity $I_v$ for events whose initial states lie within selected cells in the $I_v$-$\mathit{\Psi}_v$ (upper row) and $I_v$-$\sqrt{\mathcal{E}_v}$ (lower row) subspaces for the Fourier mode $(n_x,n_z)=(1,0)$. 
  Cells 1-3 marked in $(a)$ correspond to panels $(b.1)$-$(b.3)$, respectively, while cells 1-3 marked in $(c)$ correspond to panels $(d.1)$-$(d.3)$, respectively. 
  In each panel of $(b)$ and $(d)$, the blue, purple, and red lines show the PDFs of $I_v/u_\tau$ at elapsed time $t=0.075$, $0.150$, and $0.225\,h/u_\tau$, respectively, from the initial states in the corresponding cell. 
  Solid and dotted lines represent the evolution in real turbulence and the assumed linear evolution, respectively. The black dashed line indicates the unconditional invariant PDF of $I_v/u_\tau$ in turbulence.}
\label{fig:predictability}
\end{figure}
\begin{figure}
  \centerline{\includegraphics[width=1.00\textwidth]{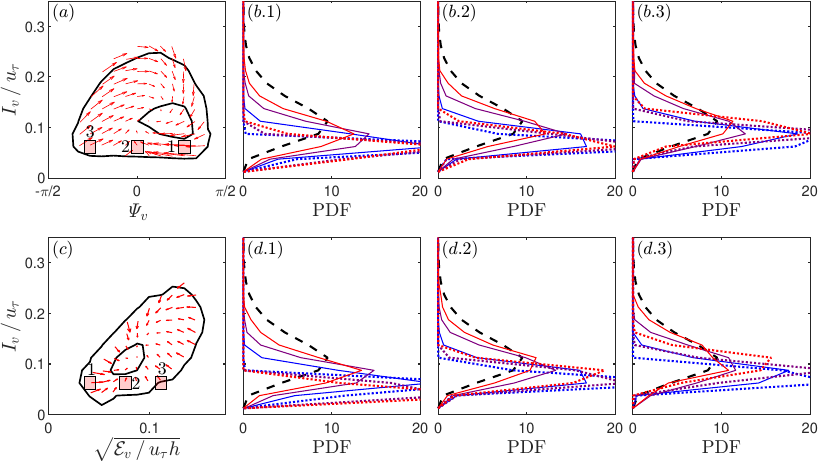}}
  \caption{Probability densities of the future intensity $I_v$ for events whose initial states lie within selected cells in the $I_v$-$\mathit{\Psi}_v$ (upper row) and $I_v$-$\sqrt{\mathcal{E}_v}$ (lower row) subspaces for the Fourier mode $(n_x,n_z)=(1,1)$. 
  All graphical conventions are the same as in figure \ref{fig:predictability}.}
\label{fig:predictability_oblq}
\end{figure}

\section{Finite-time predictability within the two-dimensional subspaces}
\label{section:lossPred}

To assess the finite-time predictive power of the summary variables $\mathit{\Psi}_v$ or $\mathcal{E}_v$ for subsequent bursting, figures \ref{fig:predictability} and \ref{fig:predictability_oblq} show the evolution of the probability density function (PDF) of $I_v$ for ensembles of events whose initial states lie within cells with the same low level of $I_v$ but different levels of $\mathit{\Psi}_v$ or $\mathcal{E}_v$. Over $0.225\,h/u_\tau$, each conditional PDF generally evolves towards the unconditional invariant PDF, which has a higher mean $I_v$. This evolution towards higher $I_v$ is more pronounced for ensembles with initially smaller $\mathit{\Psi}_v$ or larger $\mathcal{E}_v$. The same tendency is observed under both the assumed linear evolution and the real-turbulence evolution. It suggests that $\mathit{\Psi}_v$ and $\mathcal{E}_v$ have some predictive power for subsequent bursting in real turbulence, although their relative predictive power cannot be determined from the available data.

It is worth noting that predictability within these coarse-grained, two-dimensional subspaces is generally rather limited. Additional tests suggest that, regardless of the selected initial cell, the conditional PDF for the real-turbulence evolution becomes essentially indistinguishable from the unconditional invariant PDF within $0.4$-$0.5\,h/u_\tau$; that is, the predictive information provided by the initial cell is lost. This loss of predictability reflects not only the chaotic nature of turbulence but also the information lost through projection onto these highly reduced subspaces and through coarse graining. Reliably assessing the predictive power of summary variables such as $\mathit{\Psi}_v$ and $\mathcal{E}_v$ may therefore require finer graining and/or higher-dimensional subspaces, and consequently a considerably larger database.

\section{Asymptotic properties of the DJM}
\label{section:dominantJointMode}

This section presents the proof of the asymptotic properties (\ref{eq:JointModeOrder}) and (\ref{eq:CorrelationDominantJoint}) of the DJM $\hat{\eta}^\mathrm{D} = [\hat{v}_\alpha^\mathrm{D} \quad \hat{v}_\beta^\mathrm{D}]^\mathrm{T}$, which is defined as (\ref{eq:DominantJointMode}), as $\Delta t\to0$.

Using (\ref{eq:EigenExpansion}-\ref{eq:governEqnCoef}), we obtain the relation between the coefficient vectors $\boldsymbol{\kappa}_\mathrm{OS}$ of the two states in a joint sample $\hat{\eta}(t_i)=[\hat{v}(t_i)\quad\hat{v}(t_i\!+\!\Delta t)]^\mathrm{T}$ to the first order in $\Delta t$ as
\begin{equation}
\label{eq:AppKappaRelation}
    \boldsymbol{\kappa}_\mathrm{OS}^{\vphantom{*}}(t_i\!+\!\Delta t) =  \boldsymbol{\kappa}_\mathrm{OS}^{\vphantom{*}}(t_i) + \left(\, \boldsymbol{\chi}_\mathrm{OS}^{\vphantom{*}}(t_i)-\mathrm{i}k_x\mathsfbi{c}_\mathrm{OS}^{\vphantom{*}}\boldsymbol{\kappa}(t_i)\,\right)\Delta t + O(\Delta t^2)\,.
\end{equation}
Introduce the transformation 
\begin{eqnarray}
\label{eq:DefNewCoefs}
    && \boldsymbol{\sigma}(t_i) \!= \mathsfbi{M}^{1/2}\!\left(\,\boldsymbol{\chi}_\mathrm{OS}^{\vphantom{*}}(t_i)-\mathrm{i}k_x\mathsfbi{c}_\mathrm{OS}^{\vphantom{*}}\boldsymbol{\kappa}(t_i)\,\right), \nonumber \\
    && \boldsymbol{a}(t_i) = \mathsfbi{M}^{1/2}\boldsymbol{\kappa}_\mathrm{OS}^{\vphantom{*}}(t_i)\,, \ 
    \boldsymbol{a}(t_i\!+\!\Delta t) = \mathsfbi{M}^{1/2}\boldsymbol{\kappa}_\mathrm{OS}^{\vphantom{*}}(t_i\!+\!\Delta t) = \boldsymbol{a}(t_i) + \boldsymbol{\sigma}(t_i)\Delta t + O(\Delta t^2)\,, \ \ \nonumber \\
    && \boldsymbol{a}_\alpha^\mathrm{D}\ \;\; = \mathsfbi{M}^{1/2}\boldsymbol{\kappa}_\mathrm{OS\alpha}^\mathrm{D}\,, \quad \
    \boldsymbol{a}_\beta^\mathrm{D} \qquad\ \: = \mathsfbi{M}^{1/2}\boldsymbol{\kappa}_\mathrm{OS\beta}^\mathrm{D}\,. \quad
\end{eqnarray}
With (\ref{eq:DefNewCoefs}), 
the transformed DJM, i.e., $\left[(\boldsymbol{a}_\alpha^\mathrm{D})^\mathrm{T} \quad (\boldsymbol{a}_\beta^\mathrm{D})^\mathrm{T}\right]^\mathrm{T}$, is the eigenvector corresponding to the largest eigenvalue of the covariance matrix
\begin{eqnarray}
\label{eq:AppJointMatrixDef}
    \mathsfbi{K} &=& \frac{1}{2} 
    \left[\begin{array}{ll}
        \mathbb{E}\left[\boldsymbol{a}(t_i\quad\ \ \;)\,\boldsymbol{a}(t_i\quad\ \ \;)^*\right] & \mathbb{E}\left[\boldsymbol{a}(t_i\quad\ \ \;)\,\boldsymbol{a}(t_i\!+\!\Delta t)^*\right] \\
        \mathbb{E}\left[\boldsymbol{a}(t_i\!+\!\Delta t)\,\boldsymbol{a}(t_i\quad\ \ \;)^*\right] & \mathbb{E}\left[\boldsymbol{a}(t_i\!+\!\Delta t)\,\boldsymbol{a}(t_i\!+\!\Delta t)^*\right]
    \end{array}\right] \nonumber \\
    &=& \frac{1}{2} 
    \left[\begin{array}{cc}
        \mathsfbi{A} & \mathsfbi{A} \\
        \mathsfbi{A} & \mathsfbi{A}
    \end{array}\right] + \frac{1}{2} 
    \left[\begin{array}{cc}
        \mathsfbi{0}   & \mathsfbi{R} \\
        \mathsfbi{R}^* & \mathsfbi{R}+\mathsfbi{R}^*
    \end{array}\right]\Delta t + O(\Delta t^2)\,,
\end{eqnarray}
where 
\begin{equation}
\label{eq:AppDefMats}
    \mathsfbi{A} = \mathbb{E}\left[\boldsymbol{a}(t_i)\,\boldsymbol{a}(t_i)^*\right], \quad
    \mathsfbi{R} = \mathbb{E}\left[\boldsymbol{a}(t_i)\,\boldsymbol{\sigma}(t_i)^*\right].
\end{equation}
The matrix $\mathsfbi{A}$ is the covariance matrix used to extract the standard dominant mode $\hat{v}^\mathrm{D}$ defined as (\ref{eq:DominantSingleMode}). It admits the unitary diagonalisation 
\begin{equation}
\label{eq:AppDiagA}
    \mathsfbi{A}=\mathsfbi{P}\,\mathsfbi{D}\,\mathsfbi{P}^*, 
\end{equation}
where $\mathsfbi{D}=\mathrm{diag}\left[d_1\ \ d_2\ \cdots \ d_{N_\mathrm{OS}} \right]$ is a diagonal matrix containing the positive eigenvalues $d_1 \geq d_2 \geq \cdots \geq d_{N_\mathrm{OS}}>0$, and $\mathsfbi{P}=\left[\boldsymbol{p}_1\ \ \boldsymbol{p}_2\ \cdots \ \boldsymbol{p}_{N_\mathrm{OS}} \right]$ is a unitary matrix that assembles the orthonormal eigenvectors $\boldsymbol{p}_1, \boldsymbol{p}_2, \cdots, \boldsymbol{p}_{N_\mathrm{OS}}$. The eigenvector $\boldsymbol{p}_1$ corresponds to the coefficient vector $\boldsymbol{\kappa}_{\mathrm{OS}}^\mathrm{D}$ of $\hat{v}^\mathrm{D}$ via 
\begin{equation}
\label{eq:AppStandardDominantMode}
    \boldsymbol{p}_1 = \mathsfbi{M}^{1/2} \boldsymbol{\kappa}_{\mathrm{OS}}^\mathrm{D}\,,
\end{equation}
and the eigenvalue $d_1$ satisfies
\begin{equation}
\label{eq:AppMaxEigValue}
    d_1 = \mathbb{E}\left[\,\left|\,\boldsymbol{\kappa}_\mathrm{OS}^*(t_i)\,\mathsfbi{M}\,\boldsymbol{\kappa}_\mathrm{OS}^\mathrm{D}\,\right|^2\,\right].
\end{equation}
Our proof is provided for the case with $d_1>d_2$; the degenerate case with $d_1=d_2$ follows by an analogous argument and is omitted for brevity. 

Defining the unitary matrix
\begin{equation}
\label{eq:DefUnitaryMatU}
    \mathsfbi{U} = \frac{1}{\sqrt{2}}
    \left[\begin{array}{cr}
        \mathsfbi{P} & \mathsfbi{P} \\
        \mathsfbi{P} & -\mathsfbi{P}
    \end{array}\right],
\end{equation}
we make the unitary similarity transformation of $\mathsfbi{K}$:
\begin{equation}
\label{eq:AppDefKtildta}
    \Tilde{\mathsfbi{K}} = \mathsfbi{U}^*\mathsfbi{K}\,\mathsfbi{U} = \Tilde{\mathsfbi{K}}^{(0)} + \Tilde{\mathsfbi{K}}^{(1)}
    \Delta t + O(\Delta t^2)\,,
\end{equation}
where
\begin{equation}
\label{eq:AppDefTildeKandS}
    \Tilde{\mathsfbi{K}}^{(0)} = 
    \left[\begin{array}{cc}
        \mathsfbi{D} & \mathsfbi{0} \\
        \mathsfbi{0} & \mathsfbi{0}
    \end{array}\right], \quad 
    \Tilde{\mathsfbi{K}}^{(1)} = \frac{1}{2} 
    \left[\begin{array}{cc}
        \mathsfbi{S}+\mathsfbi{S}^*   & -\mathsfbi{S} \\
        -\mathsfbi{S}^* & \mathsfbi{0}
    \end{array}\right], \quad 
    \mathsfbi{S} = \mathsfbi{P}^*\mathsfbi{R}\,\mathsfbi{P}\,.
\end{equation}
The eigenvalue problem for $\Tilde{\mathsfbi{K}}$ can be solved using asymptotic expansion. Specifically, the results of the largest eigenvalue $\lambda_1$ and its associated eigenvector $\boldsymbol{\xi}_1$ are
\begin{subeqnarray}
\label{eq:AppEigJoint}
    \lambda_1 &=& d_1 + \mathrm{Re}(\boldsymbol{e}_1^*\mathsfbi{S}\,\boldsymbol{e}_1^{\vphantom{*}})\Delta t + O(\Delta t^2)\,, \\
    \boldsymbol{\xi}_1 &=& \left[\begin{array}{l}
        \boldsymbol{e}_1^{\vphantom{*}}  \\
        \boldsymbol{0} 
    \end{array}\right] + \frac{1}{2} \left[\begin{array}{c}
        \mathsfbi{W}(\mathsfbi{S}\!+\!\mathsfbi{S}^*)\,\boldsymbol{e}_1^{\vphantom{*}} \\
        -\frac{1}{d_1}\mathsfbi{S}^*\boldsymbol{e}_1^{\vphantom{*}}
    \end{array}\right]\Delta t + O(\Delta t^2)\,,
\end{subeqnarray}
respectively, where $\boldsymbol{e}_1^{\vphantom{*}}=\left[\,1\ \ 0 \ \ 0 \ \cdots \ \ 0\,\right]^\mathrm{T}\in\mathbb{C}^{N_\mathrm{OS}}$ and 
\begin{equation}
\label{eq:AppDefWmat}
    \mathsfbi{W} = \mathrm{diag}\left[\,0\quad \frac{1}{d_1-d_2}\quad\frac{1}{d_1-d_3}\ \cdots \ \frac{1}{d_1-d_{N_\mathrm{OS}}}\,\right]\in\mathbb{C}^{N_\mathrm{OS}\times N_\mathrm{OS}}.
\end{equation}
Therefore, the DJM with a unit norm under the inner product (\ref{eq:augmentedInnerProduct}) is 
\begin{equation}
\label{eq:AppDominantJointModeA}
    \left[\begin{array}{c}
        \boldsymbol{a}_\alpha^\mathrm{D}  \\
        \boldsymbol{a}_\beta^\mathrm{D} 
    \end{array}\right] = \sqrt{2}\, \mathsfbi{U}\,\boldsymbol{\xi}_1 = 
    \left[\begin{array}{c}
        \boldsymbol{p}_1^{\vphantom{*}} \\
        \boldsymbol{p}_1^{\vphantom{*}}
    \end{array}\right] + \frac{1}{2} 
    \left[\begin{array}{c}
        \mathsfbi{P} \left(\mathsfbi{W}(\mathsfbi{S}\!+\!\mathsfbi{S}^*) -\frac{1}{d_1}\mathsfbi{S}^*\right)\boldsymbol{e}_1^{\vphantom{*}} \\
        \mathsfbi{P} \left(\mathsfbi{W}(\mathsfbi{S}\!+\!\mathsfbi{S}^*) +\frac{1}{d_1}\mathsfbi{S}^*\right)\boldsymbol{e}_1^{\vphantom{*}}
    \end{array}\right]\Delta t + O(\Delta t^2)\,.
\end{equation}

With (\ref{eq:AppDominantJointModeA}), the coefficient vector of $\hat{v}_\beta^\mathrm{D} - e^{\mathcal{L}\Delta t}\hat{v}_\alpha^\mathrm{D}$ is thus 
\begin{eqnarray}
\label{eq:AppLinStateCoef}
    \boldsymbol{\kappa}_{\mathrm{OS}\beta}^\mathrm{D} - e^{-\mathrm{i}k_x\mathsfbi{c}_\mathrm{OS}^{\vphantom{*}}\Delta t} \! \boldsymbol{\kappa}_{\mathrm{OS}\alpha}^\mathrm{D}  \!\!\!\! & = & \!\!\!\! \mathsfbi{M}^{-1/2}   \boldsymbol{a}_\beta^\mathrm{D} - e^{-\mathrm{i}k_x\mathsfbi{c}_\mathrm{OS}^{\vphantom{*}}\Delta t} \mathsfbi{M}^{-1/2}   \boldsymbol{a}_\alpha^\mathrm{D} 
     \nonumber \\
    & = & \!\!\!\! \mathsfbi{M}^{-1/2}   \boldsymbol{a}_\beta^\mathrm{D} - \left( \mathsfbi{I} \!-\!\mathrm{i}k_x\mathsfbi{c}_\mathrm{OS}^{\vphantom{*}}\Delta t \right) \!\mathsfbi{M}^{-1/2}   \boldsymbol{a}_\alpha^\mathrm{D} + O(\Delta t^2) \nonumber \\
    & = & \!\!\!\! \left(\frac{1}{d_1}\mathsfbi{M}^{-1/2}\mathsfbi{P}\,  \mathsfbi{S}^*\boldsymbol{e}_1^{\vphantom{*}} \!+ \mathrm{i}k_x\mathsfbi{c}_\mathrm{OS}^{\vphantom{*}}\mathsfbi{M}^{-1/2}\boldsymbol{p}_1^{\vphantom{*}}\right)\Delta t + O(\Delta t^2)\,.\quad
\end{eqnarray}
Using (\ref{eq:DefNewCoefs}), the matrix $\mathsfbi{S}$ defined in (\ref{eq:AppDefTildeKandS}) can be rewritten as
\begin{eqnarray}
\label{eq:AppReWriteMatS}
    \mathsfbi{S} \!\!\!\! & = & \!\!\!\! \mathsfbi{P}^*\, \mathbb{E}\left[\boldsymbol{a}(t_i)\,\boldsymbol{\sigma}(t_i)^*\right] \mathsfbi{P} \nonumber \\
    & = & \!\!\!\! \mathsfbi{P}^*\mathsfbi{M}^{1/2}\,\mathbb{E}\left[\boldsymbol{\kappa}_\mathrm{OS}^{\vphantom{*}}(t_i)\left(\,\boldsymbol{\chi}_\mathrm{OS}^{\vphantom{*}}(t_i)-\mathrm{i}k_x\mathsfbi{c}_\mathrm{OS}^{\vphantom{*}}\boldsymbol{\kappa}(t_i)\,\right)^*\right]\mathsfbi{M}^{1/2}\mathsfbi{P} \nonumber \\
    & = & \!\!\!\! \mathsfbi{P}^*\mathsfbi{M}^{1/2}\!\left\{\,\mathbb{E}\left[\boldsymbol{\kappa}_\mathrm{OS}^{\vphantom{*}}(t_i)\,\boldsymbol{\chi}_\mathrm{OS}^*(t_i)\right] + \mathrm{i}k_x \mathbb{E}\left[\boldsymbol{\kappa}_\mathrm{OS}^{\vphantom{*}}(t_i)\,\boldsymbol{\kappa}_\mathrm{OS}^*(t_i)\right]\!\mathsfbi{c}_\mathrm{OS}^*\,\right\}\!\mathsfbi{M}^{1/2}\mathsfbi{P} \nonumber \\
    & = & \!\!\!\! \mathsfbi{P}^*\mathsfbi{M}^{1/2}\,\mathbb{E}\left[\boldsymbol{\kappa}_\mathrm{OS}^{\vphantom{*}}(t_i)\,\boldsymbol{\chi}_\mathrm{OS}^*(t_i)\right]\!\mathsfbi{M}^{1/2}\mathsfbi{P} + \mathrm{i}k_x \mathsfbi{D}\,\mathsfbi{P}^*\mathsfbi{M}^{-1/2}\mathsfbi{c}_\mathrm{OS}^*\mathsfbi{M}^{1/2}\mathsfbi{P}\,.
\end{eqnarray}
Substituting (\ref{eq:AppReWriteMatS}) into (\ref{eq:AppLinStateCoef}) and employing (\ref{eq:AppStandardDominantMode}-\ref{eq:AppMaxEigValue}), we finally obtain
\begin{eqnarray}
\label{eq:AppLinStateCoefFinal}
    \boldsymbol{\kappa}_{\mathrm{OS}\beta}^\mathrm{D} - e^{-\mathrm{i}k_x\mathsfbi{c}_\mathrm{OS}^{\vphantom{*}}\Delta t} \! \boldsymbol{\kappa}_{\mathrm{OS}\alpha}^\mathrm{D}  \!\!\!\! & = & \!\!\!\! \frac{1}{d_1}\mathbb{E}\left[\boldsymbol{\chi}_\mathrm{OS}^{\vphantom{*}}(t_i)\,\boldsymbol{\kappa}_\mathrm{OS}^*(t_i)\right] \mathsfbi{M}^{1/2}\boldsymbol{p}_1^{\vphantom{*}} \Delta t + O(\Delta t^2)\,.\quad \nonumber \\
    & = & \!\!\!\! \frac{\mathbb{E}\left[\boldsymbol{\chi}_\mathrm{OS}^{\vphantom{*}}(t_i)\,\boldsymbol{\kappa}_\mathrm{OS}^*(t_i)\,\mathsfbi{M}\,\boldsymbol{\kappa}_\mathrm{OS}^\mathrm{D}\right]}{\mathbb{E}\left[\,\left|\,\boldsymbol{\kappa}_\mathrm{OS}^*(t_i)\,\mathsfbi{M}\,\boldsymbol{\kappa}_\mathrm{OS}^\mathrm{D}\,\right|^2\,\right]}\Delta t + O(\Delta t^2)\,.
\end{eqnarray}

\end{appen}

\bibliographystyle{jfm}
\bibliography{jfm2}

@Article{Orr1907,
  author  = {W. M'F. Orr},
  title   = {The stability or instability of the steady motions of a perfect liquid and of a viscous liquid. Part I: A perfect liquid},
  journal = {Proc. R. Irish Acad. A},
  year    = {1907},
  volume  = {27},
  pages   = {9--68}
}

@article{Reynolds1967,
  title     = {Stability of turbulent channel flow, with application to Malkus's theory},
  author    = {Reynolds, WC and Tiederman, WG},
  journal   = {Journal of Fluid Mechanics},
  volume    = {27},
  number    = {2},
  pages     = {253--272},
  year      = {1967},
  publisher ={Cambridge University Press}
}

@Book{Drazin1981,
  author    = {P. G. Drazin and W. H. Reid},
  title     = {Hydrodynamic Stability},
  publisher = {Cambridge University Press},
  address   = {Cambridge},
  year      = {1981}
}

@Article{Kim1987,
  author  = {J. Kim and P. Moin and R. Moser},
  title   = {Turbulence statistics in fully developed channel flow at low Reynolds number},
  journal = {J. Fluid Mech.},
  year    = {1987},
  volume  = {177},
  pages   = {133--166}
}

@Article{Nagata1990,
  author  = {M. Nagata},
  title   = {Three-dimensional finite-amplitude solutions in plane Couette flow: bifurcation from infinity},
  journal = {J. Fluid Mech.},
  year    = {1990},
  volume  = {217},
  pages   = {519--527}
}

@Article{Berkooz1993,
  author  = {G. Berkooz and P. Holmes and J. L. Lumley},
  title   = {The proper orthogonal decomposition in the analysis of turbulent flows},
  journal = {Annu. Rev. Fluid Mech.},
  year    = {1993},
  volume  = {25},
  number  = {1},
  pages   = {539--575}
}

@Article{Butler1993,
  author  = {K. M. Butler and B. F. Farrell},
  title   = {Optimal perturbations and streak spacing in wall-bounded turbulent shear flow},
  journal = {Phys. Fluids A},
  year    = {1993},
  volume  = {5},
  number  = {3},
  pages   = {774--777}
}

@Article{Jimenez1994,
  author  = {J. Jim{\'e}nez},
  title   = {On the structure and control of near wall turbulence},
  journal = {Phys. Fluids},
  year    = {1994},
  volume  = {6},
  number  = {2},
  pages   = {944--953}
}

@Article{Hamilton1995,
  author  = {J. M. Hamilton and J. Kim and F. Waleffe},
  title   = {Regeneration mechanisms of near-wall turbulence structures},
  journal = {J. Fluid Mech.},
  year    = {1995},
  volume  = {287},
  pages   = {317--348}
}

@Article{Waleffe1997,
  author  = {F. Waleffe},
  title   = {On a self-sustaining process in shear flows},
  journal = {Phys. Fluids},
  year    = {1997},
  volume  = {9},
  number  = {4},
  pages   = {883--900}
}

@Article{Andersson2001,
  author  = {P. Andersson and L. Brandt and A. Bottaro and D. S. Henningson},
  title   = {On the breakdown of boundary layer streaks},
  journal = {J. Fluid Mech.},
  year    = {2001},
  volume  = {428},
  pages   = {29--60}
}

@Article{Jimenez2001,
  author  = {J. Jim{\'e}nez and M. P. Simens},
  title   = {Low-dimensional dynamics of a turbulent wall flow},
  journal = {J. Fluid Mech.},
  year    = {2001},
  volume  = {435},
  pages   = {81--91}
}

@Article{Kawahara2001,
  author  = {G. Kawahara and S. Kida},
  title   = {Periodic motion embedded in plane Couette turbulence: regeneration cycle and burst},
  journal = {J. Fluid Mech.},
  year    = {2001},
  volume  = {449},
  pages   = {291--300}
}

@Book{Schmid2001,
  author    = {P. J. Schmid and D. S. Henningson},
  title     = {Stability and Transition in Shear Flows},
  series    = {Applied Mathematical Sciences},
  volume    = {142},
  publisher = {Springer},
  address   = {New York, NY},
  year      = {2001},
  doi       = {10.1007/978-1-4613-0185-1}
}

@Article{Waleffe2001,
  author  = {F. Waleffe},
  title   = {Exact coherent structures in channel flow},
  journal = {J. Fluid Mech.},
  year    = {2001},
  volume  = {435},
  pages   = {93--102}
}

@Article{Schoppa2002,
  author  = {W. Schoppa and F. Hussain},
  title   = {Coherent structure generation in near-wall turbulence},
  journal = {J. Fluid Mech.},
  year    = {2002},
  volume  = {453},
  pages   = {57--108}
}

@article{Jimenez2005,
  title     ={Characterization of near-wall turbulence in terms of equilibrium and “bursting” solutions},
  author    = {Jim{\'e}nez, Javier and Kawahara, Genta and Simens, Mark P and Nagata, Masato and Shiba, Makoto},
  journal   = {Physics of Fluids},
  volume    = {17},
  number    = {1},
  year      = {2005},
  publisher = {AIP Publishing}
}

@Article{delAlamo2006,
  author  = {J. C. {del {\'A}lamo} and J. Jim{\'e}nez},
  title   = {Linear energy amplification in turbulent channels},
  journal = {J. Fluid Mech.},
  year    = {2006},
  volume  = {559},
  pages   = {205--213}
}

@article{delAlamo2009,
  title     = {Estimation of turbulent convection velocities and corrections to Taylor's approximation},
  author    = {J. C. {del {\'A}lamo} and J. Jim{\'e}nez},
  journal   = {Journal of Fluid Mechanics},
  volume    = {640},
  pages     = {5--26},
  year      = {2009},
  publisher = {Cambridge University Press}
}

@Article{Hwang2010,
  author  = {Y. Hwang and C. Cossu},
  title   = {Linear non-normal energy amplification of harmonic and stochastic forcing in the turbulent channel flow},
  journal = {J. Fluid Mech.},
  year    = {2010},
  volume  = {664},
  pages   = {51--73}
}

@Article{Mckeon2010,
  author  = {B. J. McKeon and A. S. Sharma},
  title   = {A critical-layer framework for turbulent pipe flow},
  journal = {J. Fluid Mech.},
  year    = {2010},
  volume  = {658},
  pages   = {336--382}
}

@Article{Farrell2012,
  author  = {B. F. Farrell and P. J. Ioannou},
  title   = {Dynamics of streamwise rolls and streaks in turbulent wall-bounded shear flow},
  journal = {J. Fluid Mech.},
  year    = {2012},
  volume  = {708},
  pages   = {149--196}
}

@Article{Jimenez2013,
  author  = {J. Jim{\'e}nez},
  title   = {How linear is wall-bounded turbulence?},
  journal = {Phys. Fluids},
  year    = {2013},
  volume  = {25},
  number  = {11},
  pages   = {110814},
  doi     = {10.1063/1.4819081}
}

@Article{Moarref2013,
  author  = {R. Moarref and A. S. Sharma and J. A. Tropp and B. J. McKeon},
  title   = {Model-based scaling of the streamwise energy density in high-Reynolds-number turbulent channels},
  journal = {J. Fluid Mech.},
  year    = {2013},
  volume  = {734},
  pages   = {275--316}
}

@Article{Thomas2014,
  author  = {V. L. Thomas and B. K. Lieu and M. R. Jovanovi{\'c} and B. F. Farrell and P. J. Ioannou and D. F. Gayme},
  title   = {Self-sustaining turbulence in a restricted nonlinear model of plane Couette flow},
  journal = {Phys. Fluids},
  year    = {2014},
  volume  = {26},
  number  = {10},
  pages   = {105112}
}

@Article{Jimenez2015,
  author  = {J. Jim{\'e}nez},
  title   = {Direct detection of linearized bursts in turbulence},
  journal = {Phys. Fluids},
  year    = {2015},
  volume  = {27},
  number  = {6},
  pages   = {065102},
  doi     = {10.1063/1.4921748}
}

@Article{Farrell2016,
  author  = {B. F. Farrell and P. J. Ioannou and J. Jim{\'e}nez and N. C. Constantinou and A. Lozano-Dur{\'a}n and M.-A. Nikolaidis},
  title   = {A statistical state dynamics-based study of the structure and mechanism of large-scale motions in plane Poiseuille flow},
  journal = {J. Fluid Mech.},
  year    = {2016},
  volume  = {809},
  pages   = {290--315}
}

@Article{Marston2016,
  author  = {J. B. Marston and G. P. Chini and S. M. Tobias},
  title   = {Generalized quasilinear approximation: application to zonal jets},
  journal = {Phys. Rev. Lett.},
  year    = {2016},
  volume  = {116},
  number  = {21},
  pages   = {214501}
}

@Article{Mckeon2017,
  author  = {B. J. McKeon},
  title   = {The engine behind (wall) turbulence: perspectives on scale interactions},
  journal = {J. Fluid Mech.},
  year    = {2017},
  volume  = {817},
  pages   = {P1}
}

@article{Jimenez2018a,
  title     = {Coherent structures in wall-bounded turbulence},
  author    = {J. Jim{\'e}nez},
  journal   = {J. Fluid Mech.},
  volume    = {842},
  pages     = {P1},
  year      = {2018},
  publisher = {Cambridge University Press}
}

@Article{Jimenez2018b,
  author  = {J. Jim{\'e}nez},
  title   = {Machine-aided turbulence theory},
  journal = {J. Fluid Mech.},
  year    = {2018},
  volume  = {854},
  pages   = {R1}
}

@Article{Encinar2020,
  author  = {M. P. Encinar and J. Jim{\'e}nez},
  title   = {Momentum transfer by linearised eddies in turbulent channel flows},
  journal = {J. Fluid Mech.},
  year    = {2020},
  volume  = {895},
  pages   = {A23}
}

@Article{Jimenez2020,
  author  = {J. Jim{\'e}nez},
  title   = {Monte Carlo science},
  journal = {J. Turbul.},
  year    = {2020},
  volume  = {21},
  number  = {9-10},
  pages   = {544--566}
}

@Article{Bae2021,
  author  = {H. J. Bae and A. Lozano-Dur{\'a}n and B. J. McKeon},
  title   = {Nonlinear mechanism of the self-sustaining process in the buffer and logarithmic layer of wall-bounded flows},
  journal = {J. Fluid Mech.},
  year    = {2021},
  volume  = {914},
  pages   = {A3}
}

@Article{Lozano2021,
  author  = {A. Lozano-Dur{\'a}n and N. C. Constantinou and M.-A. Nikolaidis and M. Karp},
  title   = {Cause-and-effect of linear mechanisms sustaining wall turbulence},
  journal = {J. Fluid Mech.},
  year    = {2021},
  volume  = {914},
  pages   = {A8}
}

@Article{Hernandez2022a,
  author  = {C. G. Hern{\'a}ndez and Q. Yang and Y. Hwang},
  title   = {Generalised quasilinear approximations of turbulent channel flow. Part 1. Streamwise nonlinear energy transfer},
  journal = {J. Fluid Mech.},
  year    = {2022},
  volume  = {936},
  pages   = {A33}
}

@Article{Hernandez2022b,
  author  = {C. G. Hern{\'a}ndez and Q. Yang and Y. Hwang},
  title   = {Generalised quasilinear approximations of turbulent channel flow. Part 2. Spanwise triadic scale interactions},
  journal = {J. Fluid Mech.},
  year    = {2022},
  volume  = {944},
  pages   = {A34}
}

@Article{Jimenez2022,
  author  = {J. Jim{\'e}nez},
  title   = {The streaks of wall-bounded turbulence need not be long},
  journal = {J. Fluid Mech.},
  year    = {2022},
  volume  = {945},
  pages   = {R3}
}

@Article{Encinar2023,
  author  = {M. P. Encinar and J. Jim{\'e}nez},
  title   = {Identifying causally significant features in three-dimensional isotropic turbulence},
  journal = {J. Fluid Mech.},
  year    = {2023},
  volume  = {965},
  pages   = {A20}
}

@Article{Jimenez2023,
  author  = {J. Jim{\'e}nez},
  title   = {A Perron--Frobenius analysis of wall-bounded turbulence},
  journal = {J. Fluid Mech.},
  year    = {2023},
  volume  = {968},
  pages   = {A10}
}

@Article{Maeyama2023,
  author  = {H. Maeyama and S. Kawai},
  title   = {Near-wall numerical coherent structures and turbulence generation in wall-modelled large-eddy simulation},
  journal = {J. Fluid Mech.},
  year    = {2023},
  volume  = {969},
  pages   = {A29}
}

@InProceedings{Martinez2023,
  author       = {C. Mart{\'i}nez-L{\'o}pez and O. Flores and J. Jim{\'e}nez},
  title        = {The shortest channels that sustain short-streak turbulence},
  booktitle    = {Division of Fluid Dynamics Annual Meeting 2023},
  year         = {2023},
  organization = {APS}
}

@Article{Ballouz2024,
  author  = {E. Ballouz and B. L{\'o}pez-D{\'o}riga and S. T. M. Dawson and H. J. Bae},
  title   = {Wavelet-based resolvent analysis of non-stationary flows},
  journal = {J. Fluid Mech.},
  year    = {2024},
  volume  = {999},
  pages   = {A53}
}

@Article{Osawa2024,
  author  = {K. Osawa and J. Jim{\'e}nez},
  title   = {Causal features in turbulent channel flow},
  journal = {J. Fluid Mech.},
  year    = {2024},
  volume  = {1000},
  pages   = {A4}
}

@Article{Young2024,
  author  = {J. D. S. Young and Z. Hao and R. Garc{\'i}a-Mayoral},
  title   = {Inter-scale causality relations in wall turbulence},
  journal = {J. Phys.: Conf. Ser.},
  year    = {2024},
  volume  = {2753},
  number  = {1},
  pages   = {012019}
}

@InProceedings{Jimenez2025,
  author       = {J. Jim{\'e}nez and C. Mart{\'i}nez-L{\'o}pez},
  title        = {The regeneration of long streaks in wall-bounded flows},
  booktitle    = {Division of Fluid Dynamics Annual Meeting 2025},
  year         = {2025},
  organization = {APS},
  note         = {Available at \url{https://meetings-archive.aps.org/dfd/2025/a17/6/}}
}

@InProceedings{Martinez2025,
  author       = {C. Mart{\'i}nez-L{\'o}pez and O. Flores and J. Jim{\'e}nez},
  title        = {Streak meandering without streaks},
  booktitle    = {Division of Fluid Dynamics Annual Meeting 2025},
  year         = {2025},
  organization = {APS},
  note         = {Available at \url{https://meetings-archive.aps.org/dfd/2025/c21/14/}}
}

@InProceedings{Nishimoto2025,
  author       = {M. K. K. Nishimoto and M.-L. Tsai and S. T. M. Dawson and H. J. Bae},
  title        = {Causal analysis of turbulent Couette-Poiseuille flow using wavelet-based resolvent analysis},
  booktitle    = {Division of Fluid Dynamics Annual Meeting 2025},
  year         = {2025},
  organization = {APS},
  note         = {Available at \url{https://meetings-archive.aps.org/dfd/2025/k25/3/}}
}



\end{document}